Universidade Federal do Amazonas – UFAM
Instituto de Computação - IComp
Programa de Pós-Graduação em Informática – PPGI

# Mapeamento Sistemático

Marco Aurélio dos Santos, Raimundo da Silva Barreto.
marrco.santos@gmail.com, Rbarreto@icomp.ufam.edu.br

**Linha de Pesquisa**
Sistemas Embarcados

# 1. Introdução

Um mapeamento sistemático é uma forma de identificar, avaliar e interpretar todas as pesquisas disponíveis relevantes para uma questão de pesquisa particular. Umas das razões para a realização de revisões sistemáticas é que esta resume as evidências existentes em relação a um tratamento ou tecnologia [KITCHENHAM, 2004].

# 2. Protocolo do Mapeamento

Kitchenham (2004) diz que o protocolo de mapeamento especifica os métodos que serão usados para realizar um mapeamento sistemático específico, fazendo que este diminua a possibilidade de viés do pesquisador.

## 2.1. Objetivo

A descrição do objetivo se encontra descrita segundo o paradigma GQM (Goal-Question-Metric, e pode ser observado na Tabela 1:

**Tabela 1 GQM**

| ANALISAR | |
|---|---|
| **Com o propósito de** | Identificar/Caracterizar |
| **Em relação a** | Identificar as técnicas/abordagens/métodos para localização indoor |
| **Do ponto de vista dos** | Pesquisadores em Sistemas Embarcados |
| **No contexto** | Sistemas Embarcados |

## 2.2. Questão de Pesquisa

Como é possível definir a posição em um ambiente INDOOR?

A questão de pesquisa pode ser refletida através da representação do PICO, descrita na tabela 2 abaixo:

**Tabela 2.** PICO.

| ANALISAR | |
|---|---|
| **População** | Localização INDOOR |
| **Intervenção** | Técnicas/abordagens/métodos para localização INDOOR |
| **Controle** | N/A |
| **Outputs(Resultados)** | Definir uma abordagem para localização INDOOR |

## 2.3. Estratégia utilizada para pesquisa dos estudos primários

Nesta seção serão descritos: o escopo da pesquisa, o idioma considerado, os termos utilizados, a string de busca e os critérios de seleção de artigos.

### 2.3.1. Escopo da Pesquisa

A fonte considerada no escopo desta pesquisa foi adotada com base no portal de periódicos da CAPES. A pesquisa dar-se-á em duas bibliotecas digitais: IEEE Xplore e Scopus.

### 2.3.2. Idiomas dos Artigos

O idioma escolhido foi o Inglês e português, pois são adotados pela grande maioria das conferências e periódicos nacionais e internacionais relacionados com tema de pesquisa.

## 2.4. Termos utilizados na pesquisa (palavras-chave)

Os termos que foram utilizados neste mapeamento foram agrupados em dois grupos que quando combinados entre si formam as strings de busca. Os termos estão descritos na língua inglesa, por esta ser o idioma utilizado pelas máquinas de buscas. Na tabela 3, são mostrados os termos de busca utilizados para esta pesquisa.

**Tabela 3. Termos utilizados**.

| Grupo 1 | Grupo 2 |
|---|---|
| Indoor localization systems | Techniques for indoor localization |
| Indoor Positioning System | Techniques for indoor localizations |
| Indoor Location Positioning | Indoor localization methods |
| Indoor navigation system | Approach for Indoor Localization |

### 2.4.1. String de Busca

((("Indoor localization systems" OR "Indoor Positioning System" OR "Indoor Location Positioning “OR "indoor navigation system" OR "Local Positioning System") **AND** ("techniques for indoor localization" OR "techniques for indoor localizations" OR "Indoor Localization Methods" OR "Approach for Indoor Localization”)))

## 2.5. Critérios de Seleção de Artigos e Procedimentos

Kitchenham (2004) diz que devem ser seguidos critérios de inclusão e exclusão para os artigos que são retornados pela string de busca. Sendo assim, foram definidos os seguintes critérios:

### 2.5.1. Critérios para Inclusão de Artigos

Os critérios de Inclusão são:

- **CI1.** Podem ser selecionadas publicações que apresentam técnicas ou métodos aplicados a Localização Indoor;
- **CI2.** Podem ser selecionadas publicações que descrevam uma customização de técnicas ou métodos aplicados a Localização Indoor;
- **CI3.** Podem ser selecionadas publicações que descrevam uma customização de técnicas ou métodos aplicados a Localização Indoor, incluindo apoio ferramental;

### 2.5.2. Critérios para Exclusão de Artigos

Os critérios de Exclusão são:

- **CE1.** Não serão selecionadas publicações que não satisfaçam a nenhum critério de inclusão;
- **CE2.** Não serão selecionadas publicações em que o idioma seja diferente do exigido;
- **CE3.** Não serão selecionadas publicações de artigos duplicados;
- **CE4.** Não serão selecionadas publicações que não apresentem população de Localização Indoor;
- **CE5.** Não serão selecionadas publicações que não tenha disponibilidade de conteúdo para leitura e análise dos dados (especialmente em casos, onde os estudos são pagos ou não disponibilizados pelas máquinas de buscas);
- **CE6.** Não serão selecionadas publicações em que o conteúdo disponha apenas conceitos.

### 2.5.3. Processo de Seleção Preliminar (1° Filtro)

Serão selecionados artigos que apresentem informações no título e no abstract relacionado à questão de pesquisa principal. Para cada estudo incluído ou excluído será apresentado um critério (Inclusão ou Exclusão).

### 2.5.4. Processo de Seleção Final (2° Filtro)

Como a leitura de duas informações (título e abstract) não é suficiente para identificar se o estudo é realmente relevante para a pesquisa realizada, torna-se necessário realizar a leitura completa dos estudos que restaram do 1° filtro. Dessa forma, esta fase do mapeamento, tem como objetivo fazer uma análise mais apurada dos estudos, identificando e extraindo dados também de acordo com os critérios de inclusão e exclusão descritos anteriormente. Para cada estudo incluído ou excluído será apresentado um critério (Inclusão ou Exclusão).

# 3. Lista de Artigos Encontrados

Nesta seção são apresentados os artigos que foram selecionados pelo 1° Filtro e 2° Filtro.

## 3.1. Lista de artigos selecionados no 1° Filtro na IEEE Xplore

Na biblioteca digital IEEE Xplore, houve um retorno de 24 artigos utilizando a string de busca apresentada na seção 2, subseção 2.4.1. Destes artigos retornados, somente 20 artigos foram selecionados no primeiro filtro, os quais são listados abaixo na tabela 4 com os seus respectivos critérios de inclusão.

**Tabela 4. Artigos selecionados no 1º filtro na IEEE Xplore.**

| Nº | Nome do artigo | Critério |
|---|---|---|
| 01 | Ullah, K. and Custodio, I.V. and Shah, N. and dos Santos Moreira, E. An Experimental Study on the Behavior of Received Signal Strength in Indoor Environment. Frontiers of Information Technology (FIT), 2013 11th International Conference on. 2013 | **CI1** |
| 02 | Agrawal, L. and Toshniwal, D. Smart Phone Based Indoor Pedestrian Localization System. Computational Science and Its Applications (ICCSA), 2013 13th International Conference on. 2013 | **CI1. CI2.** |
| 03 | Vinyals, O. and Martin, E. and Friedland, G. Multimodal Indoor Localization: An Audio-Wireless-Based Approach. Semantic Computing (ICSC), 2010 IEEE Fourth International Conference on. 2010. | **CI1. CI2.** |
| 04 | Shuai Shao and Burkholder, R.J.Item-Level RFID Tag Location Sensing Utilizing Reader Antenna Spatial Diversity. Sensors Journal, IEEE. 2013 | **CI1. CI2.** |
| 05 | Bacak, A. and Celebi, H. Practical considerations for RSS RF fingerprinting based indoor localization systems. Signal Processing and Communications Applications Conference (SIU), 2014 22nd. 2014 | **CE5.** |
| 06 | Nazari Shirehjini, A.A. and Yassine, A. and Shirmohammadi, S. An RFID-Based Position and Orientation Measurement System for Mobile Objects in Intelligent Environments. 2012 | **CI2.** |
| 07 | Shanklin, T.A. and Loulier, B. and Matson, E.T. Embedded sensors for indoor positioning. Sensors Applications Symposium (SAS), 2011 IEEE. 2011 | **CI1. CI2.** |
| 08 | Yuhang Gao and Jianwei Niu and Ruogu Zhou and Guoliang Xing. ZiFind: Exploiting cross-technology interference signatures for energy-efficient indoor localization. INFOCOM, 2013 Proceedings IEEE. 2013 | **CE5.** |
| 09 | Ye Tian and Denby, B. and Ahriz, I. and Roussel, P. and Dubois, R. and Dreyfus, G. Practical indoor localization using ambient RF. Instrumentation and Measurement Technology Conference (I2MTC), 2013 IEEE International. 2013 | **CI1. CI2.** |
| 10 | Jianglong Liu and Bao Gen Xu and Yi He Wan and Si Long Tang and Xue Ke Ding and Qun Wan. A rang-free location method based on differential evolution algorithm. Cross Strait Quad-Regional Radio Science and Wireless Technology Conference (CSQRWC), 2013. 2013 | **CI1** |
| 11 | Segura, M.J. and Mut, V.A. and Patino, H.D. Wavelet correlation TOA estimation with dynamic threshold setting for IR-UWB localization system. Communications, 2009. LATINCOM '09. IEEE Latin-American Conference on. 2009 | **CE4.** |
| 12 | Zhi Zhang and Zhonghai Lu and Saakian, V. and Xing Qin and Qiang Chen and Li-Rong Zheng. Item-Level Indoor Localization With Passive UHF RFID Based on Tag Interaction Analysis. Industrial Electronics, IEEE Transactions on. 2014 | **CI1. CI2.** |
| 13 | Wei Shi-Sue and Shiuhpyng Shieh and Bing-Han Li and Cho, M.C.Y. and Chin-Wei Tien. A framework using fingerprinting for signal overlapping-based method in WLAN. Computer Symposium (ICS), 2010 International. 2010 | **CI1. CI2.** |
| 14 | Cinaz, B. and Kenn, Holger. HeadSLAM - simultaneous localization and mapping with head-mounted inertial and laser range sensors. Wearable Computers, 2008. ISWC 2008. 12th IEEE International Symposium on. 2008 | **CI1.** |
| 15 | Redzic, M.D. and Brennan, C. and O'Connor, N.E. SEAMLOC: Seamless Indoor Localization Based on Reduced Number of Calibration Points. Mobile Computing, IEEE Transactions on. 2014 | **CI1.** |
| 16 | Olowolayemo, A. and Md Tap, A.O. and Mantoro, T. Fuzzy Logic Based Compensated Wi-Fi Signal Strength for Indoor Positioning. Advanced Computer Science Applications and Technologies (ACSAT), 2013 International Conference on. 2013 | **CI2.** |
| 17 | Tabibiazar, A. and Basir, O. Compressive sensing indoor localization. Systems, Man, and Cybernetics (SMC), 2011 IEEE International Conference on. 2011 | **CI2.** |
| 18 | Lewandowski, A. and Wietfeld, C. A comprehensive approach for optimizing ToA-localization in harsh industrial environments. Position Location and Navigation Symposium (PLANS), 2010 IEEE/ION. 2010 | **CI1. CI2.** |
| 19 | Martin, E. Multimode radio fingerprinting for localization. Radio and Wireless Symposium (RWS), 2011 IEEE. 2011 | **CI1.** |
| 20 | Chenshu Wu and Zheng Yang and Yunhao Liu and Wei Xi. WILL: Wireless Indoor Localization without Site Survey. Parallel and Distributed Systems, IEEE Transactions on. 2013 | **CI1. CI2** |
| 21 | Xiaoyan Li and Martin, R.P. A simple ray-sector signal strength model for indoor 802.11 networks. Mobile Adhoc and Sensor Systems Conference, 2005. IEEE International Conference on. 2005 | **CI1.** |
| 22 | Luo, R.C. and Chen, O. and Pei Hsien Lin. Indoor robot/human localization using dynamic triangulation and wireless Pyroelectric Infrared sensory fusion approaches. Robotics and Automation (ICRA), 2012 IEEE International Conference on. 2012 | **CI1. CI2** |
| 23 | Chenshu Wu and Zheng Yang and Yunhao Liu and Wei Xi. WILL: Wireless indoor localization without site survey. INFOCOM, 2012 Proceedings IEEE. 2012 | **CE3.** |
| 24 | Wang, B. and Zhou, S. and Liu, W. and Mo, Y. Indoor Localization based on Curve Fitting and Location Search using Received Signal Strength. Industrial Electronics, IEEE Transactions on. 2014 | **CI1.** |

### 3.1.1. Lista de artigos removidos no 2° Filtro na IEEE Xplore

A abordagem adotada nesta fase foi a leitura completa do artigo e validação pelos critérios de inclusão/exclusão, e posteriormente a extração dos dados dos aceitos. Como artigos aprovados tiveram um total de 17 artigos, os quais se listam na tabela 6 os removidos e os seus respectivos critérios de exclusão.

**Tabela 5 - Artigos removidos no 2º filtro na IEEE Xplore.**

| Nº | Nome do artigo | Critério de Exclusão |
|---|---|---|
| 10 | Jianglong Liu and Bao Gen Xu and Yi He Wan and Si Long Tang and Xue Ke Ding and Qun Wan. A rang-free location method based on differential evolution algorithm. Cross Strait Quad-Regional Radio Science and Wireless Technology Conference (CSQRWC), 2013. 2013 | **CE6.** |
| 21 | Xiaoyan Li and Martin, R.P. A simple ray-sector signal strength model for indoor 802.11 networks. Mobile Adhoc and Sensor Systems Conference, 2005. IEEE International Conference on. 2005 | **CE6.** |
| 24 | Wang, B. and Zhou, S. and Liu, W. and Mo, Y. Indoor Localization based on Curve Fitting and Location Search using Received Signal Strength. Industrial Electronics, IEEE Transactions on. 2014 | **CE6.** |

## 3.2. Lista de artigos selecionados no 1° Filtro na Scopus

Na biblioteca digital Scopus, houve um retorno de 48 artigos utilizando a string de busca apresentada na seção 2, subseção 2.4.1. Destes artigos retornados, somente 35 artigos foram selecionados no primeiro filtro, os quais são listados abaixo na tabela 6 com os seus respectivos critérios de inclusão.

**Tabela 6. Artigos selecionados no 1º filtro na Scopus**

| Nº | Nome do artigo | Critério |
|---|---|---|
| 01 | Zhang, C. and Subbu, K.P. and Luo, J. and Wu, J. GROPING: Geomagnetism and crowdsensing powered indoor navigation. IEEE Transactions on Mobile Computing. 2015 | **CI1.** |
| 02 | Li, L. and Shen, G. and Zhao, C. and Moscibroda, T. and Lin, J.-H. and Zhao, F. Experiencing and handling the diversity in data density and environmental locality in an indoor positioning service. 2014 | **CI1.** |
| 03 | Mariakakis, A.T. and Sen, S. and Lee, J. and Kim, K.-H. SAIL: Single access point-based indoor localization. MobiSys 2014 - Proceedings of the 12th Annual International Conference on Mobile Systems, Applications, and Services. 2014 | **CI1. CI2.** |
| 04 | Aomumpai, S. and Kondee, K. and Prommak, C. and Kaemarungsi, K. Optimal placement of reference nodes for wireless indoor positioning systems. 2014 11th International Conference on Electrical Engineering/Electronics, Computer, Telecommunications and Information Technology, 2014 | **CI2.** |
| 05 | Zheng, Y. and Shen, G. and Li, L. and Zhao, C. and Li, M. and Zhao, F. Travi-Navi: Self-deployable indoor navigation system. Proceedings of the Annual International Conference on Mobile Computing and Networking, MOBICOM. 2014 | **CI1. CI2.** |
| 06 | Li, W. and Li, H. and Jiang, Y. A practical RF-based indoor localization system combined with the embedded sensors in smart phone. IWCMC 2014 - 10th International Wireless Communications and Mobile Computing Conference. 2014 | **CI2. CI3.** |
| 07 | Schatzberg, U. and Banin, L. and Amizur, Y. Enhanced WiFi ToF indoor positioning system with MEMS-based INS and pedometric information. Record - IEEE PLANS, Position Location and Navigation Symposium. 2014 | **CI1.** |
| 08 | Kim, D.Y. and Yi, K.Y., RSSI-based indoor localization method using virtually overlapped visible light. Transactions of the Korean Institute of Electrical Engineers. 2014 | **CE5.** |
| 09 | Oksar, I. A Bluetooth signal strength based indoor localization method. International Conference on Systems, Signals, and Image Processing. 2014 | **CI1. CI2.** |
| 10 | Bacak, A. and Celebi, H. Practical considerations for RSS RF fingerprinting based indoor localization. 2014 22nd Signal Processing and Communications Applications Conference, SIU 2014 – Proceedings. 2014 | **CE5.** |
| 11 | Sharma, P. and Chakraborty, D. and Banerjee, N. and Banerjee, D. and Agarwal, S.K. and Mittal, S. KARMA: Improving WiFi-based indoor localization with dynamic causality calibration. 2014 11th Annual IEEE International Conference on Sensing, Communication, and Networking, SECON 2014. 2014 | **CI1. CI2. CI3** |
| 12 | Subbu, K. and Zhang, C. and Luo, J. and Vasilakos, A. Analysis and status quo of smartphone-based indoor localization systems. IEEE Wireless Communications. 2014 | **CI1.** |
| 13 | Luo, C. and Hong, H. and Chan, M.C. PiLoc: A self-calibrating participatory indoor localization system. IPSN 2014 - Proceedings of the 13th International Symposium on Information Processing in Sensor Networks (Part of CPS Week). 2014 | **CI1. CI2.** |
| 14 | Xie, H. and Gu, T. and Tao, X. and Ye, H. and Lv, J. MaLoc: A practical magnetic fingerprinting approach to indoor localization using smartphones. UbiComp 2014 - Proceedings of the 2014 ACM International Joint Conference on Pervasive and Ubiquitous Computing. 2014 | **CE5.** |
| 15 | Guo, Y. and Yang, L. and Li, B. and Liu, T. and Liu, Y. RollCaller: User-friendly indoor navigation system using human-item spatial relation. Proceedings - IEEE INFOCOM. 2014 | **CI2. CI3.** |
| 16 | Robles, J.J. Indoor localization based on wireless sensor networks. AEU - International Journal of Electronics and Communications. 2014 | **CE5** |
| 17 | Kumar, S. and Gil, S. and Katabi, D. and Rus, D. Accurate indoor localization with zero start-up cost. Proceedings of the Annual International Conference on Mobile Computing and Networking, MOBICOM. 2014 | **CI1. CI2.** |
| 18 | Zhang, C. and Luo, J. and Wu, J. A dual-sensor enabled indoor localization system with crowdsensing spot survey. Proceedings - IEEE International Conference on Distributed Computing in Sensor Systems, DCOSS 2014. 2014 | **CI1. CI2. CI3.** |
| 19 | Gu, Y. and Liu, J. and Chen, Y. and Jiang, X. Constraint Online Sequential Extreme Learning Machine for lifelong indoor localization system. Proceedings of the International Joint Conference on Neural Networks. 2014 | **CI1.** |
| 20 | Tian, Y. and Denby, B. and Ahriz, I. and Roussel, P. and Dreyfus, G. Hybrid indoor localization using GSM fingerprints, embedded sensors and a particle filter, 2014 11th International Symposium on Wireless Communications Systems, ISWCS 2014 – Proceedings. 2014 | **CI1.** |
| 21 | Zhang, Z. and Lu, Z. and Saakian, V. and Qin, X. and Chen, Q. and Zheng, L.-R. Item-level indoor localization with passive UHF RFID based on tag interaction analysis. IEEE Transactions on Industrial Electronics. 2014 | **CE5.** |
| 22 | Nirjon, S. and Liu, J. and DeJean, G. and Priyantha, B. and Jin, Y. and Hart, T. COIN-GPS: Indoor localization from direct GPS receiving. MobiSys 2014 - Proceedings of the 12th Annual International Conference on Mobile Systems, Applications, and Services. 2014 | **CI1. CI2.** |
| 23 | Chandel, V. and Choudhury, A.D. and Ghose, A. and Bhaumik, C. AcTrak - Unobtrusive activity detection and step counting using smartphones. Lecture Notes of the Institute for Computer Sciences, Social-Informatics and Telecommunications Engineering, LNICST. 2014 | **CE5.** |
| 24 | GalvÃ¡n-Tejada, C.E. and GarcÃ-a-VÃ¡zquez, J.P. and Brena, R.F. Natural or generated signals for indoor location systems? An evaluation in terms of sensitivity and specificity. CONIELECOMP 2014 - 24th International Conference on Electronics, Communications and Computers. 2014 | **CI1. CI2.** |
| 25 | Safak, I. RFID-Based indoor localization using angle-of-arrival and return time [Kapali ortamlarda geliÅŸ aÃ§isi ve zamani kullanilarak rfid tabanli konum belirleme] 2014 22nd Signal Processing and Communications Applications Conference, SIU 2014 – Proceedings. 2014 | **CE2.** |
| 26 | Dang, C. and Sezaki, K. and Iwai, M. DECL: A circular inference method for indoor pedestrian localization using phone inertial sensors. 2014 7th International Conference on Mobile Computing and Ubiquitous Networking, ICMU 2014. 2014 | **CI2. CI3.** |
| 27 | Nakib, A. and Daachi, B. and Dakkak, M. and Siarry, P. Mobile tracking based on fractional integration. IEEE Transactions on Mobile Computing. 2014 | |
| 28 | Moreno-Cano, M.V. and Zamora-Izquierdo, M.A. and Santa, J. and Skarmeta, A.F. An indoor localization system based on artificial neural networks and particle filters applied to intelligent buildings. Neurocomputing. 2013 | **CE5.** |
| 29 | Bera, R. and Kirsch, N.J. and Fu, T.S. Using prior measurements to improve probabilistic-based indoor localization methods. Proceedings of the 2013 IEEE 7th International Conference on Intelligent Data Acquisition and Advanced Computing Systems, IDAACS 2013. 2013 | **CI2. CI3.** |
| 30 | Phimmasean, S. and Chuenurajit, T. and Cherntonomwong, P. Indoor localization system based on fingerprint technique using RFID passive tag. 2013 10th International Conference on Electrical Engineering/Electronics, Computer, Telecommunications and Information Technology, ECTI-CON 2013. 2013 | **CI1.** |
| 31 | Gao, Y. and Niu, J. and Zhou, R. and Xing, G. ZiFind: Exploiting cross-technology interference signatures for energy-efficient indoor localization. Proceedings - IEEE INFOCOM. 2013 | **CI1.** |
| 32 | Tian, Y. and Denby, B. and Ahriz, I. and Roussel, P. and Dubois, R. and Dreyfus, G. Practical indoor localization using ambient RF. Conference Record - IEEE Instrumentation and Measurement Technology Conference. 2013 | **CI1. CI2.** |
| 33 | Van Velzen, J. and Zuniga, M. Let's collide to localize: Achieving indoor localization with packet collisions2013 IEEE International Conference on Pervasive Computing and Communications Workshops, PerCom Workshops 2013. 2013 | **CI1.** |
| 34 | Sen, S. and Lee, J. and Kim, K.-H. and Congdon, P. Avoiding multipath to revive inbuilding WiFi localization. MobiSys 2013 - Proceedings of the 11th Annual International Conference on Mobile Systems, Applications, and Services. 2013 | **CI2. CI3.** |
| 35 | Ha, Y.-G. and Eun, A.-C. and Byun, Y.-C. Efficient sensor localization for indoor environments using classification of link quality patterns. International Journal of Distributed Sensor Networks. 2013 | **CE5.** |
| 36 | Lv, X. and Mourad-Chehade, F. and Snoussi, H. Fingerprinting-based localization using accelerometer information in wireless sensor networks. GLOBECOM - IEEE Global Telecommunications Conference. 2013 | **CI1.** |
| 37 | Buchman, A. and Lung, C. Received signal strength based room level accuracy indoor localisation method. 4th IEEE International Conference on Cognitive Infocommunications, CogInfoCom 2013 – Proceedings. 2013 | **CI1.** |
| 38 | Bharanidharan, M. and Li, X.J. and Jin, Y. and Pathmasuntharam, J.S. and Xiao, G. Design and implementation of a real time locating system utilizing Wi-Fi signals from iPhones. IEEE International Conference on Networks, ICON. 2012 | **CI1.** |
| 39 | Li, F. and Zhao, C. and Ding, G. and Gong, J. and Liu, C. and Zhao, F. A Reliable and accurate indoor localization method using phone inertial sensors. UbiComp'12 - Proceedings of the 2012 ACM Conference on Ubiquitous Computing. 2012 | **CI1. CI2.** |
| 40 | Nandakumar, R. and Chintalapudi, K.K. and Padmanabhan, V.N. Centaur: Locating devices in an office environment. Proceedings of the Annual International Conference on Mobile Computing and Networking, MOBICOM. 2012 | **CI1. CI2. CI3.** |
| 41 | Dakkak, M. and Nakib, A. and Daachi, B. and Siarry, P. and Lemoine, J. Mobile indoor location based on fractional differentiation. IEEE Wireless Communications and Networking Conference, WCNC. 2012 | **CE5.** |
| 42 | Wang, J. and Prasad, R.V. and An, X. and Niemegeers, I.G.M.M. A study on wireless sensor network based indoor positioning systems for context-aware applications. Wireless Communications and Mobile Computing. 2012 | **CI1.** |
| 43 | Martin, E. and Bajcsy, R. Precise multimodal localization with smart phones. Communications in Computer and Information Science. 2011 | **CE5.** |
| 44 | Goswami, A. and Ortiz, L.E. and Das, S.R. WiGEM: A learning-based approach for indoor localization. Proceedings of the 7th Conference on Emerging Networking EXperiments and Technologies, CoNEXT'11. 2011 | **CI1.** |
| 45 | Shanklin, T.A. and Loulier, B. and Matson, E.T. Embedded sensors for indoor positioning. SAS 2011 - IEEE Sensors Applications Symposium, Proceedings. 2011 | **CE3.** |
| 46 | Park, J. and Lim, Y. and Lee, K. and Choi, Y.-H. A polygonal method for ranging-based localization in an indoor wireless sensor network. Wireless Personal Communications. 2011 | **CE5.** |
| 47 | Ott, A.T. and Shalaby, M. and Siart, U. and Kaliyaperumal, E. and Eibert, T.F. and Engelbrecht, J. and Collmann, R. Performance analysis of a low cost wireless indoor positioning system with distributed antennas. Advances in Radio Science. 2011 | **CI1.** |
| 48 | Martin, E. and Vinyals, O. and Friedland, G. and Bajcsy, R. Precise indoor localization using smart phones. MM'10 - Proceedings of the ACM Multimedia 2010 International Conference. 2010 | **CI1.** |

## 3.3. Lista de artigos removidos no 2° Filtro na Scopus

A abordagem adotada nesta fase foi a leitura completa do artigo e validação pelos critérios de inclusão/exclusão, e posteriormente a extração dos dados dos aceitos. Como artigos aprovados tiveram um total de 31 artigos, os quais se listam na tabela 7 os removidos e os seus respectivos critérios de exclusão.

**Tabela 7. Artigos removidos no 2º filtro na Scopus**

| Nº | Nome do artigo | Critério de Exclusão |
|---|---|---|
| 12 | Subbu, K. and Zhang, C. and Luo, J. and Vasilakos, A. Analysis and status quo of smartphone-based indoor localization systems. IEEE Wireless Communications. 2014 | **CE6.** |
| 33 | Van Velzen, J. and Zuniga, M. Let's collide to localize: Achieving indoor localization with packet collisions2013 IEEE International Conference on Pervasive Computing and Communications Workshops, PerCom Workshops 2013. 2013 | **CE6.** |
| 36 | Lv, X. and Mourad-Chehade, F. and Snoussi, H. Fingerprinting-based localization using accelerometer information in wireless sensor networks. GLOBECOM - IEEE Global Telecommunications Conference. 2013 | **CE4.** |
| 48 | Martin, E. and Vinyals, O. and Friedland, G. and Bajcsy, R. Precise indoor localization using smart phones. MM'10 - Proceedings of the ACM Multimedia 2010 International Conference. 2010 | **CE6.** |

# 4. Extração de Dados

Após aplicação dos filtros (primeiro/segundo) em ambas as bibliotecas digitais, o numero de artigos resultantes para fase de extração de dados são apresentados na tabela a seguir:

**Tabela 8 Feedback após o segundo filtro nas bibliotecas digitais**

| Maquina de Busca | Resultado Busca | Total Após o primeiro Filtro | Total Após o segundo Filtro |
|---|---|---|---|
| **IEEE** | 24 | 20 | 17 |
| **SCOPUS** | 48 | 35 | 31 |
| | | | **48** |

Cada um dos 48 artigos aprovados pelo processo de seleção final é extraído os seguintes dados, de acordo com a Tabela a seguir:

**Table 9 Template de extração dos dados**

| Dados do ARTIGO | Descrição dos dados do artigo |
|---|---|
| **Autor** | Autor (es) do artigo |
| **Ano** | Ano de publicação. |
| **Keywords** | Palavras que servem como ponto de referencia para localizar um contexto. |
| **Siglas importantes** | Abreviações que são de relevância significativa ao longo do referido artigo. |
| **Objetivo do Artigo** | Qual o Objetivo Proposto? |
| **Técnicas/Abordagens/métodos/algoritmos** | Quais as técnicas/abordagens/métodos/algoritmos que são empregados no artigo. |
| **Qual é o tipo de pesquisa?** | A pesquisa descrita no artigo pode ser do tipo:<br>• Conceitual;<br>• Empírica; |
| **Especificação do tipo da técnica** | Em que a técnica/abordagem proposta é baseada? |
| **Descrição da técnica** | Resumo detalhado da técnica/abordagem proposto pelo artigo |
| **Utiliza algum Algoritmo de Localização?** | Foi aplicado/customizado algum algoritmo de localização. |
| **Cobertura** | Qual a escala de cobertura da técnica? |
| **Contextualização** | Qual o cenário foi empregado. |
| **Tem apoio ferramental. Qual?** | Alguma ferramenta foi desenvolvida para dar suporte a abordagem proposta?<br>• Sim<br>• Não |
| **Infra** | Qual a infraestrutura utilizada? |
| **Abordagem Hibrida** | • Sim<br>• Não |
| **Qual é o resultado da pesquisa?** | O resultado da pesquisa pode ser:<br>• Ferramenta<br>• Framework<br>• Técnica<br>• Modelo<br>• Processo<br>• Método<br>• Guideline |
| **Resultados** | Quais foram às conclusões do artigo. |
| **Limitações** | Há limitações encontradas na aplicação da técnica/abordagem? |
| **Trabalhos Futuros** | Trabalhos que podem estar sendo executados baseados no referido artigo |
| **Algoritmos** | Foi disponibilizado algum trecho/método/classe de um algoritmo utilizado no artigo. |

**Legenda**

**NA: NOT available**

## 4.1. Fonte IEEExplore

1)

| Dados do ARTIGO | An Experimental Study on the Behavior of Received Signal Strength in Indoor Environment |
|---|---|
| **Autor** | Kifayat Ullah*, Igor Vitorio Custodio*, Nadir Shah**, Edson dos Santos Moreira* |
| **Ano** | 2013 |
| **Keywords** | Location Based Services (LBS), Received Signal Strength (RSS), Fingerprinting, WiFi Signal Behavior, Indoor localization, Indoor Positioning System |
| **Siglas importantes** | – Location based services (LBS);<br>– Access Points (aps);<br>– Reference Point (RP). |
| **Objetivo do Artigo** | Foi estudado o comportamento dos sinais de WIFI em um ambiente para localização Indoor baseada RSS, analisando sinais de três pontos de acesso diferentes usando a técnica de triangulação. O método é baseado no método de impressão digital. |
| **Técnicas/Abordagens/métodos/algoritmos** | RSSI-based x Fingerprinting Method. |
| **Qual é o tipo de pesquisa?** | • Empírica; |
| **Especificação do tipo da técnica** | Foi Realizado experimentos numa sala aberta com uma área total de 8x14 metros. Foram utilizados três APs (802.11n A D-Link). Os APs foram colocados em três locais diferentes |
| **Descrição da técnica** | 1. Depois de realizar o site survey e colocar os APs em localizações ideais, a próxima tarefa foi a de dividir a área testada em grids de tamanho igual. A área foi dividida em dois planos dimensionais (eixo X e do eixo Y). O tamanho de cada grid foi mantido para 1 metro x1. O objetivo da presente RP foi recolher amostras durante a fase off-line.<br>2. A base da técnica de fingerprinting é a coleção do "signal strength" (SS) das amostras. Foram coletadas amostras SS para as três APS em todos os pontos de referência.<br>3. Foi coletado amostras em quatro direções diferentes, ou seja: 0 °, 90 °, 180 ° e 270 °. Em vez de recolher uma única amostra, foram coletadas 20 amostras para cada AP em todas as direções em cada RP. Assim, um total de 80 amostras foram coletadas em cada RP para cada AP. |
| **Utiliza algum Algoritmo de Localização?** | Baseado no método “fingerprinting” |
| **Cobertura** | **NA** |
| **Contextualização** | Foi selecionado o primeiro andar do Departamento de Computação e Eletrônica da Universidade de São Paulo (USP) - São Carlos Campus. |
| **Tem apoio ferramental. Qual?** | Para o sistema operacional Windows, foi realizado o site survey com  Network Stumble tool, enquanto para o sistema operacional Ubuntu Linux foi usado IWList tool. |
| **Abordagem Hibrida** | • Não |
| **Qual é o resultado da pesquisa?** | • Método |
| **Resultados** | Os resultados experimentais revelam que o comportamento do sinal muda frequentemente. Os resultados nos conduzem que a compreensão do comportamento dos sinais é importante antes de |
| | estimar a localização atual e fornecendo diferentes LBS. |
| **Limitações** | Conclui-se que muitos fatores podem afetar o comportamento do RSS em um ambiente Indoor. Alguns desses fatores foi a inserção (planejadas) de APs, consideração de obstáculos (como paredes, móveis e pessoas) os efeitos de portas, direção do objeto (notebook), etc. O artigo afirma que, esses fatores sendo considerados antes da implementação do LBS, a precisão e granularidade de sistemas de localização interior seria aumentada. |
| **Trabalhos Futuros** | Pretende-se:<br>• Analisar os diferentes fatores, que afetam o comportamento de RSS, através do uso de alguns modelos matemáticos;<br>• Verificar a exatidão de tais sistemas, considerando os diferentes fatores que influenciam o comportamento de RSS;<br>• Estender este trabalho para o desenvolvimento da fase on-line e técnicas de estimativa de localização |
| **Algoritmos** | Este algoritmo funciona da seguinte forma: O notebook foi colocado no primeiro RP e colocado direção (0 °). Um total de 20 amostras foram coletadas neste RP para cada AP. O processo é então repetido para as restantes três direções (90 °, 180 ° e 270 °). Após a coleta das amostras para cada direção, o notebook foi movido para a segunda RF e o processo foi repetido para todos os restantes RPs. O shell script (algoritmo) permite que o usuário armazene as amostras coletadas em um arquivo de texto dentro do notebook.<br><br>**SAMPLES=20**<br>**DIRECTIONS=4**<br>**X_SIZE= 8**<br>**Y_SIZE=14**<br><br>**Function Ask_For_Confirmation (message)**<br>**{**<br>**//Press Enter when ready for collecting samples**<br>**}**<br><br>**Function Collect_A_Sample(sample_number, direction, position)**<br>**{** |
| | SS, SQ, AP_MAC<br>storeInFile (SS, SQ, APMAC, direction, position)<br>return True if “OK”<br>}<br><br>for(x=0; x<X_SIZE, x++)<br>{<br>for(y=0; <Y_SIZE, y++)<br>{<br>for(d=0; d<DIRECTIONS, d++)<br>{<br>S=0<br>While (S<SAMPLES)<br>{<br>RESP=TakeASample(x,y,d,s)<br>If RESP =“ok”<br>S++<br>}<br>Ask_for_Confirmatinon(“Change direction”)<br>}<br>Ask_for_confirmation(“Move to next Reference Point”)<br>}<br>} |

2)

| Dados do ARTIGO | Smart Phone Based Indoor Pedestrian Localization System |
|---|---|
| **Autor** | Lokesh Agrawal,Durga Toshniwal |
| **Ano** | 2013 |
| **Keywords** | Localization, Mobile Phones, Sensors, Probability, Indoor Positioning System |
| **Siglas importantes** | - Radio Signal Strength Indications (RSSI) ;<br>- Particle Filters (PFs);<br>- Access points (APs). |
| **Objetivo do Artigo** | Neste artigo é apresentado um sistema que aproveita a câmera e Wi-Fi presente nos smartphones dos usuários, para controlá-los, enquanto circulam em ambientes Indoor. Ele faz uso de um mapa de rádio de um ambiente Indoor. |
| **Técnicas/Abordagens/métodos/algoritmos** | WIFI x Smartphones's Camera |
| **Qual é o tipo de pesquisa?** | • Empírica; |
| **Especificação do tipo da técnica** | Um dos principais inconvenientes das soluções de Wi-Fi existentes baseados é que a maior parte deles não considera a mobilidade de um utilizador, e é exatamente a mobilidade a premissa deste artigo. O Uso de Smartphones rodando Android 2.0 e acima é importante porque os valores são atualizados dinamicamente RSSI como o usuário se move no interior das instalações. |
| **Descrição da técnica** | A idéia central envolve a combinação da força atual de smartphones com técnica de filtragem para estimar a localização do usuário dentro de uma premissa de um ambiente Indoor. A principal vantagem de usar esta abordagem é que ele não requer qualquer infra-estrutura de rede dedicada. Nem qualquer hardware especializado é necessário. Esta abordagem utiliza impressões digitais (fingerprints) de várias medições de potência do sinal Wi-Fi para a estimativa da posição. |
| **Utiliza algum Algoritmo de Localização?** | NA |
| **Cobertura** | NA |
| **Contextualização** | Departamento de Ciência da Computação no Instituto Indiano de Campus Tecnologia Roorkee( Computer Science Department in the Indian Institute of Technology Roorkee campus). |
| **Tem apoio ferramental. Qual?** | Inicialmente foi utilizado "insider" software de scanner de rede Wi-Fi para o Microsoft Windows a partir MetaGeek, |
| **Abordagem Hibrida** | Sim.<br>- Foi utilizado 2 recursos disponíveis em SmartPhones atualmente: Wi-Fi e câmera. |
| **Qual é o resultado da pesquisa?** | • Técnica |
| **Resultados** | Os resultados obtidos após a realização de simulações demonstram a validade e adequação do algoritmo proposto para fornecer um alto nível de desempenho em termos de precisão posicional e escalabilidade. |
| **Limitações** | Foi integrada a capacidade de detecção, juntamente com filtros para oferecer precisão/acuracidade de até 1-1,5. |
| **Trabalhos Futuros** | Adicionar um recurso extra de obter o caminho mais curto para o destino, uma vez que a pessoa é localizada. |
| **Algoritmos** | NA |

3)

| Dados do ARTIGO | Multimodal Indoor Localization: An Audio-Wireless-Based Approach |
|---|---|
| **Autor** | Oriol Vinyals, Eladio Martin, Gerald Friedland |
| **Ano** | NA |
| **Keywords** | Localization, indoor, audio, wifi, multimodal |
| **Siglas importantes** | – Radio Signal Strength Indications (RSSIs). |
| **Objetivo do Artigo** | São utilizados os microfones simultaneamente a vários sinais wireless que são detectados por smartphones para servir como entrada a abordagem. É estudado as possibilidades oferecidas por três recursos disponíveis em smartphones : rádio WiFi, rádio de comunicação celular e acelerômetro, com o objetivo de construir uma abordagem multimodal para localização. |
| **Qual é o tipo de pesquisa?** | • Empírica; |
| **Especificação do tipo da técnica** | A presença de microfones nos dispositivos que levamos em conexão com vários sinais wireless detectadas por smartphones atuais servem para indicar o local com cerca de +/- precisão 3m. Nesta abordagem propõe a utilização de técnicas recentes de aprendizado de máquina para a integração de modalidades, fornecendo maior precisão do que os trabalhos atuais na área. |
| **Descrição da técnica** | Foi projetado um algoritmo de estimativa de curta distância do acelerómetro, que não requer entrada do usuário para a calibração. |
| **Utiliza algum Algoritmo de Localização?** | NÃO |
| **Cobertura** | NA |
| **Contextualização** | Os testes para medir diferentes intensidades de sinal de rádio freqüência no interior do edifício Cory na University of California, Berkeley campus. |
| **Tem apoio ferramental. Qual?** | Para a medição dos sinais e aplicação prática do nosso aplicativo de localização, temos usado smartphones rodando em Android, em particular o G1 e os Droid . |
| **Abordagem Hibrida** | • Sim<br><br>Tenham sido previamente utilizados para a localização indoor: sinais wireless e acústica. |
| **Qual é o resultado da pesquisa?** | • Framework<br>• Técnica<br>• Modelo<br>• Método |
| **Resultados** | Este artigo indica que uma abordagem multimodal é viável e que a utilização integrada de duas modalidades prontamente disponíveis em qualquer dispositivo móvel beneficia a precisão. |
| **Limitações** | • Deve-se notar que valores abaixo de -85 dBm são geralmente demasiado inconsistente para ser aproveitado como referência. Consequentemente, os valores de RSSI acima -80 dBm são desejáveis a fim de obter resultados fiáveis.<br>• Deve-se notar que, como o número de rádios WiFi (e seus valores RSSI) que podem ser ouvidos em um decréscimo local específico (por exemplo, apenas 2 rádios WiFi com valores acima de -75 dBm RSSI), acarreta na diminuição da precisão de localização do aplicativo proposto.<br>• Um dos principais desafios impostos pelos dados Wifi é que depende de fatores externos tais condições ou a posição de objetos no ambiente. |
| **Trabalhos Futuros** | NA |
| **Algoritmos** | NA |

4)

| Dados do ARTIGO | Item-Level RFID Tag Location Sensing Utilizing Reader Antenna Spatial Diversity |
|---|---|
| **Autor** | Shuai Shao, Robert J. Burkholder |
| **Ano** | 2013 |
| **Keywords** | UHF, RFID, item-level, RSSI, localization, estimation. |
| **Siglas importantes** | – RFID (Radio Frequency IDentification). |
| **Objetivo do Artigo** | Desenvolver uma solução de localização usando apenas hardware RFID existente que é independente dos tipos de tags e suas orientações |
| **Qual é o tipo de pesquisa?** | • Empírica; |
| **Especificação do tipo da técnica** | Neste artigo é proposto uma abordagem (técnica de localização tag) em tempo real em nível de item RSSI baseado em múltiplas antenas de leitor espacialmente dispersas. Cada antena gera um número de RSSI, que depende da localização da antena relativamente à marca. Múltiplas antenas interrogando a mesma etiqueta a partir de diferentes pontos de vista fornecer um conjunto de dados de RSSI a partir do qual a localização de etiquetas pode ser estimada. A estimativa é baseada em um modelo de sinal para cada antena, que depende da localização da antena, alcance e ângulo em relação à etiqueta, e orientação da etiqueta. Um modelo simples multipath também é investigada para melhorar a precisão de localização. |
| **Descrição da técnica** | No experimento proposto, é usada uma polarização circular (CP) antena do leitor remendo padrão. Uma etiqueta RFID "ondulada" é colocada em frente do bordo da antena do leitor, a uma distância de 2,4 m. A etiqueta é girada na direção transversal plano e o valor RSSI é medida em cada ângulo. |
| **Utiliza algum Algoritmo de Localização?** | NA |
| **Cobertura** | Estimativa de precisão de 35 cm. |
| **Contextualização** | Os experimentos foram conduzidos em uma grande sala com um leitor de RFID 4-port Impinj Speedway_. Quatro antenas patch polarização circulares idênticos estão ligados ao leitor. Um padrão de radiação cosin esquared é assumido. Os Resultados experimentais são apresentados para um cenário de varejo realista com múltiplos itens com etiquetas que atingiu uma precisão da estimativa de 35 cm. |
| **Tem apoio ferramental. Qual?** | • Speedway_ 4-port RFID reader |
| **Abordagem Hibrida** | • Sim<br>Métodos de localização Wireless |
| **Qual é o resultado da pesquisa?** | • Técnica |
| **Resultados** | Os resultados experimentais mostram que estimativa de ERRO de distancia é de 35 cm para uma área cheia de itens de varejo, quando as antenas estão na mesma altura com etiquetas, e 41 cm de altura com antenas montadas perto do teto. Mesmo que a precisão é, naturalmente, de vários fatores, tais como trajetórias múltiplas, desaparecendo e bloqueio, que é suficiente para encontrar artigos etiquetados, sob condições realistas. |
| **Limitações** | O algoritmo proposto pode ser aplicado para diferentes tipos de tags, independentemente da sua orientação. É só necessário saber os locais das antenas de leitor e os seus padrões de ganho associados e polarizações. A técnica proposta também sofre de trajetórias múltiplas desconhecidas, sombreamento, bloqueio e outras interferências que têm diferentes efeitos sobre os sinais com base na estrutura da área de interrogação, a colocação de antenas do leitor, a densidade dos objetos etiquetados, a presença de obstruções, etc. |
| **Trabalhos Futuros** | NA |
| **Algoritmos** | NA |

5)

| Dados do ARTIGO | An RFID-Based Position and Orientation Measurement System for Mobile Objects in Intelligent Environments |
|---|---|
| **Autor** | Ali Asghar Nazari Shirehjini, Abdulsalam Yassine, and Shervin Shirmohammadi, |
| **Ano** | 2012 |
| **Keywords** | Ambient intelligence (AmI), location and orientation measurement, radio frequency identification (RFID), smart environment. |
| **Siglas importantes** | • Ambient intelligence (AmI) |
| **Objetivo do Artigo** | Este trabalho toma sobre o desafio de determinar a localização e orientação de objetos móveis em ambiente Indoor “byproposing” um sistema robusto, baseado em tecnologia de identificação passiva de rádio frequência (RFID). |
| **Qual é o tipo de pesquisa?** | • Empírica;<br><br>A validação da abordagem proposta foi feita por meio de  uma aplicação de prova de conceito que é utilizado para analisar o erro médio de posicionamento e orientação medições do objeto móvel. |
| **Especificação do tipo da técnica** | O sistema proposto, que consiste em tapetes RFID e vários periféricos para a interpretação dos dados do sensor, é implementado e testado através de vários periféricos para a interpretação dos dados do sensor e distribuição das informações de posicionamento. |
| **Descrição da técnica** | As etiquetas RFID são montadas em pads tapete (“carpet pads”) onde os objetos móveis são colocados. O sistema é projetado para trabalhar para objetos que estejam ligadas, direta ou indiretamente, a um ponto que está a uma curta distância do chão. As etiquetas RFID são colocadas em posições pré-definidas fixos dentro de pads tapete específicos. As tags não armazenaram qualquer informação de posição, exceto a sua linha e coluna, coordenadas dentro do tapete correspondente no qual estão |
| | montados. Em vez disso, os dados nas marcas correspondem aos números de linha e coluna dentro de uma placa de tapete que integra essas etiquetas de RFID, como uma grade. Ao fazê-lo, pode-se mover a placa tapete para qualquer lugar na sala sem a necessidade de alterar os dados armazenados. |
| **Utiliza algum Algoritmo de Localização?** | NA |
| **Cobertura** | Fornece dados de posicionamento e orientação com um erro médio de 6,5 cm, com um desvio padrão de 4,5 centímetros, para medição de posicionamento e um erro médio de 1.9◦, com um desvio padrão de 2.5◦, para medição de orientação. |
| **Contextualização** | Considere-se que um grupo de colegas de trabalho, Alice, Bob, e Jack, estão planejando se reunir e discutir o seu projeto em uma das salas de conferência da empresa. Alice quer apresentar seus slides, e Bob precisa mostrar um vídeo sobre um protótipo que ele desenvolveu. O gerente de projeto, Jack, quer discutir finanças do projeto e fornecer alguns dados estatísticos. A sala inteligente oferece um conjunto de displays e dispositivos de renderização: duas placas inteligentes, uma estação de TV móvel e dispositivos de processamento de áudio. Além de computadores pessoais do trabalhador, existe um sistema de projeção digital montado na parede. Também é considerado que a TV, o que seria a melhor escolha para o processamento de vídeo, é exposta à luz do sol, porque cortina da janela foi desmontado para ser limpo. Portanto, a utilização deste dispositivo pode não ser uma boa escolha, neste contexto específico. |
| **Tem apoio ferramental. Qual?** | Os componentes de leitor de RFID são conectados a um computador embutido através da interface serial através do qual as informações de posição e orientação é calculada com base na informação do rótulo armazenado. |
| **Abordagem Hibrida** | • Sim |
| **Qual é o resultado da pesquisa?** | • System |
| **Resultados** | Foi mostrado que o sistema proposto supera os sistemas semelhantes existentes em minimizar o erro de posicionamento. Certos números de contribuições foram feitas neste trabalho. Especificamente, têm sido demonstrado através de uma aplicação de prova de conceito e uma série de experimentos que o sistema atinge um erro médio baixo para o interior objeto de posicionamento e orientação. |
| **Limitações** | O principal desafio para a realização de tais aplicações é a correta determinação das características contextuais de mudança do ambiente. Isto é, a determinação da posição e orientação de pessoas e objectos com a finalidade de ter o melhor dispositivo adequado apresenta um dado teor de meios. No entanto, é importante que as informações armazenadas em cada posição refiram-se à marcação variável x e y em relação ao desenho<br><br>Além disso, porque a orientação é muito importante para a aplicação, dois leitores não são sempre suficientes, uma vez que não corresponde, necessariamente, com as etiquetas. Por exemplo, se a distribuição das etiquetas é muito escasso, então a probabilidade de se obter um leitor em uma zona não marcado é elevada, e assim, não recebe os dados de posicionamento. Portanto, os usos mais leitores por objeto aumenta a robustez dos sistemas e garante a qualidade da medição em termos de menor erro médio. |
| **Trabalhos Futuros** | Para trabalhos futuros, estamos planejando para estudar o efeito do uso de diferentes tipos de pisos. Isto é porque a taxa de absorção de energia RF varia de um tipo de piso para o outro (por exemplo, chão de madeira, etc.), afetando assim o erro de medição do objeto de posicionamento e orientação móvel. |
| **Algoritmos** | NA |

6)

| Dados do ARTIGO | Embedded Sensors for Indoor Positioning |
|---|---|
| **Autor** | Teresa A. Shanklin, Benjamin Loulier, and Eric T. Matson |
| **Ano** | 2011 |
| **Keywords** | NA |
| **Siglas importantes** | − Inertial Measurement Unit (IMU)<br>− Degrees of freedom (DOF).<br>− embedded accelerometer |
| **Objetivo do Artigo** | O objetivo deste artigo é desenvolver um sistema de posicionamento Indoor útil usando um baixo custo, o dispositivo disponível ao público. |
| **Qual é o tipo de pesquisa?** | • Empírica;<br>...A próxima seção inclui nossos resultados experimentais. |
| **Especificação do tipo da técnica** | Em foi utilizado o primeiro sistema, a medida e filtragem foram embutidos no iPhone. Para obter mais flexibilidade e tornar mais fácil para alterar os parâmetros de filtragem esses dados são agora enviados via pacotes UDP através da rede Wi-Fi para um computador com o LabVIEW. |
| **Descrição da técnica** | Foi fornecido processamento de  funcionalidades para determinar os parâmetros mais apropriados para filtrar e integrar. Então, pode-se detectar o fim de uma etapa e o início da seguinte, podemos assumir que durante este período, a velocidade do pé é zero. Isso nos permite calibrar a velocidade, diminuindo drasticamente o efeito da deriva (“drift”). |
| **Utiliza algum Algoritmo de Localização?** | NAO |
| **Cobertura** | Range (g): +-2<br>Acceleration noise density (_g= pHz): 218<br>Bandwidth (Hz): 25 |
| **Contextualização** | Este artigo apresenta os resultados experimentais para criar um sistema de posicionamento de interiores usando um Smartphone modelo atual: o iPhone 4 da Apple e os sensores embutidos. |
| **Tem apoio ferramental. Qual?** | Este artigo centra-se no iPhone de quarta geração, já que contém sensores embutidos: acelerômetro; giroscópio e magnetômetro |

Tabela da linha "Cobertura":

| | |
|---|---|
| Range (g) | +-2 |
| Acceleration noise density (_g= pHz) | 218 |
| Bandwidth (Hz) | 25 |
| **Abordagem Hibrida** | Os acelerómetros e giroscópios permitem-nos ter a aceleração em um referencial ligada à Terra agora enviado via pacotes UDP através da rede Wi-Fi para um computador com o LabView |
| **Qual é o resultado da pesquisa?** | • System |
| **Resultados** | Foi simular uma linha reta com o IPHONE preso firmemente ao pé do participante para reduzir a quantidade de ruído devido à complexidade do movimento humano. Em relação ao formato das curvas, temos resultados limpos que combina com o que estávamos esperando, ainda os valores que recebemos estão longe de serem precisos. |
| **Limitações** | O corpo é - felizmente - não é completamente rígido e tem muitos graus de liberdade (DOF). Isto representa um verdadeiro desafio para o sistema proposto, como muitos DOF significa vários movimentos indesejados interfiram na medida, como tremores ou vibrações.<br><br>Inicialmente, o iPhone foi preso ao pé, como uma parte do corpo com um movimento consistente e estável ao caminhar que nos permitiu usar o Velocity Zero. Atualizar princípio para remover o desvio devido à integração da velocidade. Esta (baixa freqüência de corte) fez com que o ruído seja muito baixo, mas também filtrou uma parte do sinal que se desejou atingir. Precisa-se melhorar a filtragem para obter valores precisos. |
| **Trabalhos Futuros** | Identificou-se um potencial nestes resultados e tem planos para aperfeiçoamento e implementação deste protótipo. O artigo levanta melhorias relacionadas a técnicas de filtragem, bem como determinar o melhor método para a remoção de drift |
| **Algoritmos** | NA |

7)

| Dados do ARTIGO | Practical Indoor Localization using Ambient RF |
|---|---|
| **Autor** | Ye Tian, Bruce Denby, Iness Ahriz, Pierre Roussel, |
| **Ano** | 2013 |
| **Keywords** | localization; indoor; fingerprint; transductive support vector machine |
| **Siglas importantes** | Support Vector Machine (SVM) regression. |
| **Objetivo do Artigo** | O artigo apresenta uma abordagem simples e prática para a localização Indoor, utilizando sinal recebido impressões digitais forçam a partir da rede GSM, incluindo uma análise da relação entre a intensidade do sinal e localização, e a evolução do desempenho de localização ao longo do tempo. |
| **Qual é o tipo de pesquisa?** | • Empírica;<br>… e resultados experimentais mostram que o percentual de rotulagem correta quarto pode ser de até 94%.. |
| **Especificação do tipo da técnica** | O artigo foca à busca de uma relação funcional entre GSM RSS e posição, usando Support Vector Machine (SVM) regression. A utilização de classificadores baseados em GSM RSS ambiente treinados com dados recolhidos ao longo das áreas de salas, no entanto, apresenta uma alternativa viável, e os resultados experimentais mostram que a percentagem de rotulagem quarto correto pode ser de até 94%, se o modelo é usado antes significativo conjuntos de deriva em RSS. |
| **Descrição da técnica** | A fim de lidar com a degradação do desempenho causada por RSS desvio ao longo do tempo, inferência transdutor foi introduzida para atualizar os classificadores SVM com novos dados não rotulados. Quando testado em dados coletados ao longo de nove meses, essa abordagem provou ser capaz de restaurar uma parte significativa do desempenho perdido. O uso de pequenas quantidades de dados atuais marcadas para criar "current" classificadores de ambientes também parece ser uma abordagem promissora, mesmo que o desempenho ainda precisa ser melhorado. |
| **Utiliza algum Algoritmo de Localização?** | NÃO |
| **Cobertura** | Um total de 600 scans GSM para cada módulo foram registrados ao longo de cinco dias úteis. Cada análise contém o RSS de todos os 548 operadoras nas faixas GSM900 e GSM1800, e é composto por valores que variam em RSS valor de -108dBm para -40dBm. |
| **Contextualização** | Foram coletados dois tipos de conjuntos de dados, para experiências de regressão e classificação, respectivamente, ambas registradas no 4º andar de um prédio de laboratório (paredes de armação de aço, concreto e gesso), no centro de Paris, França. |
| **Tem apoio ferramental. Qual?** | 1. O primeiro set, foi coletado em uma única de ambientes (escritório 7) usando chamada machine-to- machine ou módulos M2M, GSM / GPRS.<br>2. O segundo conjunto de dados, que será usado para a classificação SVM, o dispositivo de aquisição de dados neste caso é um telefone móvel Sony-Ericsson com software de digitalização integrado, o qual é capaz de obter uma imagem de toda a GSM900, GSM 1800 bandas de cerca de 300 milissegundos. |
| **Abordagem Hibrida** | • Sim<br><br>Um estudo de distribuição RSS ambiente em um ambiente interno por meio de regressão SVM. |
| **Qual é o resultado da pesquisa?** | • Método<br>O artigo apresenta uma abordagem simples e prática para a localização Indoor. |
| **Resultados** | Support Vector regressão Máquina aplicada a muito alto impressões digitais dimensionais não revela qualquer relação funcional suave entre as impressões digitais e posição. A classificação recorrendo Support Vector Machines, no entanto fornece resultados muito bons em diferentes quartos em um ambiente Indoor, embora com desempenho que se degrada com o tempo. Inferência transdutor, introduzido como um meio de atualizar os modelos para superar a degradação ao longo do tempo fornece dicas que a localização exata interior pode ser alcançada através da aplicação de métodos de classificação para Recebidos impressões digital intensidade de sinal celular. |
| **Limitações** | O erro de regressão é aproximadamente igual à distância média entre os 8 locais analisados, o que significa que nenhuma relação linear ou não linear entre RSS e a posição de um pequeno ambiente indoor é evidente. Isso parece se minimizado utilizando vectores de banda completa RSS GSM obtidos deste modo a interpolar entre posições fixas no interior de um método de localização. Outro ponto é representado pela degradação do desempenho ao longo do tempo é causada |
| | principalmente por confusões entre salas adjacentes. |
| **Trabalhos Futuros** | Pretende-se a investigar o uso de dados da rede W-CDMA na medição. |
| **Algoritmos** | NA |

8)

| Dados do ARTIGO | Item-Level Indoor Localization With Passive UHF RFID Based on Tag Interaction Analysis |
|---|---|
| **Autor** | Zhi Zhang, Zhonghai Lu, Vardan Saakian, Xing Qin, Qiang Chen, and Li-Rong Zheng, |
| **Ano** | 2014 |
| **Keywords** | Item-level indoor localization, passive ultrahigh frequency (UHF) radio-frequency identification (RFID), received signal strength indicator (RSSI), RSSI variance, tag interaction analysis. |
| **Siglas importantes** | NA |
| **Objetivo do Artigo** | Neste trabalho, foi proposto um método de análise sobre como a interação da tag afeta um padrão de radiação da antena e uma mudança de RSSI. A análise desta interação fornece o guia para melhorar o projeto de algoritmos de localização baseados em RSSI. |
| **Qual é o tipo de pesquisa?** | • A Empírica;<br>Para avaliar como a análise da interação da tag ajuda a melhorar os algoritmos de localização, foi proposto experimentos. |
| **Especificação do tipo da técnica** | A técnica utiliza a RSSI atenuação da intensidade do sinal transmitido para calcular a distância entre o transmissor e o receptor. É uma técnica de baixo custo e baixa complexidade. |
| **Descrição da técnica** | Neste trabalho foi:<br>• Abordado o problema de localização a partir da perspectiva da interação e da radiação padrões da tag para aplicações interiores em nível de item, em especial para os casos difíceis de localizar vários objetos próximos simultaneamente.<br>• Proposto um método de análise para a interação tag, incluindo a interação de referência de referência, a interação de referência-alvo, e interação alvo-alvo.<br>• Gerado uma análise, o que conduziu a melhorar o projeto de algoritmos de localização.<br>• Realizado experimentos e resultados do relatório para validar esta abordagem. \ |
| **Utiliza algum Algoritmo de Localização?** | Dois Algoritmos foram utilizados:<br>• O k-vizinhos mais próximos (k-NN);<br>• Simplex. |
| **Cobertura** | Os experimentos foram conduzidos a um servidor e pode operar na gama de 865-868 MHz, com a potência de transmissão de 30 dBm frequência. |
| **Contextualização** | Para avaliar como a análise da interação da tag ajuda a melhorar os algoritmos de localização, foi feito experimentos em um ambiente normal de escritório, com dimensões de 7 m × 5 m. Plataformas de desenvolvimento Impinj Indy R2000 foram usadas como leitor. Os leitores recolhem IDs de tags e informações RSSI e enviá-los para o servidor, e o servidor executa os algoritmos de localização. |
| **Tem apoio ferramental. Qual?** | • Sim<br>Ettus Research VERT900 omnidirecionais antenas de 3 dBi estão cablados quatro portas Tx / Rx de um leitor |
| **Abordagem Hibrida** | • NA |
| **Qual é o resultado da pesquisa?** | • Método<br>• Algoritmo |
| **Resultados** | Os resultados experimentais mostram que:<br>1. Os algoritmos k-NN e Simplex são robustos a diferentes números, espaços e materiais sof objetos de destino, e eles são superiores a outros sistemas de localização RFID, considerando o custo, capacidade de localização simultânea de |
| | múltiplos alvos, e os erros de estimativa de localização.<br>2. O algoritmo k-NN revisto reduz o erro de estimativa de localização do algoritmo k-NN 25-18 cm, para a localização de um único alvo e 35-21 cm, para localizar alvos múltiplos. Estes são os seus resultados globais considerando diferentes números alvo, espaçamento e materiais.<br>3. A proposta do artigo é vantajosa em relação a outros sistemas de localização RFID em consideração de custo, capacidade de localização simultânea de múltiplos alvos, e erro de estimativa de localização. |
| **Limitações** | • A atenuação não é uma função monótona de distância devido ao efeito multipath e interferência em condições práticas, particularmente em ambientes interiores .Vários pontos inesperados podem mapear para o mesmo valor de RSSI, o que leva a erros significativos de estimativa.<br>• O padrão de radiação de alterações da tag antena devido a da tag a interação, como blindagem, podem refletir a acoplamento mútuo. Isto leva a uma alteração significativa na RSSI algumas direções e, portanto, afeta a precisão da localização algoritmos baseados em RSSI. Com base na análise da interação tag, podemos melhorar os algoritmos de localização baseados em RSSI. |
| **Trabalhos Futuros** | Aplicar o método de análise de interação para análise da interação do leitor da antena, e avaliar a influência da densidade leitor antena no erro de estimativa de localização. |
| **Algoritmos** | NA |

9)

| Dados do ARTIGO | A Framework Using Fingerprinting for Signal Overlapping-Based Method in WLAN |
|---|---|
| **Autor** | Wei Shi-Sue, Shiuhpyng Shieh, Bing-Han Li, Michael Cheng Yi Cho, Chin-Wei Tien |
| **Ano** | 2010 |
| **Keywords** | Wireless Local Area Network; Receive Signal Strength Indication; RSSI Localization method; RSSI Fingerprinting method; Hybrid RSSI Localization System |
| **Siglas importantes** | – Wireless Local Area Network (WLAN);<br>– Global Positioning;<br>– System (GPS);<br>– Floor Attenuation Factor [8] (FAF);<br>– Hybrid Overlapping fingerprinting method (HOF method). |
| **Objetivo do Artigo** | Neste trabalho, foi proposto um novo método que baseada em RSSI, o qual contém as vantagens de dois tipos de métodos de localização, o RSSI Triangulação e o RSSI Fingerprinting, de maneira a melhorar algumas das desvantagens dos métodos de localização baseados em RSSI tradicionais. |
| **Qual é o tipo de pesquisa?** | • Empírica;<br>A fim de avaliar a proposta, foi implementado o sistema em ambiente real ao invés de simulações. |
| **Especificação do tipo da técnica** | O esquema de localização de rede wireless atual é baseada principalmente na RSSI, neste artigo é proposto um esquema que combina a vantagem de dois tipos de métodos baseados RSSI tradicionais, método de triangulação e método de fingerprinting RSSI. |
| **Descrição da técnica** | A fim de coletar informações relativos a conectividade de rede wireless local, foi desenvolvido um sistema de busca de topologia Indoor, que é usado para monitorar a rede wireless e para rastrear as conexões entre dispositivos sem fio.<br><br>Caso seja necessário localizar um dispositivo qualquer, é disponibilizado o mínimo de três nós de monitoramento para calcular a sua posição. Com base nisto, estes três pontos de acesso do monitor (MAP) foram inseridos em um andar de um edifício para recolher RSSI, fornecendo assim a distancia relativa baseados na força do sinal recebido.<br><br>No primeiro momento, RSSI triple serão tidas por métodos de “matching”: triangulação e Fingerprinting. Dependendo da quantidade de dados coletados nesta primeira fase, o sistema irá tomar decisões sobre a escolha do método adequado correspondente (s), caso os dados coletados sejam insuficientes, o sistema só funcionará no método da triangulação. |

<table>
<tr><td></td><td>Usando o método de triangulação método, na maioria casos, é possível reduzir a área dos dispositivos, este processo é chamado: Redução de Região. Após redução Região, o sistema irá tentar trabalhar com método Fingerprinting. O resultado é então gravado para o banco de dados como parte dos dados coletados.</td></tr>
<tr><td>Utiliza algum Algoritmo de Localização?</td><td>É empregado o algoritmo K-Nearest- Neighbor para satisfazer a melhora unidade da divisão possível.</td></tr>
<tr><td>Cobertura</td><td>NA</td></tr>
<tr><td>Contextualização</td><td>We decided to implement the entire system in the National Chiao Tung University (NCTU) instead of using Simulations. Our experiment environment is EC building 6F and MISRC 6F of NCTU; all the related dimension figures of the EC building is listed in Table I and the floor:<br><br>O sistema foi implementado na National Chiao Tung University (NCTU) em vez de por meio de simulações, sendo o ambiente de experimento é o edifício CE 6F e MISRC 6F de NCTU com as seguintes características listadas abaixo:
<table>
<tr><td>Altura</td><td>60.4 m</td></tr>
<tr><td>Largura</td><td>37.6 m</td></tr>
<tr><td>altura de cada andar</td><td>4.00 m</td></tr>
<tr><td>Material empregado</td><td>Cimento</td></tr>
</table></td></tr>
<tr><td>Tem apoio ferramental. Qual?</td><td>O equipamento consiste de pontos de acesso que são usados para monitorar e rastreiae todos os pacotes sob a rede wireless do ambiente.</td></tr>
<tr><td>Abordagem Hibrida</td><td><ul><li>Sim</li></ul></td></tr>
<tr><td>Qual é o resultado da pesquisa?</td><td><ul><li>Método</li><li>Algoritmos</li></ul>Um novo método de localização é proposto de acordo com a sobreposição de sinal em WLAN.`<br>Um algoritmo de treinamento on-line híbrido é proposta para melhorar as desvantagens de ambos os algoritmos baseado RSSI.</td></tr>
<tr><td>Resultados</td><td>No metodo FingerPrinting tradicional, não se pode evitar o grande esforço na fase de treinamento. No entanto no metodo</td></tr>
</table>

| | |
|---|---|
| | proposto não necessita de pré-coletar todos os sinais RSSI. Em outras palavras, este é um novo método que utiliza de treinamento online para ajustar com precisão na base de tempo anterior RSSI recolhido. |
| **Limitações** | na maioria dos casos método de triangulação não se pode determinar a posição do usuário d maneira precisa, conduindo a uma margem de erro significativa. Foi constatado que o metodo método fingerprinting desperdíca quantidade considerável de tempo e força de trabalho para coletar todos os possíveis RSSI em todas as regiões possíveis.<br><br>Este artigo depende principalmente da PWR para localizar os dispositivos wireless. PWR em Airodump-ng significa que o nível de sinal relatado pela placa de rede sem fio. Embora importância dependa do driver, o sinal geralmente recebe uma leitura mais elevada quando a máquina experimental se aproxima do ponto de acesso ou a estação móvel. |
| **Trabalhos Futuros** | NA |
| **Algoritmos** | NA |

10)

| Dados do ARTIGO | HeadSLAM - Simultaneous Localization and Mapping with Head-Mounted Inertial and Laser Range Sensors |
|---|---|
| **Autor** | Burcu Cinaz, Holger Kenn |
| **Ano** | 2008 |
| **Keywords** | NA |
| **Siglas importantes** | • Inertial measurement unit (IMU)3 |
| **Objetivo do Artigo** | É demostrado técnicas de robótica móvel (localização simultânea e mapeamento) que podem ser usadas para gerar automaticamente tanto a informação de localização, quanto mapas de ambiente 2D empregando “ head-mounted inertial and laser range sensors.” |
| **Qual é o tipo de pesquisa?** | • Empírica;<br>É apresentado uma abordagem inicial e os resultados de uma série de experimentos conduzidos em um ambiente de escritório com foco no mapa de ruídos causado pela ambiguidade na “forma do ambiente”, tais como corredores. |
| **Especificação do tipo da técnica** | Em primeiro lugar, é identificado como a filtragem de partículas Rao-Blackwellized funciona, a fim de resolver o problema de SLAM e, em seguida, são apresentadas soluções rápidas para a adaptação do algoritmo de mapeamento para a configuração wearable.<br><br>Analisou-se o comportamento de ruído do movimento da cabeça e dos seus efeitos sobre a estimativa da trajetória, bem como sobre o mapa. |
| **Descrição da técnica** | Foi implementado um processo simples de detecção de passos para o modelo de movimento, o qual detecta a ocorrência de passo do pedestre, observando os dados de aceleração verticais fornecidos pela IMU e, juntamente com os dados de orientação, dá uma estimativa da posição inicial do pedestre.<br><br>Foram apresentadas modificações para a configuração wearable tais como o processamento para a movimentação da cabeça instável (“unstable head motion”) e a criação de informações de odômetro. Além disso, foram investigados os efeitos de diferentes velocidades de locomoção lineares na qualidade de compilação e mapa de localização. Gravaram-se diferentes conjuntos de dados de uma estrutura mais complexa e avaliou-se a capacidade desta abordagem para realizar os fechamentos de laço (“ loop closures”). |
| **Utiliza algum Algoritmo de Localização?** | Neste trabalho, utilizou-se a implementação de GMapping Grisetti , que está disponível no OpenSlam [1], tal como a base para a aplicação. |
| **Cobertura** | O erro máximo de localização é 0,7 metros. Boas estimativas de posições de referência deve-se a existência de obstáculos |
| | distinguíveis na área de alcance máximo do scanner laser. |
| **Contextualização** | O primeiro ambiente é um corredor de 17m x 40m, com vários escritórios adjacentes. O segundo ambiente é uma sala de 8m x 7, onde quatro obstáculos diferentes foram colocados no campo de visão ao nível da cabeça.  . O corredor tinha 16 marcadores visuais colocados em locais ao longo de um caminho especificado. Para o experimento na sala, foram utilizados oito marcadores visuais. Os locais exatos de todos os marcadores de referência (ou seja, em ambos os cenários de referência) foram pesquisados manualmente, utilizando uma fita métrica |
| **Tem apoio ferramental. Qual?** | • A LADAR chamada (Detecção Laser and Ranging) mede a distância de superfícies refletoras de obstáculos. |
| **Abordagem Hibrida** | • Sim |
| **Qual é o resultado da pesquisa?** | • Técnica<br>• Modelo<br>• Método |
| **Resultados** | • Verificou-se que o movimento holonômica(“ holonomic motion”) do corpo humano leva a surgimento de ruídos nos locais previstos.<br>• Os resultados mostram uma diminuição significativa no erro de estimativas(“ pose estimation”) em comparação com a configuração inicial.  De facto, a taxa de erro é reduzida para metade a quantidade medida anteriormente.<br>• Mostrou-se que a existência de obstáculos distinguíveis na faixa do scanner,  melhoram o desempenho de verificação correspondente e, consequentemente, as estimativas. |
| **Limitações** | Os sistemas baseados em navegação inercial podem rastrear os movimentos de pedestres em edifícios. No entanto, quando esta localização é realizada em áreas previamente desconhecidas, a sua utilidade é limitada, informações tais como mapas NAO estão disponíveis.<br><br>A aquisição incremental de mapas durante a exploração de ambientes desconhecidos constitui um problema fundamental em robótica móvel. Quando diferentes obstáculos foram detectados em um corredor, o erro de localização pode ser reduzido para metade (12-6 metros). |
| **Trabalhos Futuros** | A fim de superar os problemas encontrados de correspondência de digitalização com um scanner a laser de curto alcance, uma versão modificada do "scanner" descrito no serão utilizados e avaliados em experimentos futuros. Além disso, a localização de sistemas melhorados de pedestres será integrada, a fim de proporcionar uma melhor estimativa inicial.<br><br>Equipando o pedestre com uma câmera visual também pode ser uma alternativa para obter uma melhor informação usando odometria visual 3D.<br><br>A combinação desta abordagem baseado em RFID pode superar os problemas encontrados durante os fechamentos de laço. Outro passo será a aplicação deste cenário para adquirir mapas 3D do ambiente. |
| **Algoritmos** | NA |

11)

| Dados do ARTIGO | SEAMLOC: Seamless Indoor Localization Based on Reduced Number of Calibration Points |
|---|---|
| **Autor** | Milan D. Redži´c, Conor Brennan, and Noel E. O'Connor |
| **Ano** | 2014 |
| **Keywords** | WLAN, fingerprinting, indoor localization, Bayesian approach, linearization, performance measures |
| **Siglas importantes** | - Calibration points (CPs);<br>- Access points (APs).<br>- Experimental setups (ESs): |
| **Objetivo do Artigo** | Neste trabalho é apresentado uma abordagem que requer um número reduzido de CPs (até 4 vezes menos do CPs em comparação a outras abordagens) é apresentado. |
| **Qual é o tipo de pesquisa?** | • Empírica;<br>O método foi testado e os resultados foram apresentados em duas ESs |
| **Especificação do tipo da técnica** | A redução de CPs é conseguida através da utilização de uma abordagem de interpolação que permite localizar os participantes em pontos entre CPs. |
| **Descrição da técnica** | A interpolação baseia-se na construção robusta,"range" e dependente do ângulo. Desta forma as funções de probabilidade descrevem a probabilidade de um de usuário estar na proximidade de um CP.<br><br>"Probabilistic WLAN-based localization techniques based on fingerprinting"  começam com a aquisição de observações de treinamento que consistem em informação de intensidade do sinal nos pontos de calibração (CPs) distribuídos ao longo de uma rede densa em todo o edifício. |
| **Utiliza algum Algoritmo de Localização?** | NA |
| **Cobertura** | Precisões de 2-3 metros, utilizando cerca de 70 CPs que são colocadas de maneira não uniforme, pelo menos, a cada 2,5 metros |
| **Contextualização** | Um total de 500 observações de teste foram examinadas para cada ES. Um GRID de 3 × 5m2  foi utilizado em ES1 e ES2. Os experimentos foram realizados em momentos de níveis consideráveis de presença de pessoas em todo o ambiente.Foram eles:<br>1 O primeiro experimento( ES1), cerca de 60 CPs foram distribuídos ao longo de uma grade, mesmo cobrindo 834.08m2.<br>2 O segundo experimento foi realizado no segundo andar do edifício "Applications Dublin City University Computer" composto por escritórios pessoais, de tamanho médio 9m2.  Havia em cada escritório duas CPs, produzindo um total de 32 CPs. A espessura da parede entre escritórios foi de 15 cm. |
| **Tem apoio ferramental. Qual?** | NA |
| **Abordagem Hibrida** | NA |
| **Qual é o resultado da pesquisa?** | • Comparações de abordagens em 2 cenários específicos. |
| **Resultados** | São apresentados resultados de comparações entre este e outros métodos de localização demonstrando sua robustez e melhor desempenho sobre os outros em termos de precisão. Além disso menos CPs são usados do que em outros métodos, o tornando viável e fácil de implantar. Isto reduz o tempo de calibração, enquanto o tempo de processamento é muito semelhante a outras abordagens. |
| **Limitações** | NA |
| **Trabalhos Futuros** | O trabalho futuro será investigar a possibilidade de rastreamento perfeitamente um usuário em um ambiente Indoor. |
| **Algoritmos** | NA |

12)

| | |
|---|---|
| Dados do ARTIGO | Fuzzy Logic-Based Compensated Wi-Fi Signal Strength for Indoor Positioning |
| **Autor** | Akeem Olowolayemo, Abu Osman Md Tap, Teddy Mantoro |
| **Ano** | 2013 |
| **Keywords** | NA |
| **Siglas importantes** | – Access Points (APs); |
| **Objetivo do Artigo** | Este trabalho propõe uma abordagem determinação localização Indoor que usa Aggregation Weighted Fuzzy das potencialidades do sinal recebido (RSS) de Wi-Fi com Compensado fator de atenuação Weighted (CWAF) sob a forma de qualidade de sinal Fuzzy ponderada e ruído.<br><br>É fornecido um algoritmo eficiente computacionalmente que possui acurácia para localização indoor, enquanto ao mesmo tempo elimina as calibrações com ruído e coleta de dados históricos. A abordagem fornece informação de localização em menos de dois segundos, incluindo a captura, a extração, bem como a computação. Isto reduz a possibilidade de posições erradas como resultado um atraso de tempo de cálculo de posição, que ao mesmo tempo coloca menor carga sobre o consumo de energia do dispositivo móvel. |
| **Qual é o tipo de pesquisa?** | • Empírica; |
| **Especificação do tipo da técnica** | A ideia consiste em usar a combinação de qualidade do sinal e ruído associado para melhorar os pesos das potências de sinal para melhor precisão de localização. |
| **Descrição da técnica** | Os dados capturados são coletados em sequência e bem rotulados no banco de dados com “time stamps” para garantir que as variações de sinal devido à flutuação sejam armazenadas. Isso ocorre especialmente como as atualizações de dados de sinal são por segundo e os resultados podem ser inferidos a partir dos dados capturados. Todas as flutuações e efeitos ambientais |
| | sobre o sinal em forma de WAF, FAF, objetos em movimento etc., são considerados como ruído |
| **Utiliza algum Algoritmo de Localização?** | NA |
| **Cobertura** | NA |
| **Contextualização** | O experimento foi instalado em dois laboratórios com um total de 8 pontos de acesso disponíveis. Um dos laboratórios é restrito enquanto o outro é uma área de trabalho com circulação. O software de captura de sinal foi capaz de adquirir informações sobre o sinal a partir de pontos de acesso em tempo real e registra as informações por segundo |
| **Tem apoio ferramental. Qual?** | NA |
| **Abordagem Hibrida** | NA |
| **Qual é o resultado da pesquisa?** | • Abordagem |
| **Resultados** | The results are compared with locations away from APs with actual physical measurement in the environmental location to verify accuracy. The performance of the proposed algorithm shows that if the normalized weighted signal strength is properly compensated with weighted signal quality and noise, the approach offers a more computationally efficient positioning with adequate accuracy for indoor localization. |
| **Limitações** | Os resultados são comparados com locais afastados dos pontos de acesso (APs) com medição física real na localização ambiental para verificar a precisão.  O desempenho do algoritmo proposto mostra que, se a intensidade do sinal ponderada normalizada está devidamente equilibrada com a qualidade do sinal e ruído ponderada, a abordagem oferece um posicionamento eficiente computacionalmente com precisão adequada para localização Indoor.\`<br><br>A partir dos resultados, é evidente que o número mais elevado de pontos de acesso (APs) influencia diretamente nos resultados. No entanto, pode-se perceber que a exatidão da utilização de 7 APs parece ideal. Isto é porque a maioria das vezes (75%), uma precisão de menos de 1 m é atingida, enquanto que um aumento de 8 pontos de acesso não melhora |
| | significativamente os resultados mais. |
| **Trabalhos Futuros** | Trabalho Futuros será necessário para sintonizar o algoritmo a fim de determinar todas as fontes de erros que possam ter sido geradas neste resultado inicial.  Diferentes dispositivos como dispositivos baseados em Android, smartphones etc está sendo considerados em planos futuros para avaliar o algoritmo, uma vez que serão os dispositivos a serem usados, eventualmente, se o algoritmo vier a ser desenvolvido. Isso é necessário para mitigar erros "viés de recepção do sinal", devido às variações de dispositivas sensibilidades e antenas. |
| **Algoritmos** | NA |

13)

| Dados do ARTIGO | Compressive Sensing Indoor Localization |
|---|---|
| **Autor** | Arash Tabibiazar, Otman Basir |
| **Ano** | 2011 |
| **Keywords** | compressive sensing, \`1=\`0-norm minimization, localization, wireless sensor network |
| **Siglas importantes** | – Compressive-Sensing (CS);<br>– Threshold (THR). |
| **Objetivo do Artigo** | O método de localização proposto é uma abordagem de detecção de compressão para a localização indoor com base em medições de sinal esparsas e inconsistentes onde o sinal GPS não está disponível. O problema de localização é formulado |
| | como aproximação esparsa de uma matriz “sparsifying” (dicionário), que os seus elementos são medições do sinal recebido (por exemplo, RSS ou TOA) em pontos discretizadas(“ discretized points”) ou grades(“grids”). O objetivo deste trabalho foi investigar as capacidades teóricas CS em recuperar o vetor localização na rede de sensores wireless. |
| **Qual é o tipo de pesquisa?** | • Empírica; |
| **Especificação do tipo da técnica** | Esta abordagem utiliza um pequeno número de medidas inconsistentes para encontrar o local através de uma grade espacial não simétrica de 1 dispositivo wireless. Neste método, um programa de minimização norma 1 é usado para recuperar a localização do usuário wireless. O desempenho do método proposto é avaliado por meio de nós apenas rever os conceitos básicos em sensoriamento compressão teoria (CS). Esta teoria recentemente introduzida permite a reconstrução de sinais dispersos ou compressíveis a partir de um pequeno conjunto de medidas lineares, não adaptativas que podem ser muito menor do que o número de amostras de taxa de Nyquist. |
| **Descrição da técnica** | . Dois conjuntos de medições estão disponíveis para RSS e de sinal TOA. A potência de transmissão foi de 10MW em 2.443GHz frequência central e relação sinal-ruído foi mantida superior a 25dB para compensar o ruído e interferência “ISMband”. Para medições TOA, ambos TX e RX são sincronizados por 1 pulso por segundo sinais (1 PPS). O desvio padrão da base de tempo (_2ns) foi obtido nestas medições. Todos os dispositivos estão na faixa de uns aos outros, por isso um total de 44_43_5 = 9460 as medições. |
| **SE RFID, qual abordagem utiliza e Cobertura?** | NÃO |
| **Utiliza algum Algoritmo de Localização?** | NA |
| **Cobertura** | O desempenho do método proposto na presença de ruído para diferentes relações sinal-para-ruído é fixada a 10 mW. |
| **Contextualização** | Nesta configuração de simulação, desenvolvido em MATLAB, e para o ambiente de teste é um escritório particionado com quarenta e quatro sensores localizados em coordenadas identificadas com em uma área 14m_13m. |
| **Tem apoio ferramental. Qual?** | • Sim<br>MATLAB. |
| **Abordagem Hibrida** | • Sim |
| **Qual é o resultado da pesquisa?** | • Método<br>Este trabalho começou com a formulação do problema e propor um método de localização a partir de medidas esparsas e inconsistentes com base na teoria de detecção de compressão. |
| **Resultados** | O desempenho do método depende de vários parâmetros como a razão sinal-tonoise(“ signal-tonoise ratio”), refinamento do target, o número de medições, o esquema de programação linear, a implantação do sensor e topologia da rede.<br><br>Como esperado, a potência de ruído tem como principal efeito sobre o desempenho de localização usando informações RSS. |
| **Limitações** | O desempenho do método depende de vários parâmetros, incluindo o ruído de medição _, então em sinal recuperado, pode encontrar Ci componentes diferentes de zero em todo o índice exata do alvo que deve ser considerada para aumentar o desempenho do sistema. Foram utilizados os valores normalizados dos coeficientes de teses(“theses”) para localizar o target com precisão. Um limiar (THR) foi definido na soma dos componentes normalizados Ci para selecionar n maiores componentes. |
| **Trabalhos Futuros** | Como o trabalho futuro, vai ser investigado diferentes algoritmos de otimização para recuperar o sinal a partir de medições esparsas. Também atualização matriz “sparsifying” (dicionário) dinamicamente é uma maneira interessante para melhorar o desempenho do sistema. A aplicação deste método aos dados sensoriais maciços em redes de sensores sem fio pode melhorar a utilização de recursos em muitas aplicações |
| **Algoritmos** | NA |

14)

| Dados do ARTIGO | A comprehensive approach for optimizing To A localization in harsh industrial environments |
|---|---|
| **Autor** | Andreas Lewandowski and Christian Wietfeld |
| **Ano** | 2010 |
| **Keywords** | indoor localization, weighted multilateration, optimal anchor positioning, power management, self-interference |
| **Siglas importantes** | – Chirp Spread Spectrum (CSS);<br>– Time of Arrival (ToA);<br>– Ultra Wideband (UWB);<br>– Received Signal Strength Indicator(RSSI) |
| **Objetivo do Artigo** | Este trabalho apresenta uma solução para a geração de 1 " Weighted Multilateration Position Estimation" baseada no RSSI para TOA variando medidas para melhorar a tolerância a falhas, reduzir a auto interferência do sistema de telemetria e, finalmente, aumentar a precisão a estimativa de posição. |
| **Qual é o tipo de pesquisa?** | • Empírica; |
| **Especificação do tipo da técnica** | Ambientes industriais reais constituem um desafio para os sistemas de localização via rádio de hoje, como o canal de rádio é altamente afetada pela interferência múltipla presentes no ambiente.<br><br>As seguintes medidas são baseadas no Real-Time System Nanotron ToA Localização (RTLS) [15], que usa a tecnologia CSS e utiliza um canal de 80 MHz na faixa de 2,4 GHz. A camada física é normalizada em IEEE 802.15.4a- CSS como uma execução alternativa |
| **Descrição da técnica** | Este trabalho demonstra de forma iterativa como aperfeiçoar o desempenho de sistemas de localização Toa para aplicações relativas ambientes industriais da vida real. Portanto, as seguintes etapas de otimização são executados: |
| | 1. A combinação dos dois métodos que variam - Toa e RSSI - for afetada por alterações dinâmicas do ambiente, que ocorrem por pessoas ou estoque em movimento;<br>2. Embora o posicionamento do nó âncora tenha um grande impacto sobre o desempenho dos sistemas de localização, não muitas soluções para o melhor posicionamento âncora estão disponíveis. Portanto, um algoritmo de descida coordenar, descrito neste trabalho, faz vários “requests” para encontrar uma melhor distribuição do no da âncora.<br>3. É apresentado um método para calibração do sistema em termos de potência de transmissão para mitigar auto interferências antes das principais contribuições |
| **Utiliza algum Algoritmo de Localização?** | NA |
| **Cobertura** | Foi constatado durante as medições que o desvio varia de 5m de precisão 1m. A variação da estimativa de posição é devida a variação no canal de rádio e o hardware da abertura do receptor. |
| **Contextualização** | Este trabalho faz parte de um projeto de pesquisa que está sendo realizada com maior fabricante de aço ThyssenKrupp Steel Europe da Alemanha (TKSE). Eles vão instalar a solução proposta para aumentar a segurança dos empregados da fábrica em casos de emergência. O sistema desenvolvido está integrado num rede de sensores de gás, que consiste de equipamento imóvel e móvel.<br><br>O cenário considerado é composto por uma unidade de alimentação 26.5m 7m x no porão. Oito sensores de gás fixos equipados com módulos de RF vão ser colocados no cenário como pontos de ancoragem. À medida que a estrutura do cenário é muito complexa, realizou-se um varrimento a laser 3D para gerar uma imagem digital, que pode ser aplicado na simulação “raytracing” para o planeamento da rede wireless. A fim de cumprir o processo de planejamento de rede com uma alta precisão, medição exata do cenário de aplicação é obrigatória neste caso. |
| **Tem apoio ferramental. Qual?** | A distância entre dois nós é estimada pela SDS-TWR protocolo, o que evita a necessidade de nós da rede síncrona e também aumenta a robustez, como as estimativas de distância são realizadas duas vezes (Two-Way Variando Symmetrical frente e verso). |
| **Abordagem Hibrida** | • Sim<br><br>A combinação dos dois métodos que variam - ToA e RSSI<br>. |
| **Qual é o resultado da pesquisa?** | • Técnica/Método<br><br>Este artigo apresenta uma abordagem global para aumentar a precisão de um sistema de localização baseado em ToA sem fazer modificações no sistema em profundidade. |
| **Resultados** | No final, foi demonstrado que a precisão de localização pode ser aumentada em mais de 75% por aplicação das tecnologias avaliadas ao longo deste artigo. Os resultados da avaliação do desempenho demonstram que a distribuição de um nó de ancoragem ótima afeta fortemente o desempenho do sistema de localização.<br><br>Além disso, um melhor posicionamento âncora pode, principalmente, aumentar a precisão de localização. Mostrou-se que uma distribuição ótima economiza nós âncora, mesmo com a concessão de alta precisão de localização.<br><br>Weighted ToA além disso apresenta um comportamento discreto( smooth”), sem um elevado número de outliers. Weighted ToA é sempre superior em comparação à toa em todas as etapas de otimização, embora o ajuste da potência de transmissão aumenta a precisão dos ToA claramente. Através da aplicação de todos os passos de optimização, que têm demonstrado que um ganho de 75% em comparação com ingénuos ToA é possível. |
| **Limitações** | Existem ainda algumas posições com um desvio máximo de 20m, o que significa que a localização não é possível nestes casos. |
| **Trabalhos Futuros** | Como trabalhos futuros, é considerado o desenvolvimento de um procedimento de ajuste da potência de transmissão dinâmica para readequar o sistema de localização de acordo com a mudança do cenário em avaliação.. Além disso, uma abordagem de algoritmo genético para  o posicionamento da  ótima âncora poderia melhorar o desempenho com menor custo computacional |
| **Algoritmos** | NA |

15)

| Dados do ARTIGO | Multimode Radio Fingerprinting for Localization |
|---|---|
| **Autor** | E. Martin |
| **Ano** | 2011 |
| **Keywords** | Mobile communication, radio communication, wireless LAN. |
| **Siglas importantes** | – Radio Signal Strength Indications (RSSI). |
| **Objetivo do Artigo** | Foi apresentada uma aplicação integrando as fases tanto offline como online de fingerprinting no mesmo smartphone, e fornecimento uma precisão de até 1,5 metros. |
| **Qual é o tipo de pesquisa?** | • Empírica; |
| **Especificação do tipo da técnica** | Neste artigo foi estudado o a viabilidade de aproveitar diferentes tecnologias de rádio para localização baseada em fingerprinting em uma abordagem multimodo. As duas principais abordagens para a tomada de estimativas de localização que utilização de valores de RSSI são:<br>1)"fingerprinting", em que um mapa de rádio pré-gravado da área de interesse é empregado para inferir locais através de melhor ajustamento;<br>2) "Propagation based", no qual os valores de RSSI são utilizados para calcular as distâncias através do cálculo da perda de caminho. |
| **Descrição da técnica** | Foi desenvolvida uma aplicação que representa a primeira abordagem para sistemas baseados em fingerprinting para a tomada de localização através do uso de smartphones e integrando tanto o treinamento e as fases on-line no mesmo dispositivo. Este aplicativo de localização é a primeira entrega fazendo uso de apenas o hardware incorporado dentro do telefone e integrando as fases online e offline de RSSI “fingerprinting” dentro do mesmo dispositivo.<br>Adicionalmente foi gerada uma nova abordagem para o tratamento estatístico dos dados RSSI nos telefones, superando técnicas determinísticas existentes. Além disso, foi analisada a viabilidade de uma abordagem multimodal para WiFi aproveitamento de localização baseada em impressões digitais, comunicações celulares e DTV. |
| **Utiliza algum Algoritmo de Localização?** | NAO |
| **Cobertura** | Foi entregue uma precisão de até 1,5 metros, e Droid só fornece mais alguns valores intermediários que variam de -56 dBm a -115 dBm). |
| **Contextualização** | A WiFi signal represents the most reliable approach for indoor localization in our experimental setup in buildings across the University of California, Berkeley campus.<br><br>Um sinal WiFi representa a abordagem mais confiável para a localização Indoor, neste artigo a configuração experimental foi realizada em edifícios em toda da Universidade da Califórnia - Berkeley campus. |
| **Tem apoio ferramental. Qual?** | • Sim<br><br>The RSSI values measured with a Dell Latitude laptop and those measured with a Motorola Droid cell-phone.<br><br>It was used smartphones running on Android, in particular the G1 and the Droid |
| **Abordagem Hibrida** | • Sim<br><br>Neste artigo, foi examinada a viabilidade de uma abordagem multimodal para a localização Indoor, aproveitamento três dos sinais de rádio frequência mais difundidas dentro de edifícios: - Wi-Fi, comunicações celulares e DTV. . |
| **Qual é o resultado da pesquisa?** | • Aplication |
| **Resultados** | Foi implementado uma aplicação Android para a localização, e nós testamos em locais onde 25 “WiFi rádios” em média foram escutadas (cerca de 40% deles com RSSI acima de -80 dBm), a obtenção de precisões da ordem de 1,5 metros até mesmo dentro de um mesmo ambiente e com dinamicidade em tempo real |
| **Limitações** | A fim de minimizar os erros devido a proximidade com o corpo humano, o telefone celular deve ser manuseado durante a fase de formação tão próximo quanto possível das condições normais, em que será utilizado na fase de medição. |
| **Trabalhos Futuros** | NA |
| **Algoritmos** | NA |

16)

| Dados do ARTIGO | WILL: Wireless Indoor Localization without Site Survey |
|---|---|
| **Autor** | Chenshu Wu,Zheng Yang, Yunhao Liu, Wei Xi |
| **Ano** | 2013 |
| **Keywords** | Wireless, indoor localization, fingerprint, site survey |
| **Siglas importantes** | Received Signal Strength (RSS) |
| **Objetivo do Artigo** | Uma abordagem de localização interna com base em infra-estrutura WIFI off-the-shelf e telefones celulares. |
| **Qual é o tipo de pesquisa?** | • Empírica; |
| **Especificação do tipo da técnica** | Neste método, a localização é dividida em duas as fases: Treinamento e “Serving”. Na primeira fase, os métodos tradicionais envolvem um processo de pesquisa de local, no qual engenheiros registram as RSS de “fingerprint” (por exemplo, as forças de sinal Wi-Fi a partir de vários pontos de acesso, a APS) em cada posição de uma área interessante e consequentemente construir um banco de dados de “fingerprint”. Em seguida, na fase de servir, quando um usuário envia uma consulta com o seu de localização atual de RSS fingerprint, algoritmos de localização recuperar o banco de dados de impressões digitais e |
| | retornar as fingerprint correspondentes, bem como locais correspondentes. |
| **Descrição da técnica** | Neste estudo, propomos WILL, uma abordagem de localização lógica Indoor, que explora os movimentos do usuário a partir telefones celulares. A lógica por trás WILL é que os movimentos humanos podem ser aplicados para se conectar assinaturas de rádio anteriormente independentes sob certas semântica.<br><br>WILL requer nenhum conhecimento prévio de locais AP, e os usuários não são necessários para a participação explícita de localizações correspondentes, mesmo na fase de treinamento. |
| **Utiliza algum Algoritmo de Localização?** | NA |
| **Cobertura** | WILL is deployed in a real building covering over 1600 m2 |
| **Contextualização** | Foi implantado sistema em um andar de um edifício de escritórios que cobrem mais de 1600 m2 na Universidade Tsinghua, que contém 16 escritórios, dos quais cinco são grandes salas de 142 m2, sete são pequenos com tamanhos diferentes e os outros quatro são inacessíveis. |
| **Tem apoio ferramental. Qual?** | • Sim<br><br>WILL explora acelerômetros para obter os movimentos do usuário, que será ainda mais utilizado para auxiliar a localização. Acelerômetros triaxiais fornecem evidência aparente de comportamento do andar humanos. Nós desenvolvemos o cliente de vontade por OS Android cada vez mais popular. Sinais de WiFi são gravados com a frequência de cerca de duas vezes por segundo, quando a medição.<br><br>No protótipo em dois telefones Google Nexus S, que suportam Wi-Fi e contêm sensores de acelerômetro. |
| **Abordagem Hibrida** | • Sim<br>A idéia principal é combinar as impressões digitais WiFi fingerprints com os movimentos do usuário. |
| **Qual é o resultado da pesquisa?** | • Técnica/Processo/Método<br>Neste estudo, propomos WILL, uma abordagem de localização lógica Indoor wireless. |
| **Resultados** | Os resultados dos experimentos mostram que a localização Indoor baseada em RSS pode alcançar a precisão da localização mesmo sem um “site survuey”.  A precisão média de localização ambiente, ou seja, a precisão de localização de “fingerprints” é efetivamente coletada, é de mais de 80 por cento, que é competitivo para as soluções existentes. |
| **Limitações** | NA |
| **Trabalhos Futuros** | Mapeamento do comportamento do usuário fará do WILL um importante sistema de localização indoor. |

17)

| | |
|---|---|
| Dados do ARTIGO | Indoor Robot/human Localization Using Dynamic Triangulation and Wireless Pyroelectric Infrared Sensory Fusion Approaches |
| **Autor** | Ren C. Loo, Ogst Chen2, Pei Hsien Lin |
| **Ano** | 2012 |
| **Keywords** | wireless pyroelectric infrared (WPIR) sensory fusion, pyroelectric sensor, data fusion, localization, dynamic triangulation |
| **Siglas importantes** | - System-an-Chip (SOC);<br>- Received signal strength (RSS)<br>- strength indicator (RSSI)<br>- Pyroelectric Infrared (PIR) |
| **Objetivo do Artigo** | Este sistema tem como objetivo localizar com precisão um alvo a fim de realizar serviços relevantes entre robô e humano.Um sistema de localização WPIR pode monitorar vários alvos com relativa boa resolução |
| **Qual é o tipo de pesquisa?** | • Empírica;<br>Foi desenvolvido e demonstrado experimentalmente um sistema de fusão sensorial WPIR que pode ser aplicado com sucesso na localização de "targets", tais como pessoas e robô. Com um mecanismo de localizar com precisão para a telha("tile") de ambiente Indoor, a prestação de serviços de telha adequadas para as pessoas podem ser realizados. |
| **Especificação do tipo da técnica** | Foi proposto método dinâmico de triangulação (DTN), para reduzir o erro de RF localização através da telha. Foi desenvolvido um algoritmo fusão sensorial chamado de algoritmo de inferência WPIR. Este algoritmo determina a posição consolidada, tanto do sistema de localização PIR e frequência de rádio sistema de localização de sinais que utilizam intensidade do sinal recebido (RSS). |
| **Descrição da técnica** | Neste artigo, foi desenvolvido um sistema de sensores wireless / "pyroelectric", que se integra com um módulo de comunicação wireless e um sensor de ZigBee PIR como um dispositivo WPIR, desenvolvido um método de localização DTN para reduzir o erro de método de triangulação baseado do RF. Foi desenvolvido um algoritmo de inferência WPIR para a localização de pessoas e robôs. Vários cenários foram testados como nenhuma linha de testes de visão, implantação regular, implantação aleatória, único e múltiplos alvos.<br><br>O algoritmo de inferência WPIR fornece relativamente elevada precisão do que os sistemas de localização de PIR ou DTN, respectivamente, e o limite de erro também pode ser reduzida. O sensor WPIR integrado pode ser montado no teto em casa, construção, escritório ou fábrica para localização humano e robô ou outros aplicativos de rede sensor. |
| **Utiliza algum Algoritmo de Localização?** | Foi proposto um algoritmo de inferência WPIR que pode gerar uma estimativa o posicionamento mais confiável dos targets.. |
| **Cobertura** | O erro médio de TN é 2.2m ea média de erro de DTN é I.82m. A elipse de erro é menor do que DTN TN. O erro quadrado médio do sistema de rastreamento PIR é 1,76M. |
| **Contextualização** | Foi realizado um experimento em um ambiente espaçoso, com uma área de 30mx30mx3m. O protótipo do sistema sensorial wireless / pyroelectric foi instalado no teto para o teste. |
| **Tem apoio ferramental.** | • Sim |
| **Qual?** | Para localização de freqüência de rádio, foi utilizado ZigBee chip de protocolo / CC2431 que pode gerar informações de identificação do dispositivo dentro de pacotes wireless. |
| **Abordagem Hibrida** | • Sim<br>Neste estudo, foi desenvolvido um sistema que combina sistema de localização PIR e do RF como sistema wireless pyroelectric infravermelho fusão sensorial para monitorar as informações de localização de robôs e pessoas |
| **Qual é o resultado da pesquisa?** | • Método<br>• Algoritmo |
| **Resultados** | The testing target walks through the WPIR monitoring region and the reported trace from the PIR tracking system is shown in Fig. 7. The mean square error of the PIR tracking system is 1.76m.<br><br>O taget de testes anda pela região de monitoramento do WPIR e o rastreamento relatado a partir do sistema de rastreamento PIR. O erro quadrado médio do sistema de rastreamento PIR é 1,76M . |
| **Limitações** | O sistema de detecção PIR não pode operar durante o acompanhamento de alvos múltiplos, ea média de erro para o sistema de localização DTN é 1,83 m. O erro médio do algoritmo de inferência WPIR para o sistema de monitoramento de múltiplos alvos é O.73m |
| **Trabalhos Futuros** | NA |
| **Algoritmos** | NA |

## 4.2. Fonte Scopus

18)

| Dados do ARTIGO | GROPING: Geomagnetism and cROwdsensing Powered Indoor NaviGation |
|---|---|
| **Autor** | Chi Zhang, Kalyan P. Subbu, Jun Luo, and Jianxin Wu, Member, IEEE |
| **Ano** | 2015 |
| **Keywords** | Indoor navigation, indoor localization, geomagnetism, mobile crowdsensing |
| **Siglas importantes** | – GoogleMaps Indoor (GMI);<br>– Geomagnetism and cROwdsensing Powered Indoor NaviGation( GROPING). |
| **Objetivo do Artigo** | Foi proposto "Geomagnetism and cROwdsensing Powered Indoor NaviGation" (Groping) como um completamente auto-suficiente, leve e prático para a navegação indoor. |
| **Qual é o tipo de pesquisa?** | • Empírica;<br><br>Os experimentos aplicados com Groping demonstram a sua usabilidade e também mostram a comparação favorável com os sistemas típicos de localização baseados em WIFI no apoio a navegação Indoor. |
| **Especificação do tipo da técnica** | O design se aplica ao magnetômetro e explora o geomagnetismo como a localização indicando fingerprint: é leve (apenas um vetor 3D) e muito estável, e é completamente independente de qualquer tipo de infra-estrutura sem fio. |
| **Descrição da técnica** | Apesar do grande número de propostas sobre a localização Indoor, o único sistema que está disponível para o teste publica é o Google Maps Indoor (GMI). Por isso, foi organizado um grupo de 11 pessoas para realizar um estudo detalhado sobre isso demonstrado nos passos abaixo:<br>1) A primeira parte é um estudo de campo em cinco grandes shoppings centers (acima de 10.000 m2) para testar a precisão do GMI, bem como para ter certeza se a WIFI é utilizada por GMI (que parece ser verdadeiro);<br>2) A segunda parte, assumindo de WIFI é a principal fonte de GMI, é um teste de laboratório sobre a estabilidade do |
| | "WIFI fingerprint";<br>3) A terceira parte foi realizada a avaliação da eficiência energética dos sistemas de localização baseados em de WIFI. |
| **Utiliza algum Algoritmo de Localização?** | NA |
| **Cobertura** | NA |
| **Contextualização** | Foi motivado pelo serviço de navegação incompetente do Google Maps Indoor, por objetivo de eliminar a forte dependência de uma de 1 infra-estrutura WiFI e de forma positiva a utilização de mapa do solos |
| **Tem apoio ferramental. Qual?** | • Não |
| **Infra** | Samsung Galaxy S2 |
| **Abordagem Hibrida** | • NA |
| **Qual é o resultado da pesquisa?** | • Método/sistema<br>Foi proposto tateando, um sistema " all-in-one"que inclui geração de mapas, localização e navegação |
| **Resultados** | • Utilizar "crowdsensing" para o fingerprinting magnética e para a construção de um mapa de chão de um conjunto arbitrário de curta trajetória,<br>• Realizar localização superficial e daí navegação baseada em "fingerprints" magnéticos e os mapas gerados. |
| **Limitações** | Foi demonstrada desvantagem em eficiência energética e estabilidade de fingerprint. |
| **Trabalhos Futuros** | NA |
| **Algoritmos** | NA |

19)

| Dados do ARTIGO | Experiencing and Handling the Diversity in Data Density and Environmental Locality in an Indoor Positioning Service |
|---|---|
| **Autor** | Li, L. and Shen, G. and Zhao, C. and Moscibroda, T. and Lin, J.-H. and Zhao, F. |
| **Ano** | 2014 |
| **Keywords** | Indoor Localization; Fingerprint; Model |
| **Siglas importantes** | - Fingerprintcloud (FP-Cloud);<br>- Virtual fingerprints (VFPs). |
| **Objetivo do Artigo** | Foi proposto Modellet, uma abordagem algorítmica que de forma otimizada aproxima o mapa de rádio real por unificação do ”model  based” e abordagens fingerprints. |
| **Qual é o tipo de pesquisa?** | • Empírica; |
| **Especificação do tipo da técnica** | Modellet representa o mapa de rádio usando uma nuvem (“cloud”) de fingerprint que incorpora ambos “fingerprints” (real e virtual), os quais  que são computados a partir de modelos com um suporte local, com base na chave do conceito apoio conjunto. |
| **Descrição da técnica** | Modellet consiste em um processo de construção  banco de dados off-line (FP-Cloud)  e um processo de inferência de localização online. Para gerar o FP-Cloud, um local é primeiro examinado por especialistas (“cite survey  ou |
| | Crowdsources"),para em seguida se obter os dados de treinamento. Os dados de treinamento são diretamente incorporadas ao FPCloud, e as posições das impressões digitais virtuais (VFPs) são então determinados como locais de interesse. Para cada local de interesse, o conjunto de suporte é identificado. |
| **Utiliza algum Algoritmo de Localização?** | Estabeleceram-se como diferentes famílias de algoritmos de localização apresentam resultados diferentes em diferentes densidades de dados e condições de ambiente. |
| **Cobertura** | NA |
| **Contextualização** | Modellet foi avaliada com dados coletados a partir de um prédio de escritórios, assim como em 13 grandes centros comerciais (compras shoppings e aeroportos), localizados em toda a China, EUA e Alemanha. |
| **Tem apoio ferramental. Qual?** | NA |
| **Infra** | NA |
| **Abordagem Hibrida** | • Sim |
| **Qual é o resultado da pesquisa?** | • Algoritmo. |
| **Resultados** | Comparando Modellet com duas abordagens de base, RADAR e EZPerfect, demonstra que efetivamente Modellet se adapta a diferentes densidades de dados e condições ambientais, superando substancialmente abordagens existentes. |
| **Limitações** | NA |
| **Trabalhos Futuros** | Checar o problema de manutenção no banco de dados. |
| **Algoritmos** | NA |

20)

| Dados do ARTIGO | SAIL: Single Access Point-Based Indoor Localization |
|---|---|
| **Autor** | Mariakakis, A.T. and Sen, S. and Lee, J. and Kim, K.-H. |
| **Ano** | 2014 |
| **Keywords** | Indoor location; Smartphones; Dead-Reckoning; Time-of-Flight; Sensing |
| **Siglas importantes** | - Physical layer information (PHY);<br>- Access points (APs);<br>- Channel Impulse Response (CIR). |
| **Objetivo do Artigo** | Computa a distância entre o cliente e um ponto de acesso usando o atraso de propagação do sinal entre os dois, combinando a distância com as técnicas de cálculo de posição de smartphones, e emprega métodos geométricos para se obter a localização do cliente através de um único ponto de acesso. |
| **Qual é o tipo de pesquisa?** | • Empírica;<br>Nos experimentos demonstrados, as melhorias propostas reduziram a média de cálculo de posição de erro dos regimes existentes de 7m a 4m. |
| **Especificação do tipo da técnica** | Supõe-se que o usuário tem um smartphone conectado a um AP, e desloca-se de localização A para B, ambos os quais são desconhecidos. Pode ser possível estimar a distância do utilizador a partir do AP nesses locais, usando WIFI. |
| **Descrição da técnica** | Foi projetado um algoritmo de estimativa de uma curta distância do acelerómetro base que não necessita de entrada do usuário para a calibração.<br>Foi detectar mudanças de orientação telefone aleatórios usando o acelerômetro do smartphone e aproveitar o giroscópio para |
| | identificar voltas físicas.<br>Ele foi projetado também exige que o rumo da bússola do usuário enquanto anda entre dois locais diferentes.<br>Foi demonstrado que, respondendo por anormalidades no ”magnetometer”, é possível para determinar quando a leitura da bússola é mais confiável. |
| **Utiliza algum Algoritmo de Localização?** | Esta solução usa o filtro de Kalman para reduzir o erro de estimativa da distância e ajusta suas previsões após detectar voltas(“turns”) físicas usando giroscópio do smartphone. |
| **Cobertura** | SAIL pode alcançar um erro de localização mediana de 2,3 m usando apenas um único AP |
| **Contextualização** | Foram recrutados 10 usuários para avaliar SAIL e utilizados 20 diferentes tipos de dispositivos móveis. |
| **Tem apoio ferramental. Qual?** | • Não |
| **Infra** | Esta solução usando WIFI APs e smartphones:<br>Usando HP MSM 460 APs, com chipset Atheros 9590 sintonizado na freqüência 5.805GHz usando uma largura de banda de 40MHz.<br>– Dispositivos móveis baseados no Android, como o Samsung Galaxy S3, Samsung Galaxy S4, Nexus 4, Sony Xperia Z, HTC One e iOS baseado em dispositivos móveis, como o iPhone 5 e iPhone 5s. |
| **Abordagem Hibrida** | • Não |
| **Qual é o resultado da pesquisa?** | • Modelo |
| **Resultados** | Nestes experimentos, estas melhorias reduzidas a média cálculo de posição erro de regimes existentes de 7m a 4m. As técnicas proposto no SAIL podem alcançar um erro médio de 2,3 m. |
| **Limitações** | Foi utilizado respostas ao impulso do canal e da mobilidade humana para eliminar o efeito de sinais baseado na ToF (estimativa de distância)a: por contabilização de múltiplos caminhos, constatou-se que SAIL reduz o erro médio de ToF-base que varia de 5m de 0.8m. |
| **Trabalhos Futuros** | NA |
| **Algoritmos** | NA |

21)

| Dados do ARTIGO | **Optimal Placement of Reference Nodes for Wireless Indoor Positioning Systems** |
|---|---|
| **Autor** | Aomumpai, S. and Kondee, K. and Prommak, C. and Kaemarungsi, K. |
| **Ano** | 2014 |
| **Keywords** | Indoor positioning systems; Reference node placements; System design. |
| **Siglas importantes** | − Reference nodes (rns);<br>− Binary Integer Linear Programming (BILP) problem;<br>− Reference nodes (RNs);<br>− Received signal strength(RSS);<br>− Angle-of-Arrival (AOA);<br>− Time-of-Arrival (TOA) measurement;<br>− Mobile station (MS);<br>− Wireless Local Area Network (WLAN);<br>− Signal test points (stps). |
| **Objetivo do Artigo** | Neste trabalho, foi proposta uma técnica de otimização que podem ser usados para melhorar a disposição de nós de referência e contribuir para o desempenho local**.** |
| **Qual é o tipo de pesquisa?** | Empírica; |
| **Especificação do tipo da técnica** | NA |
| **Descrição da técnica** | Foi proposta uma formulação matemática usando uma abordagem de programação Binary Integer Linear (BILP) que pode determinar o número e locais para instalar RNs ideal na área de serviço “target” do sistema de posicionamento de indoor. Os modelos propostos incorporam fatores essenciais sugeridos no framework de implantação para sistemas Wireless de posicionamento indoor. |
| **Utiliza algum Algoritmo de Localização?** | O algoritmo de estimativa de localização baseia-se no cálculo da distância Euclidiana entre padrões RSSs medidos e todas as “fingerPrints” de localização no mapa de rádio. |
| **Cobertura** | NA |
| **Contextualização** | Do ambiente considera-se a área de serviço de tamanho 63m X 10m. |
| **Tem apoio ferramental. Qual?** | • Não |
| **Infra** | NA |
| **Abordagem Hibrida** | NA |
| **Qual é o resultado da pesquisa?** | • Técnica/Algoritmo<br>Neste trabalho, foi proposta uma técnica de |
| **Resultados** | Os resultados mostram que usando a colocação óptima de nós de referência os sistemas de posicionamento Indoor podem ganhar até cinco metros de precisão com 90% de precisão. |
| **Limitações** | NA |
| **Trabalhos Futuros** | Estudar os efeitos do aumento do número de RNs sobre a precisão dos sistemas de posicionamento wireless indoor. |
| **Algoritmos** | NA |

22)

| Dados do ARTIGO | Travi-Navi: Self-deployable Indoor Navigation System |
|---|---|
| **Autor** | Yuanqing Zhengy,x, Guobin Sheny, Liqun Liy, Chunshui Zhaoy, Mo Lix, Feng Zhaoy |
| **Ano** | 2014 |
| **Keywords** | Indoor navigation, Image direction, Self-deployable system |
| **Siglas importantes** | – IMU (Inertial Measurement Unit). |
| **Objetivo do Artigo** | Travi-Navi registra imagens de alta qualidade durante o curso de uma caminhada do usuário sobre os caminhos de navegação, recolhe um rico conjunto de leituras dos sensores e empacota-los em um rastreamento de navegação. |
| **Técnicas/Abordagens/métodos/algoritmos** | • Magnetic field signals;<br>• Wi-Fi fingerprint sequences;<br>• IMU sensor (i.e., accelerometer, gyroscope, and compass) ;<br>• Based deadreckoning. |
| **Qual é o tipo de pesquisa?** | • Empírica; |
| | Implementado Travi-Navi e realizado experimentos extensivos. |
| **Especificação do tipo da técnica** | A técnica fornece o deslocamento do usuário ao longo do caminho, de maneira a rastrear a navegação, obtendo prontas instruções visuais, dicas de imagem e recebendo alertas quando eles desviam os caminhos corretos. |
| **Descrição da técnica** | Em Travi-Navi, um orientador registra pontos de referência (imagens via, leituras de sensores de sinais de rádio, etc.) ao longo de um caminho e ações dos utilizadores. Usando as instruções do Guider, a aplicação é executada no dispositivo móvel do usuário e orienta automaticamente.<br><br>O usuário, apresentando imagens via e indicando caminhos. Travi-Navi aproveita implicitamente a capacidade de reconhecimento visual do seguidor, fornecendo-lhes as imagens da via para corrigir ligeira ambiguidade de direção e melhorar a experiência de navegação do usuário. |
| **Utiliza algum Algoritmo de Localização?** | – Algoritmo bag-ofvisual-words, o qual utiliza k-means;<br>– Alforitmo ORB;<br>– A versão atual do Travi-Navi envolve codificação Java;<br>– Foi adotada a biblioteca OpenCV (versão 2.4.6) escrito em C para implementar o processamento de imagem e imagem de correspondência via JNI. |
| **Cobertura** | NA |
| **Contextualização** | Os experimentos foram realizados em um andar do prédio de escritórios e em um primeiro andar de um grande shopping, com uma área de cerca 1900m2 e 4000m2 durante diferentes momentos do dia. No geral, foram coletadas 12 caminhos de navegação que abrange todas as principais vias. O comprimento total de navegação foi de aproximadamente de 2,8 km. Participaram 4 voluntários que caminharam muitas vezes em rotas diferentes, com uma distância total de cerca de 10 km nos experimentos. |
| **Tem apoio ferramental. Qual?** | • Não |
| **Infra** | – Samsung Galaxy S2 com um processador de 1.2GHz dual-core<br>– A resolução da imagem está definida para 320x240 que equilibra a qualidade de imagem e a sobrecarga de processamento. |
| **Abordagem Hibrida** | • Sim<br>Foi incorporado distorção de campo magnético e  sinais de Wi-Fi em filtragem de partículas para garantir rastreamento que 1 usuário. |
| **Qual é o resultado da pesquisa?** | • Ferramenta<br>Foi consolidado as técnicas acima e implementar Travi-Navi na plataforma Android. |
| **Resultados** | De maneira à implicitamente alavancar a capacidade de reconhecimento visual dos usuários, foi fornecido imagens dos caminhos para corrigir direção pequena ambiguidade e melhorar experiências de navegação. |
| **Limitações** | • Como controlar o uso de energia e, ao mesmo tempo, garantir a qualidade de imagem permanece uma questão importante para se  abordar;<br>• Os caminhos de navegação são independentemente fornecidos por diferentes “guiders” sem coordenação, por isso é um desafio para encontrar atalhos eficientes entre os caminhos.<br>• Idealmente, esta proposta deve ser capaz de orientar os usuários ao longo do caminho. |
| **Trabalhos Futuros** | Aperfeiçoar Travi-Navi e reduzir o consumo de energia |
| **Algoritmos** | NA |

23)

| Dados do ARTIGO | A Practical RF-based Indoor Localization System Combined with the Embedded Sensors in Smart Phone |
|---|---|
| **Autor** | Li, W. and Li, H. and Jiang, Y. |
| **Ano** | 2014 |
| **Keywords** | Indoor Localization; wireless Propagation Model; Smart Phone Embedded Sensor |
| **Siglas importantes** | − Radio frequency (RF);<br>− Access Points (aps);<br>− Received Signal Strength Indicators (RSSIs). |
| **Objetivo do Artigo** | Foi proposto um método inovador para usar os sensores incorporados no smartphone para sugerir a localização resultante de um sistema à base de RF, e diminuir com sucesso a discrepância do resultado original. |
| **Técnicas/Abordagens/métodos/algoritmos** | Este sistema de localização indoor é baseada em rádio freqüência (RF) obtido a partir do Pontos de acesso (APs). Os Indicadores de potência do sinal recebido (RSSIs) fornecidos pelos APs um do outro são usados para construir a propagação dos modelos de ambientes distintos. |
| **Qual é o tipo de pesquisa?** | • Empírica; |
| **Especificação do tipo da técnica** | A intenção de construir um sistema de localização "à base de propagação" que utiliza o smartphone. Para aumentar a precisão do método, os sensores incorporados no aparelho são usados. |
| **Descrição da técnica** | 1. Foram utilizados os RSSIs recolhidas por APs um do outro para construir diferente modelos de propagação que se adaptam diferentes áreas do edifício;<br>2. A atenuação relevante de RSSI ao invés de sua absoluta do valor recebido pelo receptor é utilizada para calcular a distância, desta forma, elimina-se com sucesso o viés de "Recepção de Sinais" causada por diferentes terminais;<br>3. Os sensores incorporados em smartphones são usados para conseguir estimar a distância e orientação do usuário durante o processo de localização. Com esta informação extra, apenas é necessário recolher a partir RSSIs mais forte em torno de dois pontos de acesso para localizar um utilizador na maior parte dos casos, e depois de se combinar o resultado localização, nas duas extremidades do local, a discrepância diminui para 64% do nível original. |
| **Utiliza algum Algoritmo de Localização?** | NA |
| **Cobertura** | A precisão deste sistema chega a 70% dentro de 2m e 95% dentro de 3m, que pode atender a necessidades da maioria das aplicações indoor. |
| **Contextualização** | A experiência foi feita em dois edifícios de Tsinghua campus. |
| **Tem apoio ferramental. Qual?** | • Não |
| **Infra** | NA |
| **Abordagem Hibrida** | • Sim |
| **Qual é o resultado da pesquisa?** | • Processo<br>• Método |
| **Resultados** | A precisão deste sistema chega a 70% dentro de 2m e 95% dentro de 3m, que pode atender a necessidades da maioria das aplicações indoor. |
| **Limitações** | A discrepância deve ser uma metade de 1m, que é de 0,5 m. |
| **Trabalhos Futuros** | Sensores embutidos mais potentes seriam necessários para se obter RSSIs mais precisos dos APs ao longo do deslocamento dos usuários. |
| **Algoritmos** | NA |

24)

| | |
|---|---|
| Dados do ARTIGO | Enhanced WiFi ToF Indoor Positioning System with MEMS- |
| | based INS and Pedometric Information |
| **Autor** | Schatzberg, U. and Banin, L. and Amizur, Y. |
| **Ano** | 2014 |
| **Keywords** | indoor positioning, indoor localization, WiFi timeof-flight, round-trip-time, time-based range measurement, IMU, INS, inertial sensors fusion, pedestrian dead reckoning, pedometric information |
| **Siglas importantes** | – WiFi time-of-flight (ToF) range measurements;<br>– Access Points(AP);<br>– Deployment geometry (DoP);<br>– Pedometric information(PDR);<br>– Received Signal Strength Indicator (RSSI);<br>– Inertial Measurement Unit (IMU);<br>– Pedestrian Dead Reckoning (PDR);<br>– Inertial Navigation System (INS);<br>– Extended Kalman Filter (EKF);<br>– Inertial Navigation System (INS). |
| **Objetivo do Artigo** | Este trabalho tem como objetivo fornecer um Sistema de posicionamento ToF com sensores inerciais (IMU), e a partir dos dados do IMU derivar tipos diferentes de informações. |
| **Técnicas/Abordagens/métodos/algoritmos** | – Tecnologia Wi-Fi(protocolo)";<br>– Sensores inerciais MEMS;<br>– WIFI time-of-flight (TOF) medições de distância;<br>– Informações "pedometric". |
| **Qual é o tipo de pesquisa?** | • Empírica;<br>São apresentados os resultados de dois experimentos. |
| **Especificação do tipo da técnica** | O sistema desenvolvido utiliza intervalos de ToF e medições de sensores inerciais (IMU) para estimar a posição exata. Isto foi possível através da implementação de um detector de passos e um INS e agregando |
| | aos intervalos ("ranges") de IMU e ToF a um Filtro de Kalman Estendido (EKF). |
| **Descrição da técnica** | – Quando um passo é detectado, o sistema avalia a distância percorrida a partir da posição do passo anterior a posição atual, compara com o comprimento do passo assumido, e estima os erros do INS;<br>– Quando um não movimento é detectado, o sistema avalia a solução INS durante o período de não movimento, e estima seus erros em conformidade;<br>– Quando a medição do intervalo ToF é introduzido, o EKF se funde junto com a mais recente solução INS para estimar os erros INS e corrigi-los. |
| **Utiliza algum Algoritmo de Localização?** | Algoritmo WiFi ToF |
| **Cobertura** | NA |
| **Contextualização** | Foi realizado testes de vida real em uma área de escritório Intel. Este local é um ambiente de espaço aberto 10m de largura e 25m de comprimento. O espaço é dividido em cubículos; cada cubículo é de cerca de 2mx2m. A altura do compartimento é de 1,5m e suas paredes são feitas de metal. Este ambiente é considerado muito áspero (" harsh") devido à natureza perturbadora das paredes dos cubículos. |
| **Tem apoio ferramental. Qual?** | • Não |
| **Infra** | – Para gerar as medições de distância foi utilizado "WIFI Intel ToF chipset" capaz de uma largura de banda de 40 Mhz;<br>– Foi usado notebook com o chipset WIFI acima mencionado como ambos os APs e dispositivo móvel.<br>– Foi implantado quatro APs a uma altura de 2m.<br>– Definiu-se a freqüência de solicitação de intervalo ToF ser 1Hz para cada ponto de acesso. |
| **Abordagem Hibrida** | • Sim<br>Utiliza medições WIFI ToF e informações "pedometric" para restringir a solução INS |
| **Qual é o resultado da pesquisa?** | • Sistema<br>Este trabalho tem como objetivo fornecer um Sistema de posicionamento ToF com sensores inerciais (IMU), e a partir dos dados do IMU derivar tipos diferentes de informações. |
| **Resultados** | – O aumento da precisão da posição;<br>– Trajetória estimada Smoother;<br>– Os erros que não divergem ao longo do tempo;<br>– Redução da sensibilidade à AP geometria implantação<br>– Aumento da cobertura;<br>– Redução da sensibilidade a outliers;<br>– Insensibilidade a dispositivo de mudança de orientação. |
| **Limitações** | NA |
| **Trabalhos Futuros** | As seguintes áreas requerem mais investigação:<br><br>– Incorporando de estimativa do comprimento do passo para o algoritmo;<br>– A implementação de um tipo diferente de filtro de fusão não-linear, em vez do EKF;<br>– A incorporação de um modelo mais complexo de erro de sensores inerciais para o modelo de sistema EKF;<br>– Combater a falta de posição inicial e posição inicial;<br>– Problema. |
| **Algoritmos** | NA |

25)

| Dados do ARTIGO | A Bluetooth signal strength based indoor localization method |
|---|---|
| **Autor** | Oksar, I. |
| **Ano** | 2014 |
| **Keywords** | Indoor Localization; Bluetooth; Error Function; RMSE; RSSI |
| **Siglas importantes** | – RootMean- Square-Error (RMSE) metric;<br>– Received Signal Strength Indicator (RSSI);<br>– Error Function (EF). |
| **Objetivo do Artigo** | Neste trabalho um “affordable” sistema de localização indoor é proposto onde a avaliação local é realizado através da medição dos níveis de sinal dos transmissores Bluetooth (desconhecidos) em algumas estações de base conhecida. |
| **Técnicas/Abordagens/métodos/algoritmos** | Dispositivos que utilizam o protocolo Bluetooth e RSSI |
| **Qual é o tipo de pesquisa?** | • Empírica;<br>A validade do método de decisão é verificada com um experimento do mundo real. |
| **Especificação do tipo da técnica** | Dispositivos que utilizam o protocolo Bluetooth têm a vantagem de ser de baixo custo e também estar amplamente disponível.<br>A função de erro é calculada em pontos discretos na área de interesse e os pontos com valores de erro calculado baixo indicam os locais onde transmissores pode ser localizado com uma alta probabilidade. |
| **Descrição da técnica** | Uma função de erro é definida para calcular as posições dos transmissores desconhecidos. A função de erro baseia-se uma métrica RMSE modificada. Para cada ponto discreto em uma região a função de erro é calculada pela utilizando as distâncias para as estações de base e os níveis de RSSI em as estações de base. Os pontos com valores de função menor erro são estimados como as localizações dos transmissores. |
| **Utiliza algum Algoritmo de Localização?** | NA |
| **Cobertura** | A precisão dos sistemas de localização sem fio depende pesadamente em métricas de distância. |
| **Contextualização** | O experimento é feito em uma sala de 7m x 5m com móveis no interior. |
| **Tem apoio ferramental. Qual?** | • Não |
| **Infra** | Três receptores Bluetooth são colocados no quarto e eles funcionam como estações rádio base (B1 'Bz e s3,)'.Todas as estações de base são do mesmo tipo de modo que eles têm a mesma hardware e eles têm leituras RSSI semelhantes em mesmas condições. |
| **Abordagem Hibrida** | • Sim<br>A função de erro se baseia na metrica RootMean- modificado Square-Error (RMSE). A função de erro é calculada utilizando RSSIs |
| **Qual é o resultado da pesquisa?** | • Método<br>O método de localização é proposto que utiliza valores calculados de uma função de erro definida para estimar as posições dos desconhecidos transmissores. |
| **Resultados** | Foi constatado que a precisão precisa ser melhorada para implementações práticas. Os móveis e as paredes deteriora propagação lenta do sinal e resulta em reflexos. Embora o uso de três estações de base é, teoricamente, o suficiente para de estimativa localização num ambiente bidimensional como no experimento, pensa-se que a adição e a colocação adequada de uma quarta estação base diminuiria erros de localização consideravelmente. |
| **Limitações** | NA |
| **Trabalhos Futuros** | Como parte do trabalho futuro, pretende-se realizar a mesmo experimento com quatro estações de base.<br>Além disso, o otimização das colocações de estação de base também é planejada |
| **Algoritmos** | NA |

26)

| Dados do ARTIGO | KARMA: Improving WiFi-based indoor localization with dynamic causality calibration |
|---|---|
| **Autor** | Sharma, P. and Chakraborty, D. and Banerjee, N. and Banerjee, D. and Agarwal, S.K. and Mittal, S. |
| **Ano** | 2014 |
| **Keywords** | NA |
| **Siglas importantes** | – Received Signal Strength Indicators (RSSI) ;<br>– Access points (APs). |
| **Objetivo do Artigo** | Construir uma solução de localização que é robusta e escalável à ocorrência individual e conjunta dos vários fatores, mantendo a tarefa de “fingerprinting” contido? |
| **Técnicas/Abordagens/métodos/algoritmos** | – Wi-Fi;<br>– Fingerprints. |
| **Qual é o tipo de pesquisa?** | • Empírica;<br>Estudos experimentais demonstram |
| **Especificação do tipo da técnica** | Investigou-se uma metodologia dinâmica de calibração online chamado KARMA. KARMA utiliza um “fingerprinting” única de espaço e sistematicamente aplica a um conjunto de funções de calibração de causalidade em tempo real para compensar a variação dos elementos. |
| **Descrição da técnica** | Verificou-se que as estimativas de localização podem ser lançadas de forma significativa. Devido a esses fatores, foi proposto investigar o desempenho de KARMA - um modelo de compensação on-line para anular o efeito de causalidade chave fatores em valores RSSI. E posteriormente foi fornecido uma avaliação precoce de KARMA em um ambiente controlado e ambiente de escritório e enquanto discutia os benefícios do KARMA, identificaram-se as áreas onde mais pesquisas são necessárias, tais como questões de escala. |
| **Utiliza algum Algoritmo de Localização?** | NA |
| **Cobertura** | NA |
| **Contextualização** | A resolução da grade (“grid”) de 2m x 2m foi utilizada para coletados dados para todas as análises experimentais. 4 orientações relativas de usuários foram utilizados. Assim, um total de 8 mapas de radio foram coletados. |
| **Tem apoio ferramental. Qual?** | • Não |
| **Infra** | − Samsung SII;<br>− Micromax |
| **Abordagem Hibrida** | • Sim<br>Wi-Fi **X** Fingerprints. |
| **Qual é o resultado da pesquisa?** | • Pesquisa<br>Investigou-se uma metodologia dinâmica de calibração online chamado KARMA. |
| **Resultados** | Como resultado, a localização provedores podem agora reduzir significativamente o dispendioso “re-learning” dos modelos para diferentes fatores e melhorar em tempo real precisão da previsão de localização que a estratégia do KARMA, mantendo a tarefa de “fingerprinting” contido, pode melhorar a qualidade de localização por um factor de 2x, em comparação com uma abordagem típica de” fingerprinting”de um estado empregado em muitas implementações comerciais atuais. |
| **Limitações** | NA |
| **Trabalhos Futuros** | Aplicar este em uma implantação comercial, e investigar o desempenho do KARMA para lidar com factores dinâmicos de aprendizagem automática de modelos de calibração por parâmetros de registro como a orientação |
| **Algoritmos** | NA |

27)

| Dados do ARTIGO | PiLoc: a Self-Calibrating Participatory Indoor Localization System |
|---|---|
| **Autor** | Luo, C. and Hong, H. and Chan, M.C. |
| **Ano** | 2014 |
| **Keywords** | Indoor Localization, Participatory, Smartphones, Floor Plan |
| **Siglas importantes** | – Inertial Measurement Unit (IMU) sensors |
| **Objetivo do Artigo** | Neste trabalho, observa a tentativa responder à seguinte pergunta: podemos projetar um interno sistema de localização que infere o planta/terreno (“ *floor plan*”) e automaticamente calibra-se sem pontos de referência? |
| **Técnicas/Abordagens/métodos/algoritmos** | Sinal IMU e Wi-Fi |
| **Qual é o tipo de pesquisa?** | • Empírica;<br><br>PiLoc avaliação ao longo de 4 diferentes áreas internas. |
| **Especificação do tipo da técnica** | NA |
| **Descrição da técnica** | PiLoc utiliza um ou mais clientes móveis(“ mobile clients”) que rastreiam usuário ao longo de seu deslocamento. Cada cliente envia os dados de trajetória para o servidor uma vez que dados suficientes foram coletados. Um programa servidor toma como entrada os dados da trajetória do cliente, gera ou atualiza o planta e mapa de rádio. O servidor também manipula as consultas de localização de o usuário e retorna a localização atual do usuário com base no mapa de rádio construído. |
| **Utiliza algum Algoritmo de Localização?** | NA |
| **Cobertura** | Mostra que o sistema pode atingir uma média de erro de localização de 1,5 m. |
| **Contextualização** | Os dados do experimento são coletados ao longo de um período de um mês a partir de 4 diferentes áreas, que cobrem 5.528 áreas m2 no total. Os tamanhos destes 4 plantas diferentes variam de 120 m2 a 3.000 m2. a menor área de 120 m2 envolve o interior de um laboratório de pesquisa com várias partições que representam um desafio especial, devido à sua voltas e caminhadas curtas |
| **Tem apoio ferramental. Qual?** | • Não |
| **Infra** | Três modelos de telefone diferentes são utilizados: Google Galaxy Nexus, Samsung S3 e Samsung S4. Todos os smartphones utilizam OS Android. Uma média de 35 APs são detectados em cada uma das quatro zonas. |
| **Abordagem Hibrida** | • Sim<br>Sinal IMU e Wi-Fi |
| **Qual é o resultado da pesquisa?** | • Método/sistema<br>Neste trabalho, foi proposto que se PiLoc, um sistema de localização indoor que utiliza oportunisticamente os dados fornecidos pelos usuários. |
| **Resultados** | PiLoc requer nenhuma intervenção humana,  e pode alcançar alta precisão de localização com 1,5 metros de erro na média. Como PiLoc permite mínimo esforço do usuário para a calibração e manutenção, que tem potencial para implantação em larga escala. |
| **Limitações** | NA |
| **Trabalhos Futuros** | NA |
| **Algoritmos** | **Input**: Matching result $\mathcal{M}$, Trajectories set $T$ of 1 cluster $c$<br>**Output**: Updated displacement matrix $M_d$<br>Initialized displacement matrix $M_d$;<br>**for** *each matching segment pair* $(S_i, S_j)$ *in* $\mathcal{M}$ **do**<br>    // Collocate and determine displacements of<br>    // matching steps<br>    Set of collocated steps, $S_{merge}$, is initially empty;<br>    **for** *each matching step pair* $(s_m, s_n)$ *in* $(S_i, S_j)$ **do**<br>        Place $s_m$, $s_n$ into a single location;<br>        New displacement of $s_m$ and $s_n$ are average displacements of $s_m$ and $s_n$ to all points in $S_{merge}$;<br>        $S_{merge} = S_{merge} \bigcup s_n \bigcup s_m$;<br>    **end**<br>    **for** *each step p in* $T$ *or but not in* $S_{merge}$ **do**<br>        Displacement of $p$ = average displacements of $p$ to all points in $S_{merge}$;<br>    **end**<br>    Update displacement matrix $M_d$ based on all new displacements calculated;<br>**end**<br>return $M_d$; |

28)

| | |
|---|---|
| Dados do ARTIGO | RollCaller: User-Friendly Indoor Navigation System Using Human-Item Spatial Relation |
| **Autor** | Guo, Y. and Yang, L. and Li, B. and Liu, T. and Liu, Y. |
| **Ano** | 2014 |
| **Keywords** | RFID, Doppler Frequency Shift, Human-Item Spatial Relation, Indoor Navigation |
| **Siglas importantes** | – Received signal strength (RSS). |
| **Objetivo do Artigo** | O objetivo deste artigo é o de encontrar uma solução que pode localizar um item relacionado à pessoa em movimento. Especificamente, destinam-se a descobrir a localização de uma pessoa mais próxima em movimento através da passagem de prateleira de um item específico. |
| **Técnicas/Abordagens/métodos/algoritmos** | Neste estudo, foi proposta a utilizar etiquetas RFID anexadas a objetos indoor existentes e o leitor de maneira a auxiliar o deslocamento do usuário ao seu destino, sem necessidade de qualquer adicional Hardware. A chave da compreensão é que o movimento pessoal leva um impacto sobre os valores de deslocamento de frequência Doppler coletados a partir dos objetos interno, quando chegar perto da tag. |
| **Qual é o tipo de pesquisa?** | • Empírica;<br>Neste experimento, observou-se um fenômeno que, quando uma pessoa se move através da linha. |
| **Especificação do tipo da técnica** | Foi implementado um protótipo de sistema de navegação, chamado RollCaller e conduzido uma série de experimentos para analisar o seu desempenho |
| **Descrição da técnica** | 1. Em primeiro lugar, foi desenvolvido um método para medir a relação espacial entre uma pessoa e um item, utilizando o deslocamento de frequência Doppler em um sistema RFID;<br>2. Em seguida foi proposto um sistema de navegação indoor, com o nome de RollCaller, para fornecer o serviço de navegação interno para um usuário encontrar o item desejado;<br>3. A função do sistema RollCaller é como o próprio nome: quando uma pessoa quer encontrar um item, ele então "Chama de" "nome" do item.<br>4. Em seguida, o sistema RollCaller navega-o para o item que ele está procurando. Quando a pessoa caminha direto próxima ao item, o item "responde" a chamada imediatamente. |
| **Utiliza algum Algoritmo de Localização?** | Nesta implementação, a mudança de freqüência Doppler é calculada utilizando o método de diferença de |
| | fase durante a vigência de receber um pacote de uma tag. |
| **Cobertura** | Tanto a cobertura da TAG quanto dos leitores RFID são de 100 cm. |
| **Contextualização** | Foi implementado o sistema RollCaller em um ambiente de laboratório. |
| **Tem apoio ferramental. Qual?** | • Não |
| **Infra** | 1. Para implementar o sistema RollCaller, o seguinte hardware é indispensável: (1) etiquetas RFID passivas ligado por tag RFID, os itens se tornam reconhecíveis pelo contato livre com os leitores de RFID. (2) leitor RFID passiva, juntamente com múltiplas antenas do leitor para cada.<br>2. O leitor é carregado com a versão do software Octane 4,8 e capacitará os 4 antenas direcionais de leitor.. Além de “9 Alien ALN-9634 passive RFID tags”;<br>3. Além da parte RFID implantado, outro importante dispositivo exigido no sistema RollCaller é um smartphone que disponha de acelerômetro e magnetômetro. |
| **Abordagem Hibrida** | • Sim<br>Utilizar etiquetas RFID X  frequência Doppler |
| **Qual é o resultado da pesquisa?** | • Método/sistema<br>Foi projetado um sistema de navegação indoor “user-friendly” chamado RollCaller |
| **Resultados** | Os resultados experimentais mostram que sistema RollCaller pode navegar com precisão do usuário ao item que quiser com um erro médio distância inferior a 20 centímetros. |
| **Limitações** | Em ambientes que não se possa estabelecer esta relação espacial (distancia superior) entre os usuários e os itens e as tags. |
| **Trabalhos Futuros** | NA |
| **Algoritmos** | NA |

29)

| Dados do ARTIGO | Accurate Indoor Localization With Zero Start-up Cost |
|---|---|
| **Autor** | Kumar, S. and Gil, S. and Katabi, D. and Rus, D. |
| **Ano** | 2014 |
| **Keywords** | Wireless; Localization; Wi-Fi; SAR; PHY |
| **Siglas importantes** | - Synthetic Aperture Radar (SAR) |
| **Objetivo do Artigo** | Contribuição central da Ubicarse é a capacidade de executar SAR em “handheld” dispositivos “twisted” por seus usuários por caminhos desconhecidos. |
| **Técnicas/Abordagens/métodos/algoritmos** | NA |
| **Qual é o tipo de pesquisa?** | • Empírica; |
| **Especificação do tipo da técnica** | Ubicarse é um sistema de localização indoor que atinge dezenas de centímetros de precisão em dispositivos móveis de “commodity”, sem infra-estrutura ou “fingerprinting” especializada. |
| **Descrição da técnica** | Ubicarse fornece uma nova formulação de SAR que permite dispositivos móveis simulem uma matriz de antena, em que esta formulação é tolerante a tradução desconhecida(“ desconhecido”) do centro da antena de até meio metro. Ubicarse Foi implementado em um tablet “commodities” e demonstra dezenas de centímetros na precisão tanto dispositivo de localização e objeto “geotagging” em ambientes indoor complexos. |
| **Utiliza algum Algoritmo de Localização?** | “Vision algorithms for 3-D imaging”. |
| **Cobertura** | Empiricamente demonstrou um erro médio de 39 centímetros em 3-D de localização do dispositivo e 17 cm |
| | em um objeto “geotagging” em complexo ambiente indoor. |
| **Contextualização** | Foi conduzido em uma grande biblioteca da universidade, com livros dispostos em várias prateleiras. |
| **Tem apoio ferramental. Qual?** | • Não |
| **Infra** | Ubicarse foi implementado em um HP SplitX2 Tablet funcionamento Ubuntu Linux equipado com Intel 5300 placas wireless e um Yei Sensor de movimento Tecnologia (ou seja, um acelerômetro, giroscópio e bússola). |
| **Abordagem Hibrida** | NA |
| **Qual é o resultado da pesquisa?** | • Método/sistema<br>Ubicarse um sistema de localização interior preciso |
| **Resultados** | • Os resultados mostram um erro médio no ângulo de 3.4◦ através de trajetórias, demonstrando a tradução-resiliência da SAR do Ubicarse;<br>• Os resultados mostram que Ubicarse localiza o tablet com um erro médio de 39 centímetros em pleno espaço tridimensional;<br>• Os resultados mostram um erro médio de 17 cm de espaço 3-D. |
| **Limitações** | |
| **Trabalhos Futuros** | Acredita-se que a avaliação este sistema em uma ampla gama de dispositivos móveis, particularmente smartphones, é uma tarefa importante para o trabalho futuro. |
| **Algoritmos** | NA |

30)

| Dados do ARTIGO | A dual-sensor enabled indoor localization system with crowdsensing spot survey |
|---|---|
| **Autor** | Zhang, C. and Luo, J. and Wu, J. |
| **Ano** | 2014 |
| **Keywords** | Indoor Localization, Mobile Sensing |
| **Siglas importantes** | – Access points (aps); |
| **Objetivo do Artigo** | MaWi, tem o objetivo de reduzir o custo de implantação fornecendo uma melhor escalabilidade e para alcançar alta precisão de localização. |
| **Técnicas/Abordagens/métodos/algoritmos** | Sinal Wi-Fi e Campo Magnético. |
| **Qual é o tipo de pesquisa?** | • Empírica; |
| **Especificação do tipo da técnica** | Neste trabalho, foi proposto MaWi, um “ crowdsensing spot survey”  baseado a um sistema de localização indoor. MaWi faz uso de tanto do campo magnético e de sinal Wi-Fi, como “fingerprints”. |
| **Descrição da técnica** | Devido estes dois tipos de “fingerprints” apresentarem bom desempenho complementar na localização, associado ao campo magnético, o qual tem uma excelente estabilidade temporal, eles podem ser utilizados em uma forma de "dueto" de tal forma que possam contrabalançar os pontos fracos de ambos (Sinal Wi-Fi e Campo Magnético.) e, assim, melhorar a sua respectiva eficácia. Por meio uma combinação de estes dois tipos de “fingerprints”, MaWi, impõe a uma reduzida procura no volume no seu banco de dados. Portanto, MaWi alcança uma implantação escalável por empregar “crowdsensing” sem a pesquisa local intensa, |
| | enquanto recebe a precisão de localização muito competitivo em comparação com sistemas indoor “state-of-the-art”. |
| **Utiliza algum Algoritmo de Localização?** | Similarity Voronoi Algorithm (SVA). |
| **Cobertura** | NA |
| **Contextualização** | Foi implantado MaWi em três locais de teste: uma área de escritório, biblioteca e um shopping center. |
| **Tem apoio ferramental. Qual?** | Foi desenvolvido um aplicativo Android como cliente. |
| **Infra** | Empregam dois modelos de smartphones: Samsung Galaxy S2 e S3. |
| **Abordagem Hibrida** | • Sim<br>Sinal Wi-Fi e Campo Magnético. |
| **Qual é o resultado da pesquisa?** | • Método/sistema<br>Foi apresentada MaWi - sistema de localização indoor escalável de um smartphone. |
| **Resultados** | Foi utilizado tempo limite para forçar a convergência do processo de localização. Quanto maior o tempo dá a MaWi mais chances para fazer peso partícula convergem, e logo maior precisão. |
| **Limitações** | NA |
| **Trabalhos Futuros** | NA |
| **Algoritmos** | Algoritmo de localização: |

**Input**: Wi-Fi survey spot set $\mathcal{S}_w$, magnetic survey spot set $\mathcal{S}_m$, confidence coefficient $p$

**Output**: Estimated location $l_e$

**forall the** $l \in \mathcal{S}_m$ **do**

  $\mathcal{P}.add([(l, 2 \times bachward, \frac{1}{5|\mathcal{S}_m|}), (l, bachward, \frac{1}{5|\mathcal{S}_m|}), (l, static, \frac{1}{5|\mathcal{S}_m|}), (l, forward, \frac{1}{5|\mathcal{S}_m|}), (l, 2 \times forward, \frac{1}{5|\mathcal{S}_m|})])$

**while** $true$ **do**

  $F_w \leftarrow$ getOnlineWiFi(); $F_m \leftarrow$ getOnlineMagn()

  $\mathcal{A} \leftarrow \mathsf{SVA}(\mathcal{S}_w, F_w, p)$; $\omega_{in} \leftarrow 0$; $\omega_{out} \leftarrow 0$

  **forall the** $o \in \mathcal{P}$ **do**

    $o$.updateLocation()

    $o.w \leftarrow o.w \times \mathsf{similarity}(F_m, F_m(o.l))$

    **if** $o.l \in \mathcal{A}$ **then** $\omega_{in} \leftarrow \omega_{in} + o.w$ **else** $\omega_{out} \leftarrow \omega_{out} + o.w$

  **forall the** $o \in \mathcal{P}$ **do**

    **if** $o.l \in \mathcal{A}$ **then** $o.w \leftarrow \frac{o.w \times p}{\omega_{in}}$ **else** $o.w \leftarrow \frac{o.w \times (1-p)}{\omega_{out}}$

  $\mathcal{P} \leftarrow \mathsf{normalize}(\mathcal{P})$

  **if** $needResample$ **then** $\mathcal{P} \leftarrow \mathsf{resample}(\mathcal{P})$ **if** $converge$ **then**

    $o_e \leftarrow \arg\max_o (o.w | o \in \mathcal{P})$; **return** $l_e \leftarrow o_e.l$

31)

| Dados do ARTIGO | Constraint Online Sequential Extreme Learning Machine for Lifelong Indoor Localization System |
|---|---|
| **Autor** | Gu, Y. and Liu, J. and Chen, Y. and Jiang, X. |
| **Ano** | 2014 |
| **Keywords** | Wi-Fi indoor localization; fluctuation; online learning; lifelong |
| **Siglas importantes** | – Constraint Online Sequential Extreme Learning Machine ( COSELM) ;<br>– Location Based Services ( LBS) ;<br>– Received Signal Strength (RSS); |
| **Objetivo do Artigo** | Foi proposto uma fácil implementação on-line de 1 método de localização indoor: COSELM. Agregando rápida velocidade de aprendizagem e alta precisão de localização, o que pode lidar com o problema de flutuação, e ajudar a localização. |
| **Técnicas/Abordagens/métodos/algoritmos** | Wifi, Received Signal Strength. |
| **Qual é o tipo de pesquisa?** | • Empírica;<br>O desempenho do COSELM é validado em ambiente Wi-Fi indoor. |
| **Especificação do tipo da técnica** | COSELM é proposto, aprimorando os dados incrementais para atualizar modelo existente, de modo que o novo modelo possa abranger mais possibilidades de “fingerprintg” à mesma localização e refletem o ambiente melhor localização |
| **Descrição da técnica** | 1. Prepare dado treinamento inicial, e normalize as entradas características para [0,1];<br>2. Determinar os parâmetros de COSELM incluindo função de ativação g x, número de “ hiddeneurons” L ea regularização pena C<br>3. Aleatoriamente atribuir o iw peso entrada e viés ib;<br>4. Calcula-se a saída de camada oculta por 0 H (4); |
| | As etapas seguintes podem ser visualizadas no fluxograma abaixo: 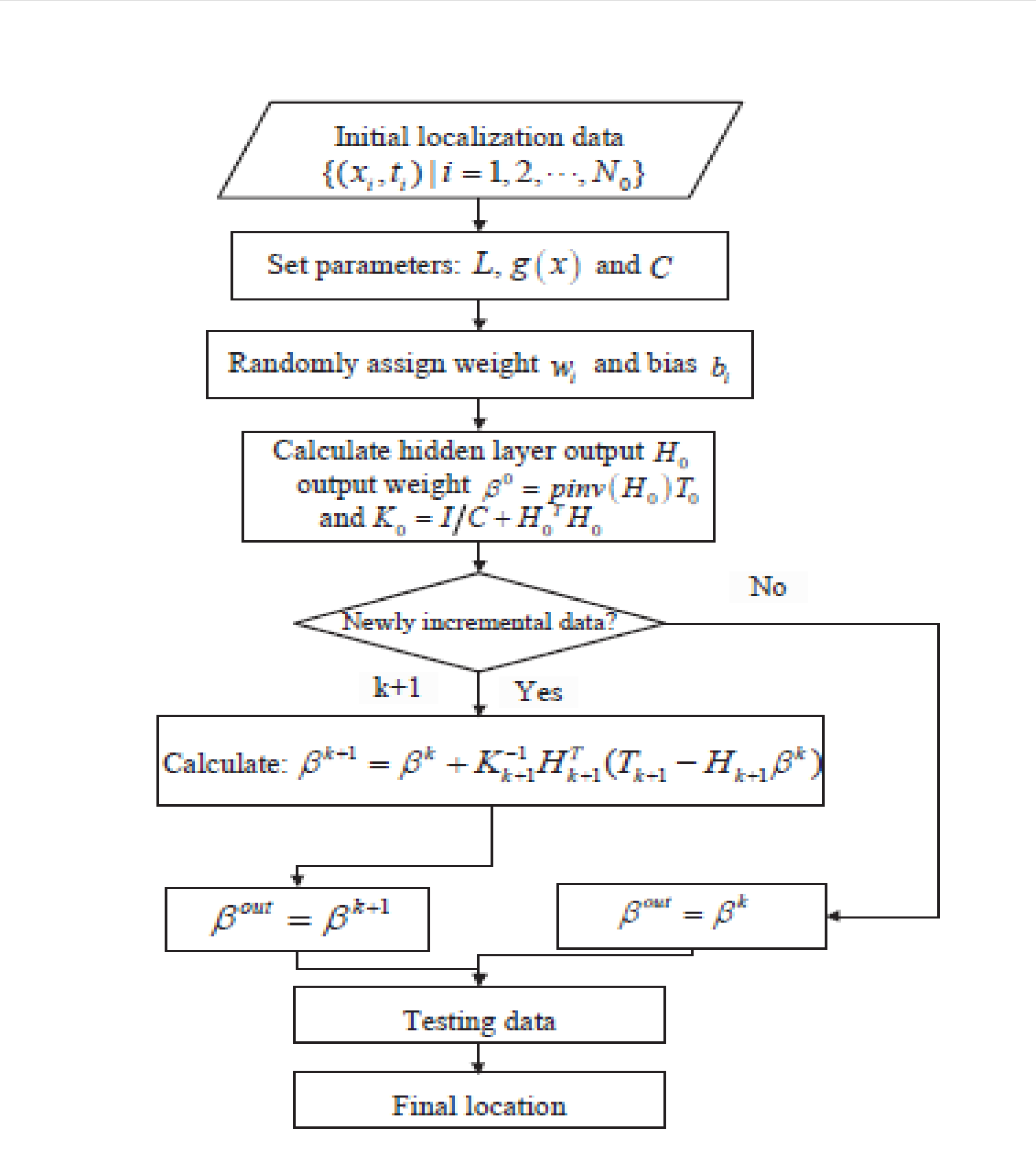  |
| **Utiliza algum Algoritmo de Localização?** | NA |
| **Cobertura** | NA |
| **Contextualização** | 1. O desempenho do COSELM foi validada em tempo real em um ambiente indoor, sendo |
| | representada em uma área de trabalho 15m * 10m na 8º andar de um prédio de pesquisa;<br>2. Os dados Wi-Fi são casualmente coletados de smartphones ao longo de 2 meses; |
| **Tem apoio ferramental. Qual?** | • Não |
| **Infra** | • 32 APs são dispostas no ambiente. |
| **Abordagem Hibrida** | • Não |
| **Qual é o resultado da pesquisa?** | • Método |
| **Resultados** | 1. Comparado com OSELM, ele pode melhorar a precisão da localização de mais do que 5%, em média;<br>2. Resultado experimental dos dois meses,  mostra que COSELM pode manter a precisão de localização em um intervalo de [80%, 95%], com menos tempo de consumo |
| **Limitações** | NA |
| **Trabalhos Futuros** | NA |
| **Algoritmos** | NA |

32)

| Dados do ARTIGO | Hybrid indoor localization using GSM fingerprints, embedded sensors and a particle filter |
|---|---|
| **Autor** | Tian, Y. and Denby, B. and Ahriz, I. and Roussel, P. and Dreyfus, G. |
| **Ano** | 2014 |
| **Keywords** | indoor localization; fingerprinting; support vector machine; sensor dead-reckoning; particle filter |
| **Siglas importantes** | - Support Vector Machine (SVM); |
| **Objetivo do Artigo** | Neste artigo, foi apresentado um filtro de partículas (“ particle”) aplicado a um sistema de localização indoor, que usa resultados do posicionamento de GSM fingerprinting para corrigir os erros cálculo de posição de inércia(“ dead-reckoning”), e ao mesmo tempo, faz uso de um mapa edifício para excluir regiões de difícil acesso e proibir movimentos não razoáveis, como atravessar uma parede. |
| **Técnicas/Abordagens/métodos/algoritmos** | GSM Received Signal Strength X fingerprintsinertial dead-reckoning erros X devices sensors |
| **Qual é o tipo de pesquisa?** | NA |
| **Especificação do tipo da técnica** | A abordagem gera coordenadas ao invés de “labels” dos ambientes, também incorpora um algoritmo de detecção passo adaptativo “stride-length”que pode lidar com mudanças de posição arbitrárias do dispositivo móvel. |
| **Descrição da técnica** | A abordagem, que usa GSM “fingerprints” e múltiplos sensores que sejam fácil de obter, devido à crescente popularidade dos smartphones. O fluxograma aplicado neste artigo se encontra representado abaixo: |
| | 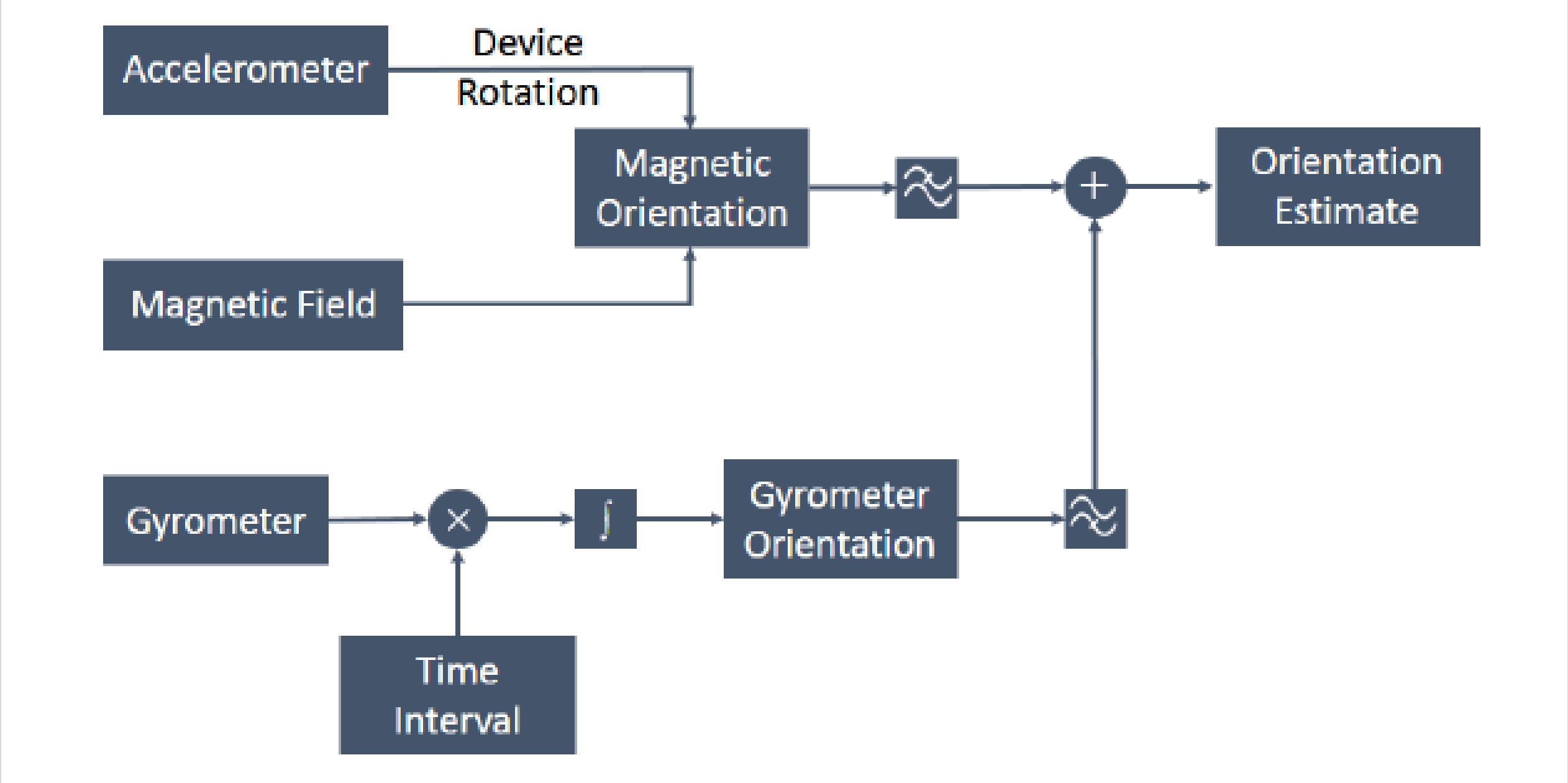 |
| **Utiliza algum Algoritmo de Localização?** | Classificadores Support Vector Machine (SVM) são usados neste trabalho são considerados adequados para lidar com o grande número de variáveis e exemplos de treinamento, devido à sua built-in mecanismo de regularização. |
| **Cobertura** | Os resultados experimentais mostram que esta abordagem pode determinar a trajetória do usuário móvel com boa precisão. |
| **Contextualização** | O teste experimental estava localizado no quarto andar de um prédio de laboratório (estrutura de aço com concreto e gesso paredes), no centro de Paris, França. |
| **Tem apoio ferramental. Qual?** | • Não |
| **Infra** | Na fase de testes, um Pocket TEMS e um tablet Nexus 7 foram utilizadoss juntos, gravando GSM |
| | fingerprinting e múltiplas leituras dos sensores simultaneamente. |
| **Abordagem Hibrida** | • Sim |
| **Qual é o resultado da pesquisa?** | • Ferramenta<br>O artigo apresenta um esquema de localização interna para dispositivos móveis |
| **Resultados** | Os erros de localização são encarados como pontos de correção através da combinação de GSM fingerprinting, sensor deadreckoning e mapear as restrições de layout. Apenas alguns erros ocorreram no início do rastreamento de, devido a posição de partida desconhecida, uma vez que o classificador SVM somente mostra a localização do “roomlevel ”, não uma posição precisa. |
| **Limitações** | NA |
| **Trabalhos Futuros** | NA |
| **Algoritmos** | NA |

33)

| Dados do ARTIGO | COIN-GPS: Indoor Localization from Direct GPS Receiving |
|---|---|
| **Autor** | Nirjon, S. and Liu, J. and DeJean, G. and Priyantha, B. and Jin, Y. and Hart, T. |
| **Ano** | 2014 |
| **Keywords** | Indoor GPS, Instant GPS, CO-GPS, COIN-GPS |
| **Siglas importantes** | – Coarse time navigation (CTN);<br>– Cloud-offloaded instant GPS (COIN-GPS). |
| **Objetivo do Artigo** | Este artigo tem o objetivo de estender o recebimento de sinal GPS aplicados ambientes indoor(s), onde podem ser utilizados, quer diretamente por dispositivos dos usuários, ou ser usado em conjunção com perfis e métodos de rastreamento como pontos de referência. |
| **Técnicas/Abordagens/métodos/algoritmos** | GPS X directional antenna,X robust acquisition algorithm |
| **Qual é o tipo de pesquisa?** | • Empírica;<br>Este sistema foi testado em 31 pontos escolhidos aleatoriamente cinco andares de um prédio. |
| **Especificação do tipo da técnica** | O projeto do COIN-GPS foi motivado por observações sobre atenuação (lento) de sinal e movimentos receptor em ambientes indoor. COIN-GPS é inspirado pelo CO-GPS que representa um recebedor de sinal. Este trabalho aborda esses desafios por meio de uma orientável/ “HighGain” antena direcional como o front-end de um receptor GPS ao longo com uma etapa de processamento de sinal robusto e uma técnica de estimativa de localização de modo a fornecer a localização indoor direta baseada em GPS. |
| **Descrição da técnica** | Esta solução GPS indoor de alta sensibilidade incorpora três tecnologias chave:<br>1. Uma antena direcional;<br>2. Um algoritmo robusto de aquisição;<br>3. Um algoritmo de estimativa de localização multi-direcional. A formulação e implementação da fórmula GPS estacionário, que diz que para direções independentes M, um total de 2M + 3 satélites são necessários para obter uma posição de sucesso. |
| **Utiliza algum Algoritmo de Localização?** | NA |
| **Cobertura** | Os experimentos mostram que o sistema é capaz de obtenção de localização dos reparos de 20 destes pontos com um erro médio de menos de 10 m, em que todos os receptores de GPS normalmente falham. |
| **Contextualização** | Foi realizado um experimento aprofundado da COIN-GPS por meio da implantação o sistema em quatro lojas em andares distintos (Starbucks, Home Depot, Fred Meyer, e Costco) e um shopping de vários andares (Bellevue Square Mall), que estão localizados no Bellevue- WA. |
| **Tem apoio ferramental. Qual?** | • Não |
| **Infra** | 1. Uma antena direcional de alto ganho;<br>2. O controlador de direção;<br>3. E um "logger" sinal gps, que está ligado a um pc que leva cerca de 3 segundos para completar um ciclo de registro de dados;<br>4. O servidor back-end que foi utilizado é um windows server 2008 r2 PC (cpu quad core @ 3.3 ghz e 16 GB de RAM) |
| **Abordagem Hibrida** | • Sim |
| **Qual é o resultado da pesquisa?** | • Técnica |
| **Resultados** | COIN-GPS obtém localização dos bem sucedida fixes em 65% dos casos, com uma localização de erro média de 17.4m e um erro médio de menos de 10m. |
| **Limitações** | A COIN-GPS não fornece uma localização "garantida"  em todos os momentos. Ao contrário, é muito provável que haverá várias zonas em um edifício onde COIN-GPS não pode adquirir o número necessário de satélite. |
| **Trabalhos Futuros** | Algumas técnicas complementares podem melhorar ainda mais a precisão da COIN-GPS. Por exemplo, técnicas de cálculo de posição de pedestres ("dead-reckoning techniques") ou abordagens híbridas como UnLoc  e GAC  podem ser usados juntamente com COIN-GPS para lidar com os casos onde COIN-GPS não recebe uma localização. |
| **Algoritmos** | NA |

34)

| Dados do ARTIGO | Natural or Generated Signals for Indoor Location Systems? An Evaluation in terms of Sensitivity and Specificity |
|---|---|
| **Autor** | GalvÃ¡n-Tejada, C.E. and GarcÃ-a-VÃ¡zquez, J.P. and Brena, R.F. |
| **Ano** | 2014 |
| **Keywords** | indoor location, feature extraction, feature selection,<br><br>multivariate models, signals, evaluation |
| **Siglas importantes** | – Forward Selection (FS) strategy;<br>– Backward elimination (BE). |
| **Objetivo do Artigo** | Foi realizada uma avaliação dos sinais naturais e gerados, a fim de ajudar os projetistas e pesquisadores de sistemas de localização indoor para decidir qual sinal e funcionalidade é mais adequado para estimar a posição indoor do usuário, considerando-se a precisão em termos de sensibilidade e especificidade. |
| **Técnicas/Abordagens/métodos/algoritmos** | – Campo magnético e luz interna/indoor;<br>– Sinais gerados para a comunicação de informações (wi-fi e bluetooth). |
| **Qual é o tipo de pesquisa?** | • Empírica;<br>Quatro sinais foram considerados nos experimentos, devido sua onipresença nos ambientes indoor(s); |
| **Especificação do tipo da técnica** | Foi assumido um sinal natural como um sinal disponível no ambiente, o que não é gerado de uma maneira artificial, como campo magnética da terra, som ambiente, entre outros; enquanto um sinal gerado |
| | geralmente contêm informações intencionalmente incorporadas para serem transmitidas e processadas, o qual é então reconstruídos  por um dispositivo(receptr). Por exemplo, o sinal de Wi-Fi produzido por um ponto de acesso. |
| **Descrição da técnica** | Os 5 passos abaixo representam a metodologia que permite a obtenção de um modelo de classificação, são eles:Coleta de dados: O processo de coleta de dados foi feita em um ambiente controlado com um smartphone comum com todos os sensores habilitados;<br><br>1. Extração de Características: consiste na realização eficiente redução de dados, preservando simultaneamente a quantidade apropriada de informação do sinal;<br>2. Processo de seleção funcionalidades: consiste a reduzir o número de funcionalidades e, em alguns casos, aumentar a precisão dos modelos com menos recursos;<br>3. Seleção Avançado (“foward” /FS): gera modelos aninhados utilizando a ranqueando das funcionalidades, adicionando maior acurácia;<br>4. Eliminação Backward(BE): envolve o teste e eliminação das funcionalidades do modelo FS; quando a eliminação de um funcionalidade do modelo FS melhora a aptidão do modelo. |
| | 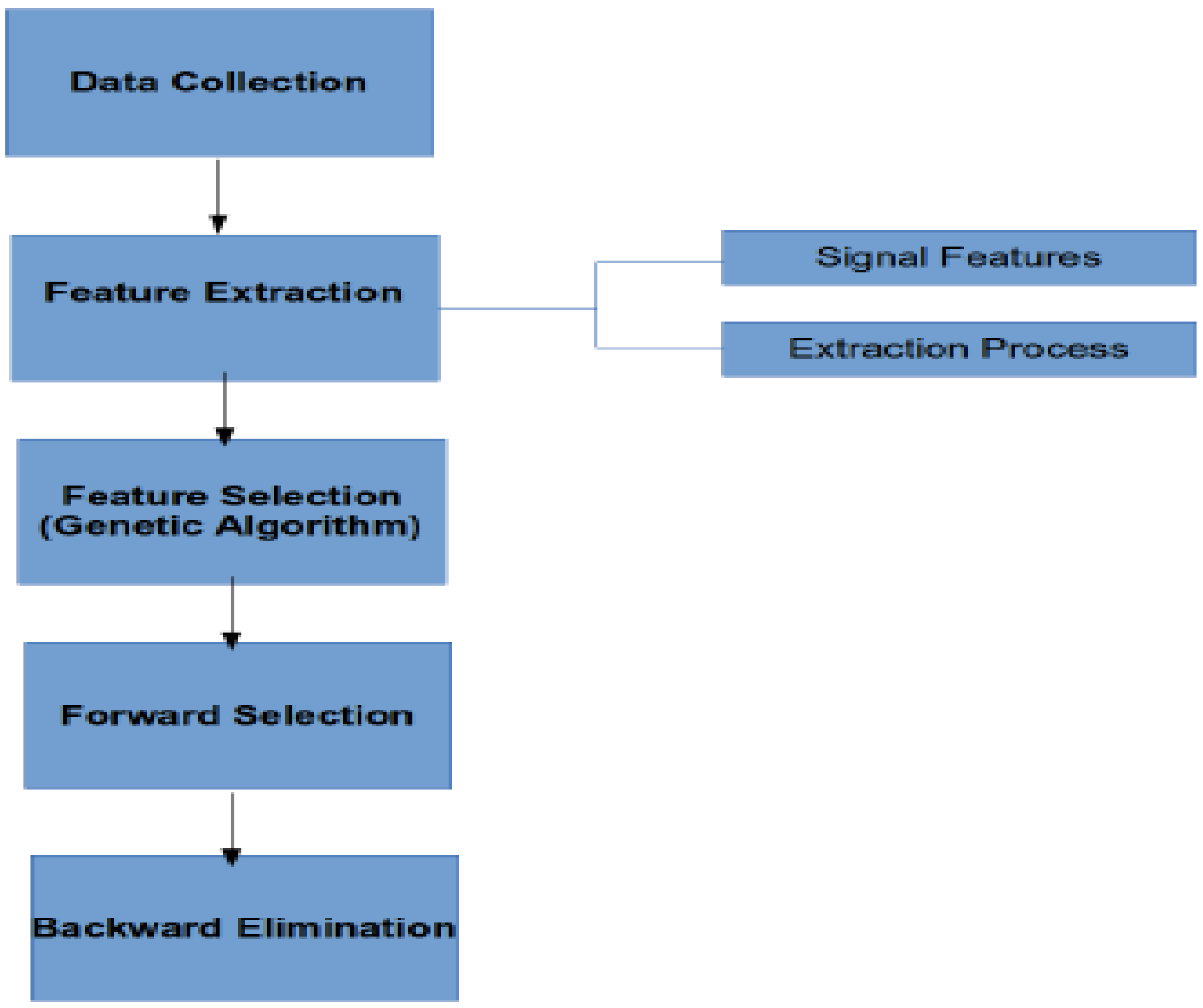 |
| **Utiliza algum Algoritmo de Localização?** | NA |
| **Cobertura** | NA |
| **Contextualização** | Os dados foram coletados em quatro cômodos: cozinha, sala de estar, casa de banho e sala de jantar. A experimentação foi feito no piso térreo de uma casa residencial. Este piso térreo é composto por 4 quartos |
| | e escadas |
| **Tem apoio ferramental. Qual?** | • Não |
| **Infra** | Os dados utilizados neste trabalho foram obtidos a partir dos sensores de um dispositivo de Smartphone (Samsung S3), pois possui sensores (sensor magnético, um sensor de luz, um Wi-Fi e Bluetooth). Foi desenvolvido um aplicativo móvel em Java usando a API Nível Google.<br>A iluminação dos quartos é dada por 7 lâmpadas normais (brancas/quentes) 107 LED SMD 3528. |
| **Abordagem Hibrida** | • Sim<br>Campo magnético e luz interna/indoor X Sinais gerados para a comunicação de informações (wi-fi e bluetooth). |
| **Qual é o resultado da pesquisa?** | • Avaliação<br>Foi realizada uma avaliação dos sinais naturais e gerados |
| **Resultados** | Os resultados indicam que a intensidade da luz superar as outras 3 fontes de informações, mas é o sinal que requeira mais informação (o modelo de classificação final requer 9 funcionalidades). |
| **Limitações** | − O sinal de luz utilizados nos experimentos foi tomado em um tempo específico do dia. Isso pode ter afetado o número de funcionalidades necessárias para modelar o sinal.<br>− Um número limitado de dispositivos transmissores geradores de sinais foram considerados. Considerou-se a apenas um transmissor neste experimento. Supõe-se que um aumento do número de transmissores pode melhorar a especificidade e sensibilidade. |
| **Trabalhos Futuros** | Foi proposto como trabalho futuro uma análise utilizando classificação diferente métodos da GA, a fim de determinar a estabilidade independentemente das características do sinal e a classificação método. Além disso, uma análise de sistemas híbridos de mesclar sinais gerados naturais e pode ser feito com o objetivo de maximizar a precisão. |
| **Algoritmos** | NA |

35)

| Dados do ARTIGO | DECL: A circular inference method for indoor pedestrian localization using phone inertial sensors |
|---|---|
| **Autor** | Dang, C. and Sezaki, K. and Iwai, M. |
| **Ano** | 2014 |
| **Keywords** | Pedestrian Localization; Online Learning; Multidimensional Optimization; Land Mark |
| **Siglas importantes** | – Pedestrian Dead Reckoning (PDR); |
| **Objetivo do Artigo** | Neste trabalho, foi proposto um método de inferência circular aplicado à aprendizagem on-line, a fim de reduzir os desvios dos erros. |
| **Técnicas/Abordagens/métodos/algoritmos** | phone-based PDR, called DECL (short for Detection, Estimation, Checking, and Learning) |
| **Qual é o tipo de pesquisa?** | • Empírica;<br>Um experimento preliminar foi realizado para identificar com exatidão a relação entre os dados de sensores inerciais (telefone) e o andar do pedestre. |
| **Especificação do tipo da técnica** | Neste artigo, foi apresentado um novo método de inferência circular a fim de resolver o problema desvio(sinais) ocorrendo por telefone(PDR).Aprendizagem on-line é usado para ajustar os parâmetros do sistema, e portanto os erros de desvio pode ser persistentemente reduzida. |
| **Descrição da técnica** | 1. Neste trabalho, foram utilizados sensores inerciais de telefone para realizar PDR, e sendo utilizadas apenas giroscópio detecção  da posição; |
| | 2. Adotou-se a abordagem de correção baseado em mapas e não usam outros tipos de sensores.<br>3. O telefone é segurado na mão ou colocado no bolso durante sua caminhada;<br>4. Supõe-se ainda que o ponto de partida é um local bem conhecido. |
| **Utiliza algum Algoritmo de Localização?** | Um algoritmo de otimização multidimensional é projetado e utilizado na fase de aprendizagem de forma eficiente a ajustar os parâmetros de estimativa. |
| **Cobertura** | De acordo com os resultados do experimento, o erro de distância média é de 1,1 metros para os casos em que p telefone foi segurado na mão e 1,5 metros para os casos em que foi colocado nos bolso. |
| **Contextualização** | Totalmente 10 pessoas participaram dos experimentos, incluindo um do sexo feminino e 9 do sexo masculino, com idade de 21 a 36. Os experimentos foram realizados em três “test-beds”, incluindo um shopping center, uma estação de metrô, e um prédio de pesquisas. Entre estes locais o shopping é o maior delas com uma área de aproximadamente 10.000 metros quadrados. |
| **Tem apoio ferramental. Qual?** | • Não |
| **Infra** | Vários tipos de smartphones foram utilizados nos experimentos, incluindo GoogleTM Nexus S, Galaxy Nexus SamsungTM / III, HTCTM Evo 3D / Sensation 4G, etc.<br>Foi desenvolvido um sistema final rodando em AndroidTM smartphones na base do desenho. A versão do usado SDK é 2.3.3. SQLiteTM foi usado para gerenciar os dados armazenados locais no aparelho |
| **Abordagem Hibrida** | • Sim<br>• |
| **Qual é o resultado da pesquisa?** | • Método<br>Neste trabalho, foi proposto um método de inferência circular aplicado à aprendizagem on-line, a fim de reduzir os desvios dos erros. |
| **Resultados** | Os resultados mostram que o sistema consegue desempenhar eficientemente a localização de com alta precisão e robustez. |
| **Limitações** | NA |
| **Trabalhos Futuros** | Planeja-se para melhorar ainda mais o sistema, incorporar outros tipos de sensores do telefone. Também se planeja usar o Wi-Fi e técnicas de “fingerprinting” de campo magnético para melhorar o sistema desempenho no futuro |
| **Algoritmos** | *dynamic_parameter_optimization* $(S, LM, V_0)$<br>**Input**: Subsequence of Locations: $S$<br>Matched Land Mark: $LM$<br>Initial Parameter Vector: $V_0$<br>**Output**: Sub-optimal Solution: $V^+$<br>**global** stop criterion of search: $C$<br>initialize search direction array $DA$ to contain unit vectors<br>$V_1 \Leftarrow V_0$<br>$f_1 \Leftarrow f_c(S, LM, V_1)$<br>$V_2 \Leftarrow V_1,\ \ f_2 \Leftarrow f_1$<br>**do**<br>$V_1 \Leftarrow V_2,\ \ f_1 \Leftarrow f_2$<br>**for** each direction vector $D$ in $DA$ **do**<br>*line search* in direction $D$ and update $f_2$ and $V_2$<br>**end for**<br>generate new direction $D' \Leftarrow (V_2 - V_1)$<br>*line search* in direction $D'$ and update $f_2$ and $V_2$<br>drop the direction in $DA$ that constitutes the minimum included angle with $D'$<br>insert $D'$ into $DA$ at the tail<br>**while** $f_1 - f_2 > C$<br>$V^+ \Leftarrow V_2$<br>**return** $V^+$ |

36)

| Dados do ARTIGO | Mobile tracking based on fractional integration |
|---|---|
| **Autor** | Nakib, A. and Daachi, B. and Dakkak, M. and Siarry, P. |
| **Ano** | 2014 |
| **Keywords** | Indoor location, path tracking, digital fractional integration, prediction filters, mobile location, Kalman filter |
| **Siglas importantes** | – Dynamic positioning (tracking) systems (DTSs);<br>– Kalman filter (KF);<br>– Digital fractional integration (DFI). |
| **Objetivo do Artigo** | Neste trabalho, foi proposto um nova técnica para melhorar o desempenho do rastreamento MT clássica em um interno ambiente. |
| **Técnicas/Abordagens/métodos/algoritmos** | Foi proposto utilizar "Digital Fractional Integration based filter" integrado a "Kalman filters". |
| **Qual é o tipo de pesquisa?** | • Empírica;<br>Os resultados experimentais mostram uma melhoria de desempenho significativa sobre a maioria dos "predictors" comuns na literatura relevante, principalmente nos casos ruidosos. |
| **Especificação do tipo da técnica** | Neste artigo, foi proposto para melhorar o desempenho dos filtros: previsão linear e kalman, para permitir o seu uso em um ambiente indoor. Para se ter uma idéia sobre a grau de previsibilidade de um percurso, é comum utilizar a função de autocorreção como métrica para medir a redundância e a relação entre as amostras de um sinal. |
| **Descrição da técnica** | Este método proposto baseia-se em duas premissas:<br>1. O problema é resolvido de associação de dados;<br>2. O caminho do terminal móvel (MT) pode ser considerado uma função positiva P (i). Então, as localizações P (i) = (X (i), Y (I)) de uma MT são separados em duas coordenadas positivas diferentes trajetórias X (i) e Y (i), em que X (i) expressa o abcissa trajetória, e Y (i) é a ordenada um ("one"). |
| | Foi explorada essa propriedade, a fim para rastrear um terminal móvel e prever sua trajetória com intervalos de confiança relativamente razoáveis. Particularmente, foi utilizado o paradigma da integração fracionário digital (DFI) para melhorar significativamente a precisão de MT padrão de mobilidade previsões em ambientes indoor. |
| **Utiliza algum Algoritmo de Localização?** | Foi utilizado o paradigma da integração fracionário digital (DFI) |
| **Cobertura** | NA |
| **Contextualização** | Foi avaliado o desempenho de esta previsão DFI proposto em dois cenários de trajetos indoor, inspirados por padrões de mobilidade do usuário típicos em condições típicas de indoor (visita ao museu e caminhar pelo hospital). |
| **Tem apoio ferramental. Qual?** | • Não |
| **Infra** | NA |
| **Abordagem Hibrida** | • Sim<br>Foi proposto utilizar “Digital Fractional Integration based filter” integrado a “Kalman filters”. |
| **Qual é o resultado da pesquisa?** | • Método |
| **Resultados** | O filtro de Kalman demonstra que os desempenhos dos integrais fraccionais melhoraram o desempenho dos preditores clássicos.<br>Constatou-se que abordagem baseada em DFI permite reduzir o tamanho do arquivo que leva em muitos casos a uma diminuição no efeito em espiral do acumulado erro de arredondamento que se aglomera ao longo de um tempo de acompanhamento longo, devido ao menor número de adendos. |
| **Limitações** | NA |
| **Trabalhos Futuros** | 1. Em trabalhos futuros incidirá sobre a concepção de um algoritmo para otimizar ainda mais o desempenho dessa trajetória baseada no algoritmo de estimação DFI.<br>2. Várias aplicações da presente abordagem proposta estão em desenvolvimento, tais como navegação de pessoas com deficiência visual Indoor, dentro da universidade campus. |
| | 3. A técnica proposta pode ser adicionada como uma camada nos dispositivos existentes |
| **Algoritmos** | NA |

37)

| Dados do ARTIGO | Using prior measurements to improve probabilistic-based indoor localization methods |
|---|---|
| **Autor** | Bera, R. and Kirsch, N.J. and Fu, T.S. |
| **Ano** | 2013 |
| **Keywords** | indoor localization; wireless sensor networks |
| **Siglas importantes** | – Received signal strength (RSS);<br>– Prior Measurement Comparison (PMC);<br>– Minimum Mean Squared Error (MMSE);<br>– Probability-based Maximum Likelihood (PML);<br>– Enhanced PML (EPML). |
| **Objetivo do Artigo** | Pelo fato que modelos baseados em probabilidade necessitam de uma prévia distribuição para o nodo desconhecido, foi proposto que medições de referência sejam realizadas no momento da implantação, de modo a melhorar o modo de distribuição. |
| **Técnicas/Abordagens/métodos/algoritmos** | |
| **Qual é o tipo de pesquisa?** | • Empírica;<br>Simulações monte Carlo forma realizadas |
| **Especificação do tipo da técnica** | Neste trabalho, foi melhorado um método probabilístico para determinar a posição de um nodo desconhecido em ambientes indoor. |
| **Descrição da técnica** | 1. Foi demonstrado que os subconjuntos podem reduzir o erro na localização de um nodo;<br>2. O método PMC primeiro compara o RSS a partir do nó desconhecido para todos RSS de todos os pontos de referência;<br>3. Chama-se o método de Comparação de "Fingerprinting" (FP);<br>4. Então um subconjunto de âncoras é escolhido para estimar a posição usando PML. |
| **Utiliza algum Algoritmo de Localização?** | |
| **Cobertura** | Foi demonstrado neste novo método reduz o erro de localização e aumenta a probabilidade de selecionar a cômodo/ambiente correto.<br>As simulações de Monte Carlo foram repetidas até que o intervalo de confiança de 95% atinge 0,2 m. |
| **Contextualização** | 1. Simulou-se uma área de 100 por 100 metros;<br>2. Separou-se em nove subconjuntos;<br>3. Cada subconjunto contém quatro quartos de tamanho 16,5 m. |
| **Tem apoio ferramental. Qual?** | • Sim<br>As simulações de Monte Carlo |
| **Infra** | NA |
| **Abordagem Hibrida** | • Não |
| **Qual é o resultado da pesquisa?** | • Técnica |
| **Resultados** | As simulações mostram que PMC+ reduz o erro de estimativa de localização e aumenta a probabilidade de selecionar o ambiente correto. |
| **Limitações** | NA |
| **Trabalhos Futuros** | O trabalho futuro irá investigar o impacto da alteração do valor de confiança a região, P, considere não uniforme arena, e incluem os resultados das medições. |
| **Algoritmos** | NA |

38)

| Dados do ARTIGO | Indoor localization system based on fingerprint technique using RFID passive tag |
|---|---|
| **Autor** | Phimmasean, S. and Chuenurajit, T. and Cherntonomwong, P. |
| **Ano** | 2013 |
| **Keywords** | Indoor localization; Fingerprint technique; RFID passive tags; Dot product; Graphic user interface |
| **Siglas importantes** | – Radio Frequency Identification (RFID);<br>– Receive Strength Signal Information (RSSI);<br>– Fingerprint locations. |
| **Objetivo do Artigo** | O estudo propõe sistema de localização indoor utilizando a técnica de “fingerprint” para encontrar a localização estimada do Target. |
| **Técnicas/Abordagens/métodos/algoritmos** | Esta técnica é comparada com a seu padrão de sinal (ou informações ID) com dados do banco de dados de sinais conhecidos. |
| **Qual é o tipo de pesquisa?** | • Empírica; |
| **Especificação do tipo da técnica** | Neste artigo, as etiquetas RFID são implantadas como o nó de referência e o leitor de RFID é empregue como o destino, outro ponto refere-se ao uso da utilização do padrão de rádio freqüência para identificação. |
| **Descrição da técnica** | Nesta pesquisa, a informação ID provindas das TAGs é aplicada como padrão de igual modo identificação humana. Esses IDs (número decimal) serão mapeados para os vectores de linha [1 x 64] e salvos no banco de dados. Da mesma maneira o ID observado também é mapeado para o vetor linha antes de comparar com a informação de "fingerprint". O produto escalar é implementado como o padrão, para apoiar a comparação entre entre dois vetores de linha (informações de "fingerprint" e informação observada) para encontrar localização estimada do target. |
| **Utiliza algum Algoritmo de Localização?** | NAO |
| **Cobertura** | A distância média de erro é de aproximadamente 75 centímetros |
| **Contextualização** | Neste artigo, o sistema consiste em transponders (tags) com o software coletando os dados recebidos pelo leitor e tags.<br>A altura do leitor é de 160 cm, o que é a altura média de povos asiáticos. |
| **Tem apoio ferramental. Qual?** | • Não |
| **Infra** | NA |
| **Abordagem Hibrida** | • Não |
| **Qual é o resultado da pesquisa?** | • Método/Sistema<br>O estudo propõe sistema de localização indoor utilizando a técnica de "fingerprint" para encontrar a localização estimada do Target. |
| **Resultados** | Os resultados mostram a média da distância de erro de 74,41 centímetros para todas as observações. Estes resultados satisfazem as expectativas, pois são menos de 120 centímetros (o intervalo entre duas TAGs). A |
| | partir do resultado, “este método pode ser aplicado em situação real”. |
| **Limitações** | NA |
| **Trabalhos Futuros** | Para o trabalho futuros, targets móveis serão aplicados em Localização indoor 3D. Esta aplicação será capaz de identificar as diferentes alturas e deslocamento do alvos em uma GUI. |
| **Algoritmos** | NA |

39)

| Dados do ARTIGO | ZiFind: Exploiting cross-technology interference signatures for energy-efficient indoor localization |
|---|---|
| **Autor** | Gao, Y. and Niu, J. and Zhou, R. and Xing, G. |
| **Ano** | 2013 |
| **Keywords** | NA |
| **Siglas importantes** | – Received Signal Strength (RSS);<br>– Digital signal processing (DSP) techniques;<br>– False positive (FP);<br>– False negative (FN) rates;<br>– Access points(AP);<br>– Common Multiple Folding (CMF); |
| **Objetivo do Artigo** | Este artigo apresenta um novo sistema de localização indoor chamado ZiFind que explora a interferência da “cross” tecnologia na freqüência( espectro) não licenciada de 2,4 GHz. |
| **Técnicas/Abordagens/métodos/algoritmos** | Neste trabalho, foi explorada a interferência de WIFI em tecnologias de baixo consumo de energia, como a ZigBee e Bluetooth. |
| **Qual é o tipo de pesquisa?** | • Empírica;<br>Foi implementado ZiFin em “TelosB motes platform” e avaliadas tal desempenho por meio de experimentos. |
| **Especificação do tipo da técnica** | O projeto de ZiFind explora a interferência da tecnologia “cross” na freqüência( espectro) não licenciada de 2,4 GHz. A adoção de rádio de 2,4 GHz espectro conduziu a tecnologias comerciais “ baratas” como: WiFi, Bluetooth e ZigBee. Essa coexistência de múltiplas tecnologias sem fio na mesma freqüência banda é considerada como uma "maldição", uma vez que muitas vezes provoca uma interferência significativa. |
| **Descrição da técnica** | Neste trabalho, foi explorada a interferência de WIFI em tecnologias de baixo consumo de energia, como a ZigBee e Bluetooth. Assim, usuário móvel utiliza uma interface de ZigBee para detectar as assinaturas de interferência únicas induzida pela infra-estrutura WIFI e utiliza-los como “fingerprints” para estimar a localização atual.<br>Para lidar com o ruído nos “fingerprints”, foi projetado um novo algoritmo de aprendizado chamado R-KNN que pode melhorar a precisão da localização através da atribuição de pesos diferentes características da “fingerprint” de acordo com a sua importância. |
| **Utiliza algum Algoritmo de Localização?** | Foi projetado um novo algoritmo de aprendizado chamado R-KNN que classifica a localização do utilizador com base numa base de dados de impressões digitais com etiquetas com locais conhecidos.<br><br>CMF algoritmo. |
| **Cobertura** | NA |
| **Contextualização** | Foi conduzido os experimentos no 10º andar do CSE edifício da Universidade Beihang, Beijing, China, o |
| | qual possui:<br>• Tem uma metragem de 16.000 metros quadrados;<br>• Há 34 quartos no total e foi usado 28 deles para o experimento;<br>• Cada quarto tem uma área de cerca de 3,75 por 8 m2. |
| **Tem apoio ferramental. Qual?** | • Não |
| **Infra** | A rede Wi-Fi se encontrava disponível no edifício. A pesquisa aconteceu no 10º andar com cobertura de mais de 100 APs, e em cada sala e o número de APs detectados varia desde 13 a 41. Dentre 16 canais ZigBee, em 2,4 GHz espectro, clientes ZiFind foram configuradas para escutar canal 17, que possuia uma sobreposição significativa com o canal de Wi-Fi 6.<br><br>A infra foi composta por 2 clientes, 4 mapeadores e um servidor. Os clientes são laptops Lenovo ThinkPad T400 integrado com motes TelosB. Laptops Lenovo ThinkPad L421 foram usados como mapeadores e servidor. Todos os laptops rodar o Linux com kernel versão 2.6.32. O servidor e “mapeadores” conectar à produção de rede Wi-Fi através de suas interfaces Wi-Fi no canal 6. |
| **Abordagem Hibrida** | • Sim |
| **Qual é o resultado da pesquisa?** | • Método<br>Este artigo apresenta um novo sistema de localização indoor chamado... |
| **Resultados** | Os resultados mostraram que ZiFind conduz a economia de energia significativa em comparação com as abordagens existentes com base na interface Wi-Fi, e produz precisão de localização satisfatória num intervalo de realista definições. |
| **Limitações** | NA |
| **Trabalhos Futuros** | NA |
| **Algoritmos** | **Algorithm IV.1** ***Alignment*** **Algorithm**<br>**Input:** $T_A$-timing information from client, $T_B$-combined timing information from mappers, $BSSID_B(t)$-combined BSSID information from mappers, $O$-maximum alignment offset.<br>**Output:** $BSSID_A(t)$-BSSID information of $T_A$.<br><br>Generate $RSS_A$ from $T_A$, $l$=len($RSS_A$)<br>Generate $RSS_B$ from $T_B$, len($RSS_B$)=$l$<br>**for all** $i \in (0, O)$ **do**<br>  /*find the partition with the minimum sum of LCMs*/<br>  $s = \{\sum_{x=1}^{l-i} RSS_A(x) * RSS_B(x+i)\}/(l-i)$<br>  $S \leftarrow S \cup s$<br>**end for**<br>/*argmax represents the argument of the maximum*/<br>$Offset = argmaxS(x)$<br>$BSSID_A(t)$=$BSSID_B(t+Offset)$<br>return |

40)

| Dados do ARTIGO | Practical indoor localization using ambient RF |
|---|---|
| **Autor** | Tian, Y. and Denby, B. and Ahriz, I. and Roussel, P. and Dubois, R. and Dreyfus, G. |
| **Ano** | 2013 |
| **Keywords** | localization; indoor; fingerprint; transductive support vector machine |
| **Siglas importantes** | – Location Based Services (LBSs);<br>– Received Signal Strength (RSS);<br>– Support Vector Machine (SVM) regression. |
| **Objetivo do Artigo** | O artigo apresenta uma abordagem simples e prática para localização indoor utilizando RSS de “fingerprints” a partir da rede GSM, incluindo uma análise do relacionamento entre a força do sinal e localização, e da evolução do desempenho de localização ao longo do tempo. |
| **Técnicas/Abordagens/métodos/algoritmos** | GMS RSS **X** SVM |
| **Qual é o tipo de pesquisa?** | • Empírica;<br>Foram utilizados os dados coletados durante as “caminhadas aleatórias" que exploraram uma área inteira. Foram coletados dois tipos de conjuntos de dados, para a regressão e experimentos de classificação. |
| **Especificação do tipo da técnica** | É investigada a relação funcional entre “GSM RSS” e posicionamento, usando SVM. |
| **Descrição da técnica** | 1. Foram utilizados os dados coletados durante as “caminhadas aleatórias" que exploraram uma área inteira;<br>2. Foram coletados dois tipos de conjuntos de dados, para a regressão e experimentos de classificação;<br>3. Foi apresentado testes de longa duração que mostra que o método de localização indoor apresenta desempenho que se degrada ao longo do tempo;<br>4. O que foi introduzido inferência “transdutivo”, que utiliza novos dados não rotulados de entrada para atualizar classificadores SVM, como um meio de reduzir a degradação do desempenho causada por RSS “drift”;<br>5. A utilização de pequenas quantidades de novos dados rotulados como um regime de atualização modelo também é aqui explorado. |
| **Utiliza algum Algoritmo de Localização?** | Utilizados regressão linear e não-linear (“Gaussian” e “Polynomial kernels”). |
| **Cobertura** | NA |
| **Contextualização** | Foram coletados dois tipos de bases de dados, para a regressão e experimentos de classificação, respectivamente, ambos registados na 4º andar de um edifício de laboratório (estrutura de aço, concreto e paredes de gesso), no centro de Paris, França. |
| **Tem apoio ferramental. Qual?** | • Não |
| **Infra** | Cada "scan" contém o RSS de todas as 548 operadoras nas bandas GSM900 e GSM1800, consistindo em valores que variam em RSS valor de -108dBm para -40dBm. Todas as verificações foram marcadas manualmente com localização de 0m a 4.2m indicando onde o experimento foi realizado.<br><br>O dispositivo de coleta de dados foi um telefone celular Sony-Ericsson com software de digitalização embutido, que é capaz de obter uma imagem de toda as bandas GSM900 e GSM1.800 em cerca de 300 milissegundos. |
| **Abordagem Hibrida** | • Sim |
| **Qual é o resultado da pesquisa?** | • Método<br>O artigo apresenta uma abordagem simples e prática para localização indoor utilizando RSS de "fingerprints" a partir da rede GSM... |
| **Resultados** | O erro de regressão é aproximadamente igual à distância média entre os locais , o que significa relação não linear entre RSS e a posição em um ambiente pequeno é claro.<br>A utilização de classificadores "GSM RSS-based" com dados coletados em todas as áreas de quartos, apresentam uma alternativa viável, e os resultados experimentais mostram que a percentual de rotulagem correta quarto pode ser de até 94%, se a modelo for utilizado antes dos " significant RSS drift sets in". |
| **Limitações** | NA |
| **Trabalhos Futuros** | Pretende-se investigar o uso de dados da rede W-CDMA em medições. |
| **Algoritmos** | NA |

41)

| Dados do ARTIGO | Avoiding multipath to revive inbuilding WiFi localization |
|---|---|
| **Autor** | Sen, S. and Lee, J. and Kim, K.-H. and Congdon, P. |
| **Ano** | 2013 |
| **Keywords** | Wireless, Localization, Cross-Layer, Application, Indoor positioning |
| **Siglas importantes** | − Access point (AP);<br>− Physical layer (PHY);<br>− Angle of the direct path(ANDP);<br>− channel state information (CSI);<br>− Angle-of-arrival (AoA). |
| **Objetivo do Artigo** | Esta solução, CUPID, identifica e “harnesses” apenas o caminho direto, evitando o efeito de reflexões ”multipath”. |
| **Técnicas/Abordagens/métodos/algoritmos** | A principal constatação é que a mobilidade natural do ser humano, quando combinado com informações da |
| | camada PHY, pode ajudar na estimativa com precisão o ângulo e distância de um móvel dispositivo a partir de um ponto de acesso sem fio (AP). |
| **Qual é o tipo de pesquisa?** | • Empírica;<br>Experimentos indoor real utilizando “off-the-shelf” chipsets wireless confirmar a viabilidade do CUPID . |
| **Especificação do tipo da técnica** | CUPID utiliza a camada física de informações (PHY) para extrair a força do sinal e o ângulo de somente o caminho direto evitando com êxito o efeito de reflexo de “multipaths”. |
| **Descrição da técnica** | Este sistema, CUPID distingue o caminho direto a partir de reflexo de múltipaths, aproveitando informações da camada PHY junto com mobilidade humana natural. CUPID determina com precisão a distância e o ângulo do dispositivo móvel do seu AP, em última análise, produzindo sua localização. CUPID não requer “crowdsourcing”, ou hardware adicional, mas depende de múltiplas antenas presentes em APs wireless.<br><br>“Dead reckoning” envolve computar o deslocamento do usuário a partir acelerómetro, e rastreando a direção de movimento utilizando a bússola. Para calcular o deslocamento físico do usuário, CUPID identifica o deslocamento do a partir das leituras do acelerômetro combinadas com a estimativa AoA da mobilidade padrão do usuário, para calcular o ângulo do cliente (ANDP). Explorando ainda mais o giroscópio do usuário para lidar com a ambiguidade lado introduzido pela colocação linear das antenas. |
| **Utiliza algum Algoritmo de Localização?** | NA |
| **Cobertura** | CUPID é capaz de localizar um dispositivo, quando apenas um único AP está presente. Quando mais alguns APs estão disponíveis, Cupido pode melhorar o erro de localização mediana de 2,7 m, o que é comparável aos esquemas que dependem de “fingerprinting” ou infra-estrutura adicional. |
| **Contextualização** | Foram realizados experimentos em um ambiente de escritório com 5 APs instalados, e ANDP de um cliente, e os envia para o servidor local. O servidor sabe a localização dos APs, e pode combinar as informações coletadas a partir de vários APs para estimar o local do cliente. A distancia percorrida arbitrariamente no prédio por uma hora durante o horário de expediente normal, cobrindo aproximadamente 4500m2. |
| **Tem apoio ferramental. Qual?** | • Não |
| **Infra** | CUPID foi implementado usando laptops com Atheros 9390 wireless e telefone Google Nexus S. O telefone foi sincronizado para timestamping e fisicamente conectado ao laptop. |
| **Abordagem Hibrida** | • Sim |
| **Qual é o resultado da pesquisa?** | • Técnica<br>• Método |
| **Resultados** | Experimentos utilizando off-the-shelf chipsets Atheros, e Telefones Google Nexus S confirmam a viabilidade, demonstrando precisão comparável ao de técnicas baseadas "fingerprinting". |
| **Limitações** | 1. O experimento foi executado com o telefone em mãos;<br>2. Enquanto estimativa do ângulo só pode depender da propagação wireless, o calculo a distancia usando a equação de perda do percurso, não é independente da potência de transmissão do cliente ("txpower");<br>3. Desempenho do Cupido certamente pode melhorar se antenas adicionais forem adicionadas. |
| **Trabalhos Futuros** | O objetivo é o de identificar heurísticas que coletem multipaths e informações diretamente do LoS de cada pacote CSI recebido, em vez de tentar sintonizar sofisticado de caminho-lossmodels em cada local diferente e meio ambiente. |
| **Algoritmos** | NA |

42)

| Dados do ARTIGO | Received signal strength based room level accuracy indoor localisation method |
|---|---|
| **Autor** | Buchman, A. and Lung, C. |
| **Ano** | 2013 |
| **Keywords** | received signal strength; WLAN; WiFi; covariance; |
| **Siglas importantes** | – Received signal strength (RSS);<br>– Interim room templates (IRT);<br>– Room templates (RT). |
| **Objetivo do Artigo** | O objetivo é alcançar acurácia de localização de um target em um ambiente (sala) indoor, utilizando WLAN já implantado infra-estrutura, mas sem realmente interferir com a rede. |
| **Técnicas/Abordagens/métodos/algoritmos** | matched filtering |
| **Qual é o tipo de pesquisa?** | • Conceitual;<br>O método precisa de mais testes, mas os resultados atuais são promissores. |
| **Especificação do tipo da técnica** | O método em suma é o seguinte:<br><br>1. Registe a leitura RSSI em cada sala de interesse;<br>2. Crie "interim room templates" (IRT)';<br>3. Normalize IRTs para a obtenção de modelos de sala (RT);<br>4. Programe o Target para transmitir leituras RSSI no formato RT;<br>5. Calcule a covariância de RT do alvo com RT de todos os quartos;<br>6. A localização do target é o quarto com a maior covariância. |
| **Descrição da técnica** | Este "covariance based fingerprinting" é derivado de um método de processamento de sinal: "matched filtering". Um "matched filtering." correlaciona-se um modelo de sinal, com o sinal de entrada, a fim de detectar a presença de um template no sinal de entrada. A decisão é feita através da comparação da saída do dispositivo de correlação para "threshold". Se o threshold for superior a um se pode presumir um match entre |
| | o sinal de entrada e o sinal do template. No nosso caso o sinal de entrada seria uma n-tuplo de leituras da RSSI do dispositivo rastreado. Isto é para ser correlacionado com o conjunto de templates de ambientes. Os coeficientes de correlação resultantes são ordenados e o modelo que produz a potência máxima é designado para ser o “match” correto. |
| **Utiliza algum Algoritmo de Localização?** | Algoritmo kNN |
| **Cobertura** | Um erro médio de 2 a 3 metros. |
| **Contextualização** | NA |
| **Tem apoio ferramental. Qual?** | • Não |
| **Infra** | NA |
| **Abordagem Hibrida** | • Não |
| **Qual é o resultado da pesquisa?** | • Método<br>O método precisa de mais testes, mas os resultados atuais são promissores. |
| **Resultados** | Como conclusão, estes resultados parecem promissores, mas ainda um grande número de testes deve ser realizado. |
| **Limitações** | Como conclusão, estes resultados parecem promissores, mas ainda um grande número de testes podem ser realizados, com diferentes tipos de dispositivos em diferentes ambientes ”indoors”. |
| **Trabalhos Futuros** | NA |
| **Algoritmos** | NA |

43)

| Dados do ARTIGO | Design and implementation of a real time locating system utilizing Wi-Fi signals from iPhones |
|---|---|
| **Autor** | Bharanidharan, M. and Li, X.J. and Jin, Y. and Pathmasuntharam, J.S. and Xiao, G. |
| **Ano** | 2012 |
| **Keywords** | IPhone; OpenWrt; Socket; Cross Compilation. |
| **Siglas importantes** | – Real time locating system (RTLS)<br>– Received signal strength indication (RSSI)<br>– Received signal strength (RSS)<br>– Access points (AP);<br>– know k-nearest neighbor (kNN). |
| **Objetivo do Artigo** | Neste trabalho foi estudado e implementado uma abordagem alternativa de utilizar APs em modo de monitoramento para coletar o RSSIs para sinais de Wi-Fi de um dispositivo móvel. |
| **Técnicas/Abordagens/métodos/algoritmos** | • Ao coletar o RSSIs para sinais de Wi-Fi a partir de um dispositivo móvel.<br>• Através da aplicação de um algoritmo de localização semelhante ao kNN. |
| **Qual é o tipo de pesquisa?** | • Empírica;<br>Os resultados experimentais mostram que esta abordagem pode atingir precisão de localização melhor do que a abordagem convencional. |
| **Especificação do tipo da técnica** | A abordagem proposta é estabelecida usando infra-estruturas de Wi-Fi de redes locais wireless aplicados a protocolos padrão e sua relação com os valores intensidade do sinal recebido, buscando, assim, a localização de um usuário de IPHONE em ambientes indoors.<br><br>Foi implementado a biblioteca de captura de pacotes, por meio da qual foi obtida a valor de intensidade de sinal recebido dos pacotes recebidos de a estação móvel para os APs sem fio. |
| **Descrição da técnica** | |
| | A idéia por trás desse sistema é fazer com que o IPHONE transmita os pacotes de Wi-Fi para todos os APs wireless próximo.<br><br>1. Os APs, que estão em modo de monitoramento, serão identificados por esses pacotes e uma vez que tenhamsido capturados, assim serão recuperados os campos obrigatórios a partir dos pacotes de entrada da iPhone. Os campos são recuperados RSSIs dos pacotes de entrada, o endereço MAC do IPHONE, hora de chegada e número de seqüência dos pacotes;<br>2. Esses valores serão enviados dos APs para um servidor, que tem um MySQL server instalado;<br>3. Um algoritmo de MATLAB chamado “Region based fingerprinting” é utilizado neste projeto para estimar a localização do usuário desconhecido;<br>4. Uma vez feito isso, o servidor precisa enviar as coordenadas de localização (x, y) de volta para o usuário. Ele envia a resposta através mesmo soquete da rede, que foi criado quando o IPHONE queria se comunicar com o servidor de localização;<br>5. O cliente (IPHONE) pode desligar a qualquer momento a partir deste soquete e pode conectar-se a mesma sempre que necessário. |
| **Utiliza algum Algoritmo de Localização?** | Através da aplicação de um algoritmo de localização semelhante ao kNN. |
| **Cobertura** | NA |
| **Contextualização** | NA |
| **Tem apoio ferramental. Qual?** | • Sim<br>Um algoritmo de MATLAB chamado “Region based fingerprinting” é utilizado neste projeto para estimar a localização do usuário desconhecido. |
| **Infra** | NA |
| **Abordagem Hibrida** | • Sim<br>Neste trabalho foi estudado e implementado uma abordagem alternativa de utilizar APs em modo de monitoramento para coletar o RSSIs para sinais de Wi-Fi de um dispositivo móvel. |
| **Qual é o resultado da pesquisa?** | • Método/Abordagem |
| **Resultados** | Os resultados experimentais mostram que esta abordagem pode conseguir uma melhor acurácia a localização do que a abordagem convencional. |
| **Limitações** | NA |
| **Trabalhos Futuros** | Em pesquisas futuras, as coordenadas podem ser melhoradas pela adição banco de dados. |
| **Algoritmos** | NA |

44)

| Dados do ARTIGO | Centaur: Locating devices in an office environment |
|---|---|
| **Autor** | Nandakumar, R. and Chintalapudi, K.K. and Padmanabhan, V.N. |
| **Ano** | 2012 |
| **Keywords** | Indoor localization, WiFi, acoustic ranging, Bayesian inference |
| **Siglas importantes** | – Radio Frequency (RF);<br>– Acoustic Ranging (AR);<br>– Inter-device Distance Constraints (IDC) |
| **Objetivo do Artigo** | A contribuição fundamental deste trabalho é Centaur, um framework que une RF e AR, técnicas de localização baseadas em uma única estrutura sistemática que é baseada em inferência bayesiana. |
| **Técnicas/Abordagens/métodos/algoritmos** | Como podemos combinar Abordagens **RF** e **AR** baseadas em sinergia para localizar uma ampla gama de dispositivos, aproveitando os benefícios de ambas as abordagens? |
| **Qual é o tipo de pesquisa?** | • Empírica;<br><br>Neste experimento foi guiado pela seguinte questão: Na ausência de quaisquer âncoras (dispositivos com locais conhecidos), pode o desempenho de localização WIFI ser melhorada usando medições de distâncias inter-dispositivo? |
| **Especificação do tipo da técnica** | Centaur é agnóstico às técnicas RF ou AR específica utilizada, dando aos usuários a flexibilidade de escolher os seus regimes de RF ou AR preferenciais. Entre as contribuições adicionais, se encontram: fazer AR mais robusto em relação a configurações non-line-of-sight (EchoBeep) e adaptando AR para localizar dispositivos "speaker-only" (DeafBeep). Também se pode avaliar o desempenho de melhorias AR graças ao framework Centaur através microbenchmarks e implantação em um ambiente de escritório. |
| **Descrição da técnica** | O software Centaur compreende um servidor e um cliente.<br><br>1. Os clientes foram implantados em dispositivos como laptops e desktops PCs. Os clientes se conectam ao servidor sempre que os dispositivos estão ligados e comunicam ao servidor para saber se eles têm um WIFI-card, alto-falantes e / ou um microfone.<br>2. Em caso de nodos âncora relatam seus localizações conhecidas para o servidor. O servidor Centaur então, dependendo das capacidades do dispositivo, programa os dispositivos para realizar medições WIFI, transmitir "Chirp sequências", e gravar sons ambientes para EchoBeep e Medições DeafBeep.<br>3. Os clientes transmitem todos os dados gravados (Medições Wifi e amostras de som gravadas) para o servidor. Centaur então usa EchoBeep e DeafBeep para converter os dados recebidos em estimativa da localização do device. |
| **Utiliza algum Algoritmo de Localização?** | Técnicas como a BeepBeep pode ser usada para estimar a distância entre eles, proporcionando um IDC para cada tal par.<br>Na primeira etapa, a etapa de inferência parcial, Centaur realiza uma inferência exata usando o algoritmo de Pearl's, acomodando tanta evidência quanto possível sem ter loops no gráfico. |
| **Cobertura** | NA |
| **Contextualização** | Através de extensos experimentos em um ambiente de escritório, dados de treinamento de defini-las foi gerado através da recolha de medições RSS em 117 locais de grade que abrangem todo o piso, com um valor aproximado espaçamento entre os locais de 3m e cerca de 12000 beacons reunidos por AP em cada local. Para o teste, foi medido WIFI RSS em um conjunto separado de 91 locais de teste. |
| **Tem apoio ferramental. Qual?** | • Não |
| **Infra** | Configuração CPU:<br>– Intel Xeon CPU E5405 com 2.0 GHz<br>– 2 processadores, 8 GB RAM and a 64 bit |
| **Abordagem Hibrida** | • Sim |
| **Qual é o resultado da pesquisa?** | • Framework<br>A contribuição fundamental deste trabalho é Centaur, um framework que une RF e AR, técnicas de localização baseadas em uma única estrutura sistemática que é baseada em inferência bayesiana. |
| **Resultados** | Centaur não só melhora a acuracidade da localização ao longo do que obtido graças às qualquer um dos “schemes”, também permite a localização de dispositivos, tais como “ speaker-only PCs”, que até agora não eram passíveis de localização ou à base de Wi-Fi ou acústica variando. |
| **Limitações** | NA |
| **Trabalhos Futuros** | NA |
| **Algoritmos** | NA |

45)

| Dados do ARTIGO | A Reliable and Accurate Indoor Localization Method Using Phone Inertial Sensors |
|---|---|
| **Autor** | Li, F. and Zhao, C. and Ding, G. and Gong, J. and Liu, C. and Zhao, F. |
| **Ano** | 2012 |
| **Keywords** | Indoor Localization, Inertial Tracking, Pedestrian Model |
| **Siglas importantes** | – Inertial measurement unit (IMU). |
| **Objetivo do Artigo** | Neste trabalho, foi desenvolvido um sistema de localização prática indoor que se baseia em apenas sensores de smartphones. A melhor desta abordagem é que representa o primeiro sistema a ser capaz de maneira confiável fornecer com a precisão de posicionamento de usuários de smartphones sem qualquer infra-estrutura adicional. |
| **Técnicas/Abordagens/métodos/algoritmos** | – IMU; |
| **Qual é o tipo de pesquisa?** | • Empírica;<br>Foi conduzido experimentos extensos |
| **Especificação do tipo da técnica** | Foi desenvolvido algoritmos para uma confiável detecção de passos e designação de direcionamento, além de 1 precisa estimativa e personalização do comprimento do passo do usuário. Foi construído um sistema “end-to-end” de localização integrando esses módulos e um mapa de chão indoor, sem a necessidade de infra-estruturas adicionais. |
| **Descrição da técnica** | O sistema interage com o usuário para obter a localização inicial através de inputs do usuário, e fornece uma estimativa de posição atual em um mapa indoor.  Além disso, um módulo sensor fornece leituras IMU para estes "estimadores" continuamente. O sistema funciona como a seguir:<br><br>1. O mapa de andar é dividido em pequenos azulejos e cada azulejo é rotulado como caminho, sala espaço, ou parede.<br>1.1. O mapa é carregado para o telefone na forma de um arquivo XML, que contém a imagem do mapa e associado aos metadados.<br>1.2. Toda vez que um usuário abre o aplicativo, uma lista de mapas irá aparecer pedindo que o usuário selecione um.<br>1.3. Após o mapa seja selecionado é exibido na tela, o usuário toca e mantém a entrada do local atual do sistema;<br>2. Em seguida, inicia-se automaticamente para rastrear o usuário e atualizações acerca do deslocamento do usuário.<br>2.1. O usuário pode também definir parâmetros como a taxa de amostragem do sensor e de "particle number", com um tradeoff entre a precisão de posicionamento e custo. |
| **Utiliza algum Algoritmo de Localização?** | Foi desenvolvido algoritmos para a confiável detecção de posição e indicações de passos. |
| **Cobertura** | O resultado da avaliação mostrou que o sistema pode atingir uma precisão média de 1.5m para o caso em mãos e 2m para o caso em moderadoras em um 31m X 15m área de testes. |
| **Contextualização** | Foi testado este sistema em um prédio de escritórios, com o mapa do chão ("floor"). Os dados foram recolhidos a partir de 13 indivíduos com perfis físicos diferentes.<br><br>1. Cada indivíduo caminhava ao longo de uma área de 31 X 15m, levando dois telefones com um na mão e outro no bolso da calça da trilha de caminhada a cada 50 centímetros.<br>2. Em seguida, as coordenadas e timestamp foram rotulado e associado a cada passo. O erro de posição é calculado como a distância prevista entre a posição estimada e a de verdade de campo rotulados ao longo da caminhada a pé. |
| **Tem apoio ferramental. Qual?** | • Não |
| **Infra** | Foi construído um protótipo no Windows Phone 7.5 e testado no que o HTC Mazza. |
| **Abordagem Hibrida** | • Sim<br>IMU X mapeamento do terreno X Client-server side |
| **Qual é o resultado da pesquisa?** | • Método/sistema/algoritmo<br>Foi desenvolvido algoritmos para uma confiável detecção de passos e designação de direcionamento. |
| **Resultados** | Na avaliação, a interferência magnética é uma das principais causas para o erro. Esta sendo exploradas maneiras de calibrar as leituras automaticamente e métodos adicionais para compensar o erro.<br>Embora esta implementação atual não dependa de infra-estrutura, o sistema pode beneficiar da assistência de infra-estrutura quando é possível. |
| **Limitações** | Cada vez que o sistema não conseguiu detectar um usuário, o mesmo teve que reiniciar o sistema para obter o sistema de volta. Com apenas o modelo genérico, o sistema reporta 17 falhas. |
| **Trabalhos Futuros** | Foi planejado para melhorar este sistema de várias maneiras. Designação direção de inferência ainda é um problema em aberto. |
| **Algoritmos** | NA |

46)

| | |
|---|---|
| Dados do ARTIGO | A study on wireless sensor network based indoor positioning systems for context-aware applications |
| **Autor** | Wang, J. and Prasad, R.V. and An, X. and Niemegeers, I.G.M.M. |
| **Ano** | 2012 |
| **Keywords** | indoor positioning; RSSI; Tmote; context-awareness |
| **Siglas importantes** | – Minimum mean square error (mmse);<br>– Received signal strength indicator (rssi);<br>– Wireless sensor networks (wsn). |
| **Objetivo do Artigo** | Considerando-se que a intensidade do sinal pode ser fortemente distorcida devida para “multipaths” e sombreamento, foi proposto aos regimes de pesagem (“weighing schemes”) para alavancar a credibilidade dos valores RSSI medidos. |
| **Técnicas/Abordagens/métodos/algoritmos** | NA |
| **Qual é o tipo de pesquisa?** | • Empírica;<br>Realizou-se um experimento simples para saber como a orientação da antena afeta os valores de RSSI. |
| **Especificação do tipo da técnica** | Para acompanhamento on-line, foi proposta a seleção limite e digitalização da rede local para diminuir o tempo de procura, e da divulgação de dados RSSI e esquemas de coleta de modo a diminuir o overhead. As avaliações foram feitas em ambas as medições de campo e com um gerador de RSSI. |
| **Descrição da técnica** | No curso da concepção do sistema, tentou-se aproveitar o trade-off entre a simplicidade e precisão, e também levado em conta o tempo real exigência de muitas aplicações sensíveis ao contexto. Foram escolhidas técnicas baseadas em RSSI de posicionamento por implantar wireless simples sensores (Tmote Sky).Ele foi desenvolvido e implementado tanto algoritmo baseado em intervalo e algoritmo de free-range, que são considerados como as duas principais abordagens da Técnicas baseadas em RSSI, em nosso banco de teste |
| **Utiliza algum Algoritmo de Localização?** | Neste experimento, três algoritmos baseados em alcance, CMMSE, M-MMSE, e W-MMSE foram comparados. |
| **Cobertura** | Para W-MMSE:<br>– 25% de posições estimadas estão dentro de uma distância de erro absoluto de 0,6m. |
| | – 75% dentro de estimativas do erro de 1,7 m. Os resultados mostram que W-MMSE dá uma precisão superior. |
| **Contextualização** | O “test-bed” foi construído no 19º andar edifício do corpo docente. A principal parte do piso a ser utilizado consiste de onze salas de escritório e uma (grande) dos alunos. A dimensão de uma sala de escritório é de 5 × 4 m, a sala de estudante é 5 × 12 m. O corredor no meio é 2m de largura. |
| **Tem apoio ferramental. Qual?** | • Não |
| **Infra** | NA |
| **Abordagem Hibrida** | • Não |
| **Qual é o resultado da pesquisa?** | • Método<br>Considerando-se que a intensidade do sinal pode ser fortemente distorcida devida para “multipaths” e sombreamento, foi proposto aos regimes de pesagem (“weighing schemes”) para alavancar a credibilidade dos valores RSSI medidos. |
| **Resultados** | O resultados mostram que sistemas de posicionamento com uma precisão adequada pode ser construído baseados no esquemas proposto.<br>– O teste de campo mostra a satisfação resultados. Em 80% dos casos estimativas no corredor estava dentro e o erro (distância) de 2m usando o algoritmo de intervalo-basedW-MMSE.<br>– No quarto, no qual a pessoa pode ficar, estimaram-se corretamente na maioria dos casos, com um erro de um quarto vizinho, em alguns casos, e com um erro de dois quartos vizinhos muito raramente por tanto em alcance e algoritmos free-range. |
| **Limitações** | NA |
| **Trabalhos Futuros** | O “test-bed” corrente é ainda relativamente em pequena escala, dentro de um único andar de um prédio de escritórios, Pretende-se estender o “test-bed” para vários andares, assim estimativa tridimensional de localização torna-se viável. Outra tarefa que é digno de exploração é implantar o participante sistema de rastreamento nas aplicações ”location-aware” para ganhar mais insights sobre sua aplicabilidade prática. |
| **Algoritmos** | NA |

47)

| Dados do ARTIGO | WiGEM : A Learning-Based Approach for Indoor Localization |
|---|---|
| **Autor** | Goswami, A. and Ortiz, L.E. and Das, S.R. |
| **Ano** | 2011 |
| **Keywords** | NA |
| **Siglas importantes** | – Expectation Maximization (EM);<br>– Gaussian Mixture Model (GMM);<br>– Received signal strength( RSS). |
| **Objetivo do Artigo** | Foi proposto um "baseado em aprendizagem" abordagem, WiGEM, aonde a intensidade do sinal recebido é modelada com um GMM. A maior contribuição deste trabalho é desenvolver um algoritmo, WiGEM, que elimina a fase caro "treinamento". |
| **Técnicas/Abordagens/métodos/algoritmos** | RSSI |
| **Qual é o tipo de pesquisa?** | • Empíria;<br>Em nossos experimentos, vamos utilizar RSSI em unidades dB. |
| **Especificação do tipo da técnica** | Usando pacotes dinâmicos a fim de capturar a estimativa de parâmetros, WiGEM pode fornecer estimativas de localização que são muito mais robustas em a face do nível de potência do dispositivo e variabilidades, mobilidade, e mudanças e reconfiguração dos espaços indoor  que muitos sistemas à base de formação são susceptíveis a. |
| **Descrição da técnica** | WiGEM aproveita a abordagem infra-estrutura baseada enquanto elimina qualquer esforço "pré-implantação". Transmissões de pacotes feitas por um cliente são recebidas no “stationary sniffers” (ou APs) |
| | que extraem RSS MAC ID do cliente-alvo e relatam esta informação para um servidor de localização central. Utilizando esta informação, WiGEM constrói um modelo para o dispositivo target e fornece uma estimativa de localização. A estimativa pode ser disponibilizada para o cliente por meio de uma simples aplicação baseada na Web, por exemplo, dependendo da aplicação desejada. |
| **Utiliza algum Algoritmo de Localização?** | NA |
| **Cobertura** | NA |
| **Contextualização** | Dois "testbeds indoors" diferentes são utilizados para a validação.<br>1. O primeiro edifício, de agora em diante chamado CEWIT, é um centro de pesquisa e desenvolvimento, em Stony Brook University com uma dimensão de 65 metros x 50 metros.<br>2. O segundo edifício, de agora em diante chamado CSD, faz parte da construção de habitação do Departamento de Ciência da Computação Stony Brook University. Este piso em forma de retangular tem uma dimensão de 20 metros x 30 metros, e também tem paredes, várias divisórias e mobiliário de escritório. |
| **Tem apoio ferramental. Qual?** | • Não |
| **Infra** | − A sniffers executar Pyramid Linux (versão 2.6.16-metrix).<br>− Um laptop é um Dell Inspiron 1545 rodando Ubuntu V9.04.<br>− 1 Telefone Android é um Nexus One Google.<br>− 1 iphone iphone 3GS (ios versão 4.2.1).<br>− 1 notbook é um Dell Latitude 2110 rodando Ubuntu v9.10. |
| **Abordagem Hibrida** | • Não |
| **Qual é o resultado da pesquisa?** | • Algoritmo<br>A maior contribuição deste trabalho é desenvolver um algoritmo, WiGEM, que elimina a fase caro "treinamento". |
| **Resultados** | As avaliações de desempenho com uma gama de diferentes dispositivos WiFi em dois "testbeds indoors" diferentes demonstram que WiGEM tem um desempenho melhor do que as técnicas baseadas em modelos |
| | e no par ou melhor do que baseada em mapa RF "state-of-the-art" técnicas. De particular importância é o desempenho superior da WiGEM quando dispositivos heterogêneos são usados e quando as técnicas baseadas em mapas de RF tem mais grosseira locais de treinamento. |
| **Limitações** | NA |
| **Trabalhos Futuros** | – Como trabalho futuro pode explorar semelhantes restrições durante o tempo de execução aumenta a de localização de precisão.<br>– Este framework também poderia ser usado para fazer um processo de treinamento mais eficiente, pelo que o modelo de propagação de rádio é substituído por alguns cuidadosamente medições feitas. Neste caso, incluindo a potência da fonte para o modelo pode aumentar a robustez do o método e fazer o trabalho para vários dispositivos.<br>– Como trabalho futuro, foi planejado para fazer de localização adaptativa fazendo aprendizagem e utilizando a adjacência dos locais como informações para acompanhar como o movimento poderia evoluir. |
| **Algoritmos** | NA |

48)

| Dados do ARTIGO | Performance analysis of a low cost wireless indoor positioning system with distributed antennas |
|---|---|
| **Autor** | Ott, A.T. and Shalaby, M. and Siart, U. and Kaliyaperumal, E. and Eibert, T.F. and Engelbrecht, J. and Collmann, R. |
| **Ano** | 2011 |
| **Keywords** | NA |
| **Siglas importantes** | – Indoor positioning systems (IPS) |
| **Objetivo do Artigo** | Neste artigo a avaliação do desempenho de localização de um sistema de antenas distribuídas relacionadas com as propriedades de um cabo onda com vazamento é relatado. |
| **Técnicas/Abordagens/métodos/algoritmos** | A ideia é a utilização de uma avaliação de detecção de nível de potência dos sinais recebidos em receptores de comunicação para localizar participantes. |
| **Qual é o tipo de pesquisa?** | • Empírica;<br>Um sistema experimental foi construído e também testado usando uma frequência automatizada de configuração de medição de domínio.<br>O desempenho do sistema de posicionamento indoor wireless é analisado por meio de simulações e medições. |
| **Especificação do tipo da técnica** | Nas simulações do processo de rastreamento de raios foi iniciado para frequências discretas com uma largura de 1 MHz ao longo de todo faixa de freqüência em cada posição do transmissor. |
| **Descrição da técnica** | Uma comparação detalhada entre os vários cenários dentro diferentes ambientes fornece uma visão profunda sobre fundamentais limitações de precisão de posicionamento e confiabilidade. Especialmente dentro de grandes edifícios como salas de concerto ou estações de trem. Assim, uma ferramenta de simulação, que permite planejar todo o sistema foi desenvolvida.<br><br>Para estimar a fidelidade máxima que pode ser alcançada por o sistema de antena que identificaram o impacto das reflexões no ambiente e no próprio sistema de antena como essencial efeito de distorções do sinal.<br><br>Para obter precisa e simulação de confiança resulta do recebimento elementos de antena foram |
| | caracterizados pelo ângulo e funções de transferência dependentes frequência semelhante à Kunisch. |
| **Utiliza algum Algoritmo de Localização?** | NA |
| **Cobertura** | Como esperado a partir de simulações a precisão de localização é significativamente aumentada com largura de banda. Dentro da necessária localização de 3m é alcançado a uma largura de banda de sistema de 70 MHz. |
| **Contextualização** | Um protótipo com quatro elementos de antena foi testada num edifício de 6m com um comprimento de 12m e uma largura de 10 m. o espaçamentos dos elementos de antena individuais foi escolhido para 3 m |
| **Tem apoio ferramental. Qual?** | • Não |
| **Infra** | NA |
| **Abordagem Hibrida** | • Não |
| **Qual é o resultado da pesquisa?** | • Simulação |
| **Resultados** | Medição resultados relativos a largura de banda do sistema para uma precisão de posicionamento mostram boa concordância com as investigações teóricas.<br><br>Ambas as simulações de sistemas numéricos e medidas práticas em um ambiente interno exemplar mostram que localização exatidão de 3m pode ser alcançado com os sinais com uma largura de bandade, pelo menos, 70 MHz. |
| **Limitações** | NA |
| **Trabalhos Futuros** | Outra melhoria pode ser esperada por aplicação de filtros de rastreamento estatísticos para o parâmetro de estimativa. Destina-se a avaliar o desempenho do sistema também com os sinais do domínio de tempo. |
| **Algoritmos** | NA |

# 5. Publicações de Resultados

Como resultado deste mapeamento sistemático apresenta-se a tabela 10, a qual reflete o objetivo descrito na tabela e através de sua analise pode-se responder a questão de pesquisa proposta:

- Como é possível definir a posição em um ambiente INDOOR?

**Table 10 Resultados**

| | Técnicas | | | | | | | | | | | | | Abordagens/ Métodos/ Algoritmos | | | | | | | | | | | | | |
|---|---|---|---|---|---|---|---|---|---|---|---|---|---|---|---|---|---|---|---|---|---|---|---|---|---|---|---|
| Artigos | RSSI-based | Calibration points (CPS) | WIFI | RFID | 2D laser-range scanners | Pyroelectric Infrared (PIR) | Radio Frequency | Bluetooth | GPS-based | Digital fractional Integration | Smartphones's Camera | Microfones signals(audio) | Embedded accelerometer | Support Vector Machine | K-NN algorithm | Simplex algorithm | RSSI Fingerprinting | RSSI Triangulation | Fuzzy Weighted Average | Compressive-Sensing model. | Time of Arrival(TOA) | Google Maps Indoor (GMI) | Smartphone Dead-reckoning. | Wifi time-of-flight (TOF) | Doppler frequency | Magnetic field | Acoustic ranging |
| 1 | X | | | | | | | | | | | | | | | | X | | | | | | | | | | |
| 2 | | | X | | | | | | | | X | | | | | | | | | | | | | | | | |
| 3 | X | | X | | | | | | | | | | | | | | | | | | | | | | | | |
| 4 | X | | | X | | | | | | | | X | | | | | | | | | | | | | | | |
| 5 | | | | X | | | | | | | | | | | | | | | | | | | | | | | |
| 6 | | | X | | | | | | | | | | X | | | | | | | | | | | | | | |

| | Técnicas | | | | | | | | | | | | | Abordagens/ Métodos/ Algoritmos | | | | | | | | | | | | | |
|---|---|---|---|---|---|---|---|---|---|---|---|---|---|---|---|---|---|---|---|---|---|---|---|---|---|---|---|
| **Artigos** | RSSI-based | Calibration points (CPS) | WIFI | RFID | 2D laser-range scanners | Pyroelectric Infrared (PIR) | Radio Frequency | Bluetooth | GPS-based | Digital fractional Integration | Smartphones's Camera | Microfones signals(audio) | Embedded accelerometer | Support Vector Machine | K-NN algorithm | Simplex algorithm | RSSI Fingerprinting | RSSI Triangulation | Fuzzy Weighted Average | Compressive-Sensing model. | Time of Arrival(TOA) | Google Maps Indoor (GMI) | Smartphone Dead-reckoning. | Wifi time-of-flight (TOF) | Doppler frequency | Magnetic field | Acoustic ranging |
| **7** | X | | X | | | | | | | | | | | X | | | | | | | | | | | | | |
| **8** | X | | | X | | | | | | | | | | | X | X | | | | | | | | | | | |
| **9** | X | | | | | | | | | | | | | | X | | X | X | | | | | | | | | |
| **10** | | | | | X | | | | | | | | | | | | | | | | | | | | | | |
| **11** | | X | X | | | | | | | | | | | | | | | | | | | | | | | | |
| **12** | X | | X | | | | | | | | | | | | | | | | X | | | | | | | | |
| **13** | X | | X | | | | | | | | | | | | | | | | | X | | | | | | | |
| **14** | X | | | | | | | | | | | | | | | | | | | | X | | | | | | |
| **15** | X | | X | | | | | | | | | | | | | | X | | | | | | | | | | |
| **16** | X | | X | | | | | | | | | | | | | | X | | | | | | | | | | |
| **17** | X | | | | | X | X | | | | | | | | | | | | | | | | | | | | |
| **18** | | | X | | | | | | | | | | | | | | | | | | | X | | | | | |
| **19** | | | X | | | | | | | | | | | | | | X | | | | | | | | | | |
| **20** | | | X | | | | | | | | | | X | | | | | | | | | | X | | | | |

| | Técnicas | | | | | | | | | | | | | Abordagens/ Métodos/ Algoritmos | | | | | | | | | | | | | |
|---|---|---|---|---|---|---|---|---|---|---|---|---|---|---|---|---|---|---|---|---|---|---|---|---|---|---|---|
| **Artigos** | RSSI-based | Calibration points (CPS) | WIFI | RFID | 2D laser-range scanners | Pyroelectric Infrared (PIR) | Radio Frequency | Bluetooth | GPS-based | Digital fractional Integration | Smartphones's Camera | Microfones signals(audio) | Embedded accelerometer | Support Vector Machine | K-NN algorithm | Simplex algorithm | RSSI Fingerprinting | RSSI Triangulation | Fuzzy Weighted Average | Compressive-Sensing model. | Time of Arrival(TOA) | Google Maps Indoor (GMI) | Smartphone Dead-reckoning. | Wifi time-of-flight (TOF) | Doppler frequency | Magnetic field | Acoustic ranging |
| **21** | | | X | | | | | | | | | | | | | | | | | | | | | | | | |
| **22** | | | X | | | | | | | | | | X | | | | | | | | | | X | | | X | |
| **23** | X | | X | | | | X | | | | | | | | | | | | | | | | | | | | |
| **24** | | | X | | | | | | | | | | | | | | | | | | | | | X | | | |
| **25** | X | | | | | | | | | | | | | | | | | | | | | | | | | | |
| **26** | | | X | | | | | | | | | | | | | | X | | | | | | | | | | |
| **27** | X | | | | | | X | | | | | | | | | | | | | | | | | | | | |
| **28** | | | | X | | | | | | | | | | | | | | | | | | | | | X | | |
| **29** | | | X | | | | X | | | | | | | | | | | | | | | | | | | X | |
| **30** | | | X | | | | | | | | | | | | | | | | | | | | | | | | |
| **31** | | | X | | | | | | | | | | | | | | | | | | | | | | | | |
| **32** | X | | | | | | | | | | | | | X | | | X | | | | | | X | | | | |
| **33** | | | | | | | | | X | | | | | | | | | | | | | | | | | | |
| **34** | | | X | | | | | X | | | | | | | | | | | | | | | | | | X | |

| Artigos | Técnicas | | | | | | | | | | | | | Abordagens/ Métodos/ Algoritmos | | | | | | | | | | | | | |
|---|---|---|---|---|---|---|---|---|---|---|---|---|---|---|---|---|---|---|---|---|---|---|---|---|---|---|---|
| | RSSI-based | Calibration points (CPS) | WIFI | RFID | 2D laser-range scanners | Pyroelectric Infrared (PIR) | Radio Frequency | Bluetooth | GPS-based | Digital fractional Integration | Smartphones's Camera | Microfones signals(audio) | Embedded accelerometer | Support Vector Machine | K-NN algorithm | Simplex algorithm | RSSI Fingerprinting | RSSI Triangulation | Fuzzy Weighted Average | Compressive-Sensing model. | Time of Arrival(TOA) | Google Maps Indoor (GMI) | Smartphone Dead-reckoning. | Wifi time-of-flight (TOF) | Doppler frequency | Magnetic field | Acoustic ranging |
| **35** | | | | | | | | | | | | | X | | | | | | | | | | X | | | | |
| **36** | | | | | | | | | | X | | | | | | | | | | | | | | | | | |
| **37** | X | | | | | | | | | | | | | | | | | | | | | | | | | | |
| **38** | | | | X | | | | | | | | | | | X | | X | | | | | | | | | | |
| **39** | X | | X | | | | | | | | | | | | | | | | | | | | | | | | |
| **40** | X | | | | | | | | | | | | | X | | | | | | | | | | | | | |
| **41** | X | | X | | | | | | | | | | | | | | X | | | | | | | | | | |
| **42** | X | | X | | | | | | | | | | | | | | | | | | | | | | | | |
| **43** | X | | X | | | | | | | | | | | | X | | X | | | | | | | | | | |
| **44** | | | | | | | X | | | | | | | | | | | | | | | | | | | | X |
| **45** | | | | | | | | | | | | | X | | | | | | | | | | | | | | |
| **46** | X | | X | | | | | | | | | | | | | | | | | | | | | | | | |
| **47** | X | | | | | | | | | | | | | | | | | | | | | | | | | | |
| **48** | X | | X | | | | | | | | | | | | | | | | | | | | | | | | |

# 5 Referências Bibliográficas